\documentclass{emulateapj}

\usepackage{xspace}
\usepackage{graphics,graphicx}
\usepackage{natbib}
\usepackage{multirow}

\newcommand{\as}{\mbox{\ensuremath{.\!\!^{\prime\prime}}}}
\newcommand{\amm}{\mbox{\ensuremath{.\!\!^{\prime}}}}
\newcommand{\asn}{$^{\prime\prime}$\xspace}
\newcommand{\am}{$^{\prime}$\xspace}
\newcommand{\nH}{$N_{\rm H}$\xspace}
\newcommand{\PL}{$\Gamma$\xspace}
\newcommand{\Msun}{$M_{\odot}$\xspace}
\newcommand{\Cdof}{$C/dof$\xspace}
\newcommand{\chisq}{$\chi^2$\xspace}

\newcommand{\Chandra}{{\it Chandra}\xspace}
\newcommand{\HST}{{\it HST}\xspace}
\newcommand{\XMM}{{\it XMM-Newton}\xspace}
\newcommand{\lum}{erg s$^{-1}$\xspace}
\newcommand{\flux}{erg s$^{-1}$ cm$^{-2}$\xspace}
\newcommand{\Mv}{$M_V$\xspace}
\newcommand{\wavdetect}{\texttt{wavdetect}\xspace}
\newcommand{\AEx}{\texttt{AE}\xspace}
\newcommand{\pns}{{\it pns}\xspace}
\newcommand{\fluxratio}{flux$_{\rm max}$/flux$_{\min}$\xspace}
\newcommand{\dflux}{$\Delta$flux\xspace}
\newcommand{\lognlogs}{log$N$-log$S$\xspace}

\shorttitle{NGC~300 X-ray Point Source Catalog}
\shortauthors{Binder et al.}

\begin{document}

\title{The {\it Chandra} Local Volume Survey: The X-ray Point Source Catalog of NGC~300}
\author{B. Binder\altaffilmark{1}, 
B. F. Williams\altaffilmark{1}, 
M. Eracleous\altaffilmark{2},
T. J. Gaetz\altaffilmark{3},
P. P. Plucinsky\altaffilmark{3},
E. D. Skillman\altaffilmark{4},
J. J. Dalcanton\altaffilmark{1},
S. F. Anderson\altaffilmark{1},
D. R. Weisz\altaffilmark{1},
A. K. H. Kong\altaffilmark{5}
}
\altaffiltext{1}{University of Washington, Department of Astronomy, Box 351580, Seattle, WA 98195}
\altaffiltext{2}{Department of Astronomy \& Astrophysics, The Pennsylvania State University, 525 Davey Lab, University Park, PA 16802}
\altaffiltext{3}{Harvard-Smithsonian Center for Astrophysics, 60 Garden Street Cambridge, MA 02138, USA}
\altaffiltext{4}{University of Minnesota, Astronomy Department, 116 Church St. SE, Minneapolis, MN 55455}
\altaffiltext{5}{Institute of Astronomy and Department of Physics, National Tsing Hua University, Hsinchu 30013, Taiwan}

\begin{abstract}
We present the source catalog of a new \Chandra ACIS-I observation of NGC~300 obtained as part of the \Chandra Local Volume Survey. Our 63 ks exposure covers $\sim88$\% of the $D_{25}$ isophote ($R\approx6.3$ kpc) and yields a catalog of 95 X-ray point sources detected at high significance to a limiting unabsorbed 0.35-8 keV luminosity of $\sim10^{36}$ \lum. Sources were cross-correlated with a previous \XMM catalog, and we find 75 ``X-ray transient candidate'' sources that were detected by one observatory, but not the other. We derive an X-ray scale length of 1.7$\pm$0.2 kpc and a recent star formation rate of 0.12 \Msun yr$^{-1}$, in excellent agreement with optical observations. Deep, multi-color imaging from the {\it Hubble Space Telescope}, covering $\sim32$\% of our \Chandra field, was used to search for optical counterparts to the X-ray sources, and we have developed a new source classification scheme to determine which sources are likely X-ray binaries, supernova remnants, and background AGN candidates. Finally, we present the X-ray luminosity functions (XLFs) at different X-ray energies, and we find the total NGC~300 X-ray point source population to be consistent with other late-type galaxies hosting young stellar populations ($\lesssim50$ Myr). We find XLF of sources associated with older stellar populations has a steeper slope than the XLF of X-ray sources coinciding with young stellar populations, consistent with theoretical predictions.
\end{abstract}

\section{Introduction}
The X-ray emission from normal, non-active galaxies originates from a mixture of stellar sources and hot, diffuse gas. In spiral galaxies, this X-ray emission is dominated by X-ray binaries (XRBs) comprising either a neutron star (NS) or a black hole (BH) accreting from a stellar companion. The fast evolutionary timescale of massive stars ($\sim10^7$ years) makes X-ray emission from high-mass XRBs (HMXBs) nearly simultaneous with their formation \citep{Shty+07}, while the longer-lived low-mass XRB (LMXB) systems trace older underlying stellar populations \citep{Kong+02,Soria+02,Trudolyubov+02}; thus, the age and star formation history (SFH) of the host galaxy should correlate with the shape of the X-ray luminosity function (XLF) \citep{Grimm+03, Belczynski+04,Eracleous+06}. Dependencies of the XRB population properties on host galaxy metallicity may provide useful constraints on binary evolution and population synthesis models.

With its excellent angular resolution ($\sim$0\as5) and positional accuracy, the {\it Chandra X-ray Observatory} is the only X-ray telescope capable of separating the X-ray point source populations of nearby galaxies from diffuse emission. When combined with deep, optical {\it Hubble Space Telescope} (\HST) imaging, reliable source identification and optical counterpart identification may be carried out even at distances of a few Mpc. The \Chandra Local Volume Survey (CLVS, P.I. Benjamin Williams) is a deep, volume-limited X-ray survey of five nearby galaxies (NGC~55, NGC~300, NGC~404, NGC~2403, and NGC~4214) with matched \HST observations down to \Mv$\sim$0 \citep{Dalcanton+09}. When combined with the already well-studied disks of M~31 \citep{Kong+03} and M~33 \citep{Williams+08,Plucinsky+08,Tullmann+11}, these galaxies contain $\sim$99\% of the stellar mass and $\sim$90\% of the recent star formation out to a distance of $\sim$3.3 Mpc \citep{Tikhonov+05,Maiz+02,Freedman+Madore88}. Additionally, these galaxies span a representative sample of disk galaxies with a range of masses, metallicities, and morphologies. 

In this paper, we present the results of a new \Chandra observation of the nearby SA(s)d spiral galaxy NGC~300, located at a distance of 2.0 Mpc \citep{Dalcanton+09} in the Sculptor Group. NGC~300 is a near-optical twin of the Local Group galaxy M~33, and has a low foreground Galactic absorption of 4.09$\times10^{20}$ cm$^{-2}$ \citep{Kalberla+05} and is viewed at an inclination angle $i=42^{\circ}$; Table~\ref{galbasic} lists some basic properties of NGC~300, as well as M~33 and the Milky Way for direct comparison. 

\begin{table*}[ht]
\centering
\caption{Summary of NGC~300 Properties}
\begin{tabular}{cccc}
\hline \hline
Property		& NGC~300	& M~33	&	Milky Way \\
(1) & (2) & (3) & (4) \\
\hline
Distance			& 2.0 Mpc$^a$					& 800 kpc$^b$					& 8.5 kpc \\
Type				& SA(s)d$^c$					& SA(s)cd$^c$					& SBc \\
$M_B$			& -17.66$^a$					& -18.4$^d$					& -20.3$^h$ \\
Scale length (kpc)	& 1.3$^e$						& 1.4$^e$						& 3.00$\pm$0.22$^h$ \\
Circular velocity 	& 97 km s$^{-1}$$^f$			& 130 km s$^{-1}$$g$			& 239$\pm$5 km s$^{-1}$$^h$ \\
Estimated stellar mass	& $4.3\times10^{9}$ \Msun$^f$	& $4.5\times10^{9}$ \Msun$^g$	& $6.43\pm0.63\times10^{10}$ \Msun$^h$\\
\hline \hline
\label{galbasic}
\end{tabular}
\tablecomments{$^a$\cite{Dalcanton+09}; $^b$\cite{Williams+09}; $^c$NED; $^d$\cite{VilaCostas+92}; $^e$\cite{Munoz+07}; $^f$\cite{Puche+90}; $^g$\cite{Corbelli+00}; $^h$\cite{McMillan11}}
\end{table*}

The first X-ray population studies of NGC~300 were performed with {\it ROSAT} between 1991 and 1997 \citep{Read+01}, identifying 29 bright X-ray sources including NGC~300 X-1, a highly variable supersoft source, and other bright sources coincident with known supernova remnants (SNRs) and \ion{H}{2} regions. More recently, an \XMM survey of the the X-ray source population of NGC~300 was presented by \cite{Carpano+05} down to a limiting luminosity of $\sim3\times10^{35}$ \lum in the 0.3-6 keV band. A total of 163 sources were detected in the energy range of 0.3-6 keV, and the 86 sources falling within the $D_{25}$ optical disk were further characterized using hardness ratios, X-ray fluxes, and the available ground-based optical imaging. A brief summary of previous X-ray NGC~300 studies is provided in Table~\ref{previous_studies}.

\begin{table*}[ht]
\centering
\caption{Summary of Prior X-ray Surveys of NGC~300}
\begin{tabular}{cccc}
\hline \hline
Observatory	& Number of Detected 		& Limiting Unabsorbed Luminosity	& References 	\\
	 		& Sources (within $D_{25}$)	& ($10^{36}$ \lum)$^a$			& 			\\
(1) 			& (2) 					& (3) 						& (4) 		\\
\hline
{\it ROSAT} 	& 47 (29) 					& $\sim4.0$ ($\sim$7.7)			& \cite{Read+01} \\
\XMM 		& 163 (86)	 			& $\sim0.3$ ($\sim$0.4) 			& \cite{Carpano+05} \\
\Chandra		& 95 (77)					& 1.5							& this work		\\
\hline \hline
\end{tabular}
\tablecomments{$^a$The limiting unabsorbed luminosities correspond to the 0.1-2.4 keV and 0.3-6.0 keV energy ranges for {\it ROSAT} and \XMM, respectively. In parenthesis we provide expected 0.35-8 keV limiting unabsorbed luminosity assuming a power law with \PL=1.9 and \nH=$4.09\times10^{20}$ cm$^{-2}$.}
\label{previous_studies}
\end{table*}

Before the publication of the full \Chandra X-ray point source catalog, our collaboration employed this new observation of NGC~300 to study two individual objects of interest: the ``supernova impostor" SN 2010da, which we argued is consistent with a Be-X-ray binary (BeXB, comprised of a young, rapidly rotating B-type star with strong emission lines and a NS) origin \citep{Binder+11a}, and the BH + Wolf-Rayet binary NGC~300 X-1 \citep{Binder+11b}. These sources are included as part of the X-ray source catalog, but are not discussed in detail in this work.

The organization of this paper is as follows: in Section~\ref{obs} we provide the details of our new observation, data reduction procedure, and the technique used to generate the X-ray source catalog. An analysis of the X-ray source catalog, including comparisons to previous studies, hardness ratios, and the radial source distribution is given in Section~\ref{xray_results}. In Section~\ref{classifications}, we present a new X-ray/optical source classification scheme, identify candidate optical counterparts to our X-ray sources, and assign likely source classifications (HMXB, LMXB, SNR, or background AGN) to each source. The X-ray luminosity function, and the possible role of transient sources in determining the luminosity function shape, is discussed in Section~\ref{XLF}. The paper concludes with a summary and the main conclusions in Section~\ref{end}. 

\section{Observations and Data Reduction}\label{obs}
\subsection{Observations}\label{obsimg}
NGC~300 was observed on 2010 September 25 for 63 ks using ACIS-I during the \Chandra X-ray Observatory Cycle 12, observation ID 12238. Data reduction was carried out with CIAO v4.3 and CALDB v4.4.2 using standard reduction procedures. Images were created with bin sizes of 1, 2, 3, and 4 pixels for each of the following energy bands: 0.35-8, 0.35-1, 1-2, and 2-8 keV. Exposure maps were constructed using \texttt{mkinstmap} and \texttt{mkexpmap} in CIAO. Exposure-corrected images were generated using our exposure maps and instrument maps. The instrument maps were created assuming spectral weights appropriate for a power law spectrum typical of XRBs and background AGN (\PL = 1.9). The average foreground column density \nH was used.

\begin{figure*}
\centering
\includegraphics[width=0.7\linewidth,clip=true,trim=0 4.5cm 0 4.5cm]{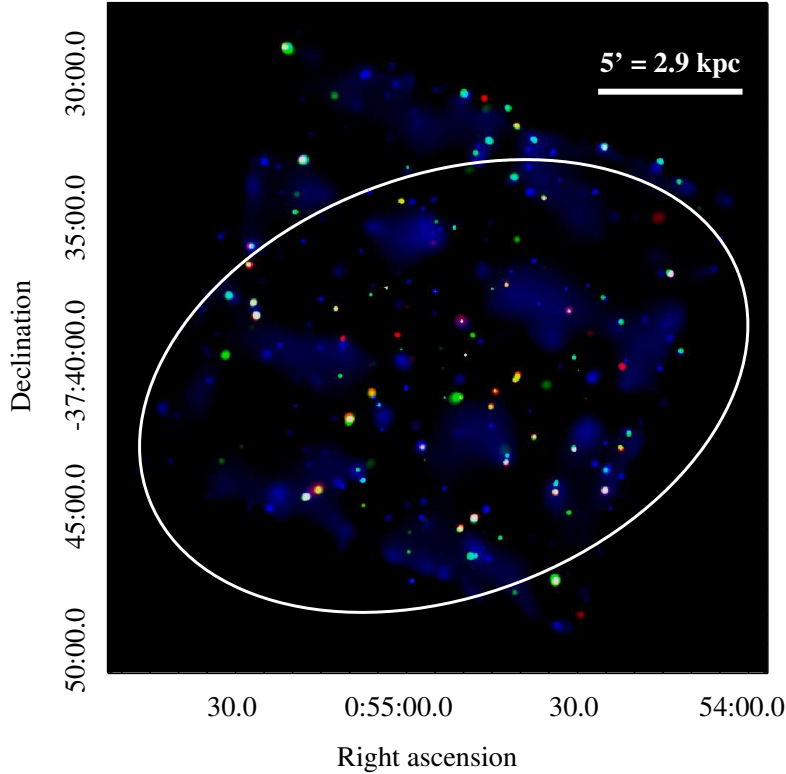}
\caption{An RGB rendering of our \Chandra exposure of NGC~300. Soft X-ray emission (0.35-1 keV) is shown in red, medium X-ray emission (1-2 keV) is shown in green, and hard X-ray emission (2-8 keV) is in blue. The white outline shows the $D_{25}$ isophote.}
\label{RGB}
\end{figure*}

Figure \ref{RGB} shows an RGB-rendered X-ray image of NGC~300, with numerous X-ray sources visible. Soft X-ray emission (0.35-1 keV) is shown in red, medium X-ray emission (1-2 keV) is shown in green, and hard X-ray emission (2-8 keV) is in blue. The $D_{25}$ isophote is superimposed in white.

Nine \HST fields were used to search for optical counterparts of each of our X-ray point sources. Three \HST Advanced Camera for Surveys \citep[ACS, ][]{Ford+98} observations of NGC~300 were taken as part of the ACS Nearby Galaxy Survey Treasury (ANGST, GO-10915) forming a radial strip from the center of the galaxy into the disk, hereafter referred to as `WIDE1-3.' Six additional archival ACS images scattered across the NGC~300 disk (GO-9492) are also included, and labeled as `NGC300-1-6.' Detailed descriptions of the \HST star catalogs, including photometric measurements and quality cuts, are given in \cite{Williams+09} and \cite{Dalcanton+09}, while a thorough discussion of the survey completeness and star formation histories (SFH) of NGC~300 are presented in \cite{Gogarten+10}.

\begin{figure*}
\centering
\includegraphics[width=0.7\linewidth,clip=true]{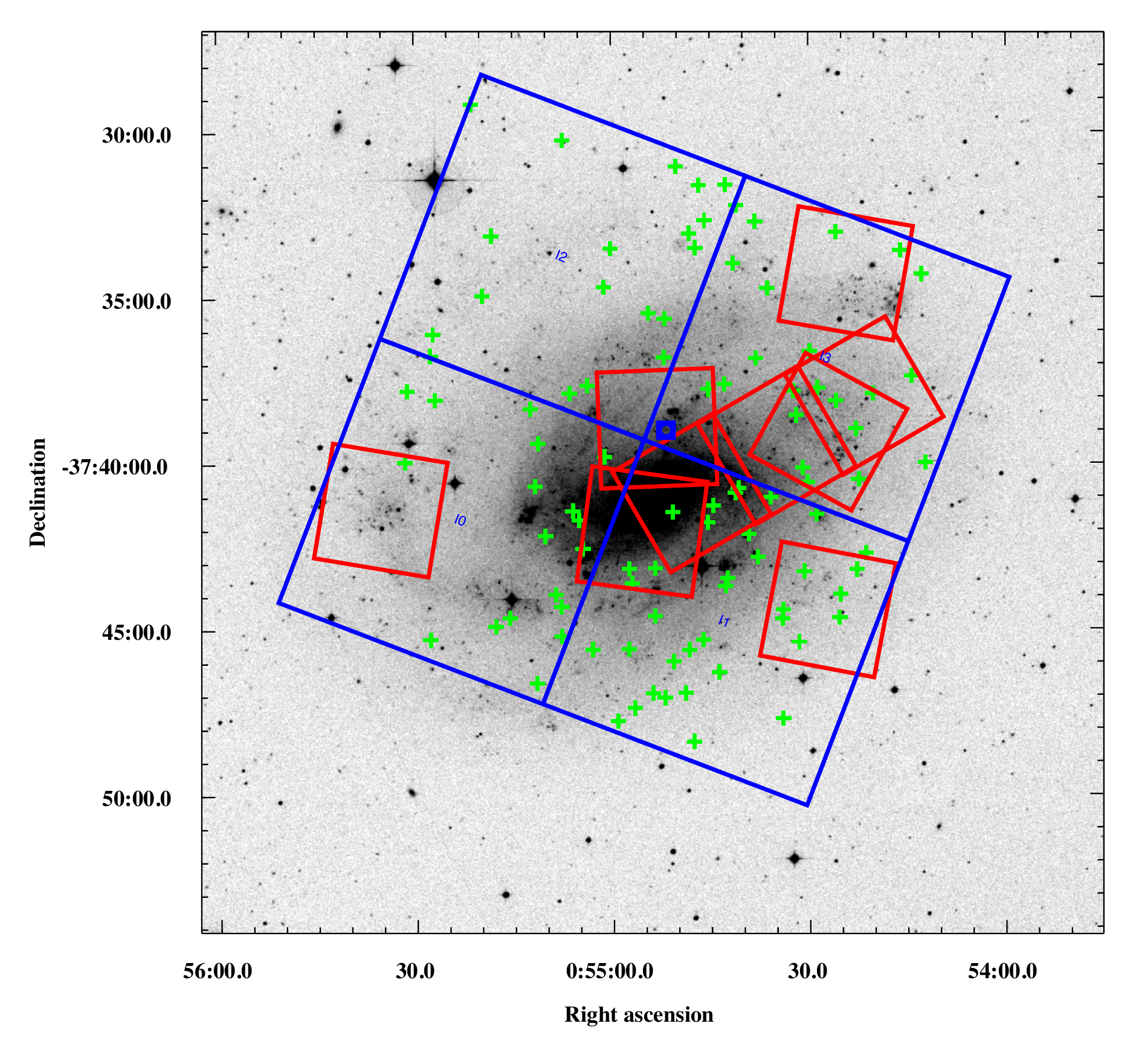}
\caption{A POSS image of NGC~300, with the \HST fields shown in red and our \Chandra field of view superimposed in blue. The locations of our X-ray point sources are shown by green crosses.}
\label{fields}
\end{figure*}

Figure~\ref{fields} shows a ground-based Palomar Observatory Sky Survey (POSS) image of NGC~300, with the locations of the \HST fields shown in red and our \Chandra field superimposed in blue. The locations of the \Chandra-detected X-ray point sources (see \S\ref{xray_results}) are shown by green crosses. 

\subsection{Image Alignment}
To classify optical counterparts to our \Chandra X-ray sources, we place both the \Chandra and \HST frames onto the International Celestial Reference System (ICRS) by finding matches between stars or background galaxies in the Two Micron All Sky Survey (2MASS) Point Source Catalog \citep{Skrutskie+06}. We performed relative astrometry by individually aligning each \HST and \Chandra field to the same reference image frame and coordinate system.

For our reference image, we retrieved a publicly-available ground-based $R$ band image \citep[see ][for more details]{Larsen+00} of NGC~300 from NED. Foreground 2MASS stars with ground-based $R$-band counterparts were identified, and we used the IRAF task \texttt{ccmap} to compute the plate solution and update the image header with corrected WCS information. By using the 2MASS stars as reference, the astrometric solution yielded root-mean-square ($r.m.s.$) residuals of 0\as00773 in right ascension and 0\as0225 in declination relative to the 2MASS astrometry for the ground-based $R$-band image.

We were next able to align each of our \HST frames and our \Chandra frame to the 2MASS-aligned ground-based $R$-band image. For each \HST field, we identified potential matches between the ground-based $R$-band image and the $F814W$ frames; we typically were able to match 3-8 stars per \HST frame. The plate solutions were again computed using \texttt{ccmap}, with $r.m.s.$ residuals typically less than a few hundredths of an arcsecond in both right ascension and declination. The \Chandra field was searched for X-ray sources that matched the positions of foreground stars or background galaxies in 2MASS. Five sources were identified, and we obtained $r.m.s.$ residuals of 0\as13 in right ascension and 0\as26 in declination. The total alignment error for each field was computed by summing in quadrature the $r.m.s$ residuals in the individual fields and the ground-based $R$-band image. The results of our 2MASS astrometry are summarized in Table~\ref{align}.

\begin{table*}[ht]
\centering
\caption{Alignment of \Chandra and \HST Images to 2MASS}
\begin{tabular}{cccccccc}
\hline \hline
Description	& Observation ID	& R.A. 	& Decl.    	& \# Sources Used	& $r.m.s.$,	& $r.m.s.$,	& \# of X-ray \\
			&		 	 	& (J2000) 	& (J2000) & For Alignment	& R. A.$^a$	& Decl.$^a$	& Sources \\
 (1)			& (2)				& (3)      	& (4)	 	& (5) 			& (6)			& (7)			& (8)		 \\
\hline
ground-based $R$ &  &  &  & 4 & 0\as00773 & 0\as0225 &  \\
\hline
\Chandra & 12238 & 00:54:53.05 & -37:38:42.1 &  5 & 0\as1332 (0\as1330) & 0\as2580 (0\as2570) & 95 \\
Wide-1	& 10915 	& 00:54:21.37 & -37:37:56.3 	& 5 & 0\as0518 (0\as0512) & 0\as0400 (0\as0327) & 5 \\
Wide-2	& 10915 	& 00:54:34.7 & -37:39:25.2 	& 4 & 0\as0171 (0\as0152) & 0\as0342 (0\as0258)  & 6 \\
Wide-3 	& 10915  & 00:54:47.7 & -37:40:51.6	& 4 & 0\as0171 (0\as0152) & 0\as0535 (0\as0485) & 5 \\
NGC~300-1$^b$ & 9492 & 00:55:35.66 & -37:41:25.2	& 3 & 0\as0340 (0\as0331) & 0\as0421 (0\as0356) & 1 \\
NGC~300-2 & 9492	& 00:54:52.96 & -37:38:59.8 	& 6 & 0\as0772 (0\as0768) & 0\as0435 (0\as0372) & 2 \\
NGC~300-3 & 9492	& 00:54:55.33 & -37:42:04.5 	& 4 & 0\as0179 (0\as0162) & 0\as0263 (0\as0137) & 5 \\
NGC~300-4 & 9492	& 00:54:24.42 & -37:34:20.8	& 5 & 0\as0528 (0\as0522) & 0\as0449 (0\as0388) & 2 \\  
NGC~300-5 & 9492	& 00:54:26.89 & -37:39:01.4	& 8 & 0\as0230 (0\as0217) & 0\as0441 (0\as0379) & 9 \\ 
NGC~300-6 & 9492	& 00:54:27.12 & -37:44:23.2 	& 4 & 0\as0253 (0\as0241) & 0\as0400 (0\as0331) & 7 \\
\hline \hline
\end{tabular}
\tablecomments{\newline
$^a$Alignment errors listed show the uncertainties in each individual field and the ground-based $R$-band image added in quadrature. The error in parentheses is the uncertainty reported by \texttt{ccmap} for the individual field.\newline
$^b$Used \texttt{rxyscale}.}
\label{align}
\end{table*}

\subsection{Source Catalog Creation}
We employed the same source detection strategy described by \cite{Tullmann+11} in the \Chandra ACIS Survey of M~33 (ChASeM33) source catalog. We present a summary of the approach here; the reader is referred to ChASeM33 work for further details. The CIAO task \wavdetect \citep{Freeman+02} was first used to create a list of potential source candidates. The positions of these potential sources was then used as input to \texttt{ACIS-Extract} \citep[\AEx;][]{Broos+10}, a source extraction and characterization tool. \AEx was used to determine various source properties (source and background count rates, detection significances, etc.) in multiple, user-defined energy bands. Additionally, \AEx extracts spectra (with appropriate response matrices) and examines source light curves for temporal variability \citep[see][for details]{Broos+10}. The Poisson probability of not being a source ({\it prob\_no\_source}; hereafter \pns) is provided by \AEx for each source (taking the local background uncertainty into account). We use these \pns values as our source significance threshold criteria when constructing the NGC~300 X-ray point source catalog.

Our initial source list was deliberately generated with many more sources than we anticipated being statistically significant. We then applied an iterative procedure to remove sources with increasing significance thresholds, until only a small number of false sources remained in the final source list. Each time the source list was filtered, the output was visually examined for potentially lost (real) sources. We used the same images described in Section~\ref{obsimg} using a variety of binning and energy ranges to create the initial source list: \wavdetect was run on each energy and bin combination, using {\it scale} sizes ranging from 1 to 16 in a power-of-two sequence.  To generate the initial list of candidate sources, we ran \wavdetect with the following parameters: the source significance ({\it sigthresh}) was set to 10$^{-4}$ (to catch all real sources and many false ones), the cleansing threshold ({\it bkgsigthresh}) was set to 10$^{-2}$ \citep{Freeman+02}, and we used maximum number of cleansing iterations  ({\it maxiter}) of 5. Default values were used for all other \wavdetect parameters. The resulting candidate lists were merged and culled of duplicate sources, resulting in an initial source list comprised of 799 candidate sources.

Low rates of false detections using the \AEx \pns threshold criteria result from values below $\sim10^{-6}$ \citep{Nandra+05,Georgakakis+08,Tullmann+11}. We adopt a value of \pns = $4\times10^{-6}$ as our criteria for selecting those sources that are highly probable to be genuine. However, due to the high number of source candidates in our initial list, source crowding posed a serious issue and made defining background regions problematic. To address this issue, the output from our first \AEx iteration was filtered with \pns = $10^{-2}$. The filtered source list contained revised source properties and improved background estimates. All sources failing to meet this initial \pns threshold were visually examined, and any obviously missed sources were added back to the candidate source list. The source list resulting from our first iteration contained 224 sources. We then ran \AEx again on the filtered source list, this time raising the \pns threshold to 10$^{-3}$. Source crowding was again reduced and spurious sources were filtered out, yielding an output list of 129 potential sources. Again, sources failing to meet our new \pns criteria were visually examined and returned to the source list if necessary.

\AEx was run a final time on the resulting source list, and the \pns value was computed in each of the following nine energy bands (in keV): 0.5-8, 0.5-2, 2-8, 0.5-1, 1-2, 2-4, 4-8, 0.35-1, and 0.35-8. To be included in the final NGC~300 catalog, a sources was required to have a \pns value less than $4\times10^{-6}$ in {\it any} of the nine energy bands; if only the 0.35-8 keV band were considered, $\sim$4\% of significant sources would have been lost. The final CLVS source catalog for NGC~300 contains 95 sources\footnote{Source catalogs are available in FITS format at: \url{http://www.astro.washington.edu/users/bbinder/CLVS/}}. Their positions and properties, such as detection significance (\pns value), net counts, and photon flux (in eight different energy bands) are listed in Tables~\ref{srclist}-\ref{fluxes}.

\begin{table*}[ht]
\centering
\caption{NGC~300 Source List}
\begin{tabular}{cccccccc}
\hline \hline
Source & Source ID & R.A. (J2000) & Decl. (J2000) & Positional   & Total exp. map      & $R_{\rm src}$ & $\theta$  \\
No.        &                    & ($^{\circ}$)    & ($^{\circ}$)      & Error (\asn) & value (s cm$^2$) & (sky pixel)        & (\am)       \\
(1)          & (2)              & (3)                  & (4)                    & (5)                 & (6)                          & (7)                      & (8)            \\
\hline
1 & 005411.98-373951.8 &    13.549937 &    -37.664408 &  0.70 &   1.589$\times10^7$ &  13.43 &  8.2 \\
2 & 005412.31-373359.6 &    13.551319 &    -37.566583 &  0.93 &   1.434$\times10^7$ &  16.66 &  9.3 \\
3 & 005413.98-373710.5 &    13.558268 &    -37.619588 &  0.36 &   1.588$\times10^7$ &  12.14 &  7.9 \\
4 & 005415.57-373316.4 &    13.564907 &    -37.554563 &  0.59 &   1.210$\times10^7$ &  15.25 &  9.2 \\
5 & 005419.92-373744.5 &    13.583001 &    -37.629032 &  0.61 &   1.711$\times10^7$ &   8.49 &  6.6 \\
\hline \hline
\end{tabular}
\tablecomments{Column 2: the source ID also contains the source coordinates (J2000.0). Column 5: the positional uncertainty is a simple error circle (in \asn) around R.A. and decl. Column 6: sum of mean exposure map values in the source region. Column 7: the average radius of the source extraction region (1 sky pixel = 0\as392). Column 8: off-axis angle.\newline
Only the first five entries are shown.}
\label{srclist}
\end{table*}

\begin{table*}[ht]
\centering
\caption{Logarithmic \pns Values for Different Energy Bands}
\begin{tabular}{ccccccccc}
\hline \hline
Source & log(\pns[1])	& log(\pns[2])	& log(\pns[3])	& log(\pns[4])	& log(\pns[5])	& log(\pns[6])	& log(\pns[7])	& log(\pns[8])                 \\
No.        & (0.5-8.0 keV) & (0.5-2.0 keV) & (2.0-8.0 keV) & (0.35-8.0 keV) & (0.35-1.1 keV) & (1.1-2.6 keV) & (2.6-8.0 keV) & (0.35-2.0 keV)    \\
(1)          & (2)                  & (3)                   & (4)                    & (5)                      & (6)                      & (7)                   & (8)                   & (9)                          \\
\hline
1        &  -9.32 &  -8.60 &  -3.16 &  -9.13 &  $<$-10 & $<$-10 &  -2.05 &  -2.05 \\
2        &  -4.54 &  -3.57 &  -2.17 &  -4.81 &  -0.42 &  -2.80 &  -2.77 &  -2.77 \\
3        & $<$-10 & $<$-10 & $<$-10 & $<$-10 & $<$-10 & $<$-10 &  -6.03 &  -6.03 \\
4        & $<$-10 & $<$-10 &  -6.94 & $<$-10 &  -3.46 & $<$-10 &  -4.76 &  -4.76 \\
5        &  -5.48 &  -4.58 &  -2.15 &  -6.04 &  -3.20 &  -3.49 &  -1.41 &  -1.41 \\
\hline \hline
\end{tabular}
\tablecomments{The (logarithmic) \pns value is the Poisson probability of not being a source. Log(\pns) values smaller than -10 have been replaced by $<$-10. All these cases are highly significant detections.\newline
Only the first five entries are shown.}
\label{pns}
\end{table*}

\begin{table*}[ht]
\centering
\caption{Total Net Counts in Different Energy Bands}
\begin{tabular}{ccccccccc}
\hline \hline
Source & {\it net\_cnts}[1]  & {\it net\_cnts}[2] & {\it net\_cnts}[3] & {\it net\_cnts}[4] & {\it net\_cnts}[5] & {\it net\_cnts}[6] & {\it net\_cnts}[7] & {\it net\_cnts}[8] \\
No.        & (0.5-8.0 keV)     & (0.5-2.0 keV)      & (2.0-8.0 keV)     & (0.35-8.0 keV)  & (0.35-1.1 keV)   & (1.1-2.6 keV)     & (2.6-8.0 keV)     & (0.35-2.0 keV)    \\
(1)          & (2)                       & (3)                        & (4)                       & (5)                       & (6)                        & (7)                       & (8)                       & (9)                          \\
\hline
1 &    22.9$^{+ 6.6}_{- 5.5}$ &    13.3$^{+ 5.0}_{- 3.8}$ &     9.6$^{+ 5.0}_{- 3.9}$ &    22.8$^{+ 6.6}_{- 5.5}$ &   $<1.1$ &    17.1$^{+ 5.5}_{- 4.3}$ &     6.4$^{+ 4.5}_{- 3.3}$ &    13.2$^{+ 5.0}_{- 3.8}$ \\
2 &    17.3$^{+ 6.5}_{- 5.4}$ &     8.3$^{+ 4.5}_{- 3.3}$ &     9.1$^{+ 5.4}_{- 4.3}$ &    18.2$^{+ 6.6}_{- 5.5}$ &  0.7$^{+ 2.7}_{- 1.3}$ &     7.5$^{+ 4.5}_{- 3.3}$ &     9.9$^{+ 5.3}_{- 4.1}$ &     9.1$^{+ 4.6}_{- 3.4}$ \\
3 &    70.9$^{+ 9.8}_{- 8.8}$ &    45.6$^{+ 7.9}_{- 6.8}$ &    25.3$^{+ 6.6}_{- 5.5}$ & 70.8$^{+ 9.8}_{- 8.8}$ & 16.5$^{+ 5.2}_{- 4.1}$ & 40.3$^{+ 7.6}_{- 6.5}$ &    14.0$^{+ 5.4}_{- 4.2}$ &    45.6$^{+ 7.9}_{- 6.8}$ \\
4 &    45.3$^{+ 8.5}_{- 7.4}$ &    25.5$^{+ 6.3}_{- 5.2}$ &    19.8$^{+ 6.3}_{- 5.2}$ &    45.1$^{+ 8.5}_{- 7.4}$ & 6.0$^{+ 3.8}_{- 2.6}$ & 24.6$^{+ 6.2}_{- 5.1}$ & 14.5$^{+ 5.7}_{- 4.6}$ &    25.3$^{+ 6.3}_{- 5.2}$ \\
5 &    11.2$^{+ 4.9}_{- 3.7}$ &     6.3$^{+ 3.8}_{- 2.6}$ &     4.9$^{+ 3.8}_{- 2.6}$ &    12.1$^{+ 5.0}_{- 3.8}$ & 3.7$^{+ 3.2}_{- 1.9}$ & 5.2$^{+ 3.6}_{- 2.4}$ & 3.2$^{+ 3.4}_{- 2.2}$ &     7.2$^{+ 4.0}_{- 2.8}$ \\
\hline \hline
\end{tabular}
\tablecomments{Only the first five entries are shown.}
\label{netcnts}
\end{table*}

\begin{table*}[ht]
\centering
\caption{Photon Flux (photons cm$^{-2}$ s) in Different Energy Bands}
\begin{tabular}{ccccccccc}
\hline \hline
Source & log({\it flux}[1])	& log({\it flux}[2])	& log({\it flux}[3])	& log({\it flux}[4])	& log({\it flux}[5])	& log({\it flux}[6])	& log({\it flux}[7])	& log({\it flux}[8])        \\
No.        & (0.5-8.0 keV) & (0.5-2.0 keV) & (2.0-8.0 keV) & (0.35-8.0 keV) & (0.35-1.1 keV) & (1.1-2.6 keV) & (2.6-8.0 keV) & (0.35-2.0 keV) \\
(1)          & (2)                   & (3)                   & (4)                   & (5)                      & (6)                      & (7)                   & (8)                   & (9)                       \\
\hline
1        & -5.75 & -6.11 & -6.09 & -5.74 &    & -6.07 & -6.25 & -6.07 \\
2        & -5.84 & -6.28 & -6.08 & -5.81 & -6.83 & -6.40 & -6.04 & -6.20 \\
3        & -5.26 & -5.57 & -5.67 & -5.25 & -5.54 & -5.69 & -5.91 & -5.54 \\
4        & -5.28 & -5.65 & -5.60 & -5.27 & -5.77 & -5.74 & -5.73 & -5.61 \\
5        & -6.07 & -6.45 & -6.39 & -6.03 & -6.23 & -6.60 & -6.57 & -6.35 \\
\hline \hline
\end{tabular}
\tablecomments{Only the first five entries are shown.}
\label{fluxes}
\end{table*}

\subsection{Sensitivity Maps}\label{sensmapsection}
\AEx provides an estimate of the photon flux ({\it flux2}) based on the net number of source counts ({\it net\_cnts}), the exposure time, and the mean ARF in the given energy band. To convert this photon flux into an energy flux, knowledge of the intrinsic X-ray spectrum is required -- however, less than 20\% of the NGC~300 X-ray sources have a sufficiently large number of counts that can be used to constrain a spectral model. 

To enable a uniform comparison between NGC~300 and M33, we adopt the same approach as employed by \cite{Tullmann+11}: we assume an absorbed power law spectrum, with \PL = 1.9 and \nH = 4.09$\times10^{20}$ cm$^{-2}$, to estimate the energy flux for each source. This model is hereafter referred to as the ``standard model,'' and appears to be reasonable for both XRBs associated with NGC~300 as well as for background AGN (especially at low flux levels). We note that the energy flux estimate for individual sources may be off by as much as a factor of two if the true spectrum is different from our assumed one, although the net effect of this assumption is not expected to systematically bias the \lognlogs relation to higher or lower fluxes.

The following analyses require, for each point in our \Chandra exposure, a sensitivity map which provides the energy flux level at which a source would be detectable. The CIAO task \texttt{lim\_sens} was used to compute limiting sensitivity maps from the background image, exposure map, and ACIS-I PSF image. Our 90\% complete limiting luminosities in the 0.35-8 keV, 0.5-2 keV, and 2-8 keV ranges are 1.5$\times10^{36}$ \lum, 8.8$\times10^{35}$ \lum, and 4.1$\times10^{36}$ \lum, respectively. Table~\ref{sens} summarizes the limiting luminosities for the 70\%, 90\%, and 95\% completeness levels for our total energy range, the soft band, and the hard band.

\begin{table*}[ht]
\centering
\caption{Limiting Luminosities for our NGC~300 Observation}
\begin{tabular}{cccc}
\hline \hline
Energy &  70\% & 90\% & 95\% \\
 (keV)   &	($10^{35}$ \lum) & ($10^{35}$ \lum)	& ($10^{35}$ \lum) \\
(1) & (2) & (3) & (4) \\
\hline
0.35-8	& 7.7 & 15 & 45 \\
0.5-2		& 4.5 & 8.8 & 27 \\
2-8		& 21 & 41 & 120 \\
\hline \hline
\end{tabular}
\label{sens}
\end{table*}

\section{The X-ray Source Catalog}\label{xray_results}
Various properties of individual X-ray sources, such as their temporal variability or spectral shape, can be used to constrain the physical origin of the X-ray emission. Additionally, bulk population properties (i.e., radial source distribution and the XLF) allow us to discriminate between those sources associated with NGC~300 and background AGN in a statistical sense. We expect the observable X-ray point source population of NGC~300 to be comprised of HMXBs, LMXBs, background AGN, and a few bright supernova remnants (SNRs). To match each point source to its most likely physical origin, we use our catalog to identify time-variable sources, analyze hardness ratios (for faint sources) and spectra (for sufficiently bright sources), and create XLFs. The additional matched \HST optical observations allow us to identify candidate optical counterparts and measure correlations between the X-ray emission and the underlying stellar age as a function of radius of NGC~300.

\subsection{Comparison with \XMM and Temporal Variability}\label{xmmcat}
Some classes of XRBs, as well as background AGN, exhibit long-term X-ray variability on timescales of months to years, while SNRs and some HMXBs will show persistent X-ray emission. To search for long-term variability, we compared our \Chandra X-ray catalog to the \XMM survey conducted by \cite{Carpano+05}, which contained 86 sources within the $D_{25}$ isophote of NGC~300. We consider sources with X-ray positional offsets $<$ 3\as5 to be matches.

To correct for the differences in energy ranges between \Chandra (0.35-8 keV) and \XMM (0.3-6 keV), we converted all \XMM fluxes into \Chandra fluxes assuming our standard model. Due to the differences in energy resolution, detector efficiencies, etc., we cannot compare \Chandra and \XMM hardness ratios to search for changes in spectral shape (see Section~\ref{hrsec} for further discussion of source hardness ratios). \cite{Tullmann+11} defined a variability threshold as $\eta$ = (flux$_{\rm max}$ - flux$_{\rm min}$)/\dflux, where the flux error \dflux is calculated using the Gehrels approximation \citep[appropriate for low count data;][]{Gehrels86}, to identify sources with potential long-term variability. Any source with $\eta\geq5$ is considered potentially variable; according to this criteria, 8 sources in the NGC~300 catalog exhibit long-term variability. Table~\ref{XMMcross} lists the temporal variability measurements for the 52 sources observed by both \Chandra and \XMM, with sources showing evidence of variability listed in boldface.

\begin{table*}[ht]
\centering
\caption{NGC~300 Cross Identifications with {\it XMM-Newton} Sources}
\begin{tabular}{cccccccc}
\hline \hline
CLVS 		&  {\it XMM-Newton} & Positional 	& $\xi$ & $\eta$ & \fluxratio  \\
Source No.        &  Source No. ID       & Offset (\asn)   &          &              &                  \\
(1)         		& (2)          			& (3)         		& (4)    & (5)      & (6)            \\
\hline
1 & 41 &  0.76 &  0.7958 &  0.21 &   1.41 \\
3 & 17 &  1.38 &  0.9152 &  0.79 &   2.49 \\
6 & 45 &  0.98 &  0.5160 &  0.44 &   1.69 \\
7 & 39 &  0.60 &  0.5590 &  2.79 &   4.30 \\
8 & 42 &  1.70 &  0.2267 &  0.14 &   1.26 \\
\hline \hline
\end{tabular}
\tablecomments{Columns 1 and 2 give the CLVS and {\it XMM-Newton} source numbers, respectively. Column 3: the difference (in arcseconds) between the CLVS and the {\it XMM-Newton} source positions. Sources with positional offsets less than 3\as5 are considered matches. Column 4 is the K-S probability of the source being constant within our \Chandra observation (short-term variability). Sources with K-S probabilities of $\xi\leq3.7\times10^{-3}$ are likely to show short-term variability and are shown in boldface. Column 5: $\eta$ is the variability index defined by \cite{Tullmann+11} and is sensitive to long-term variability; sources with variability indices of $\eta\geq5$ are likely to show long-term variability and are shown in boldface. Column 6: variability factor expressed as the ratio of \fluxratio.\newline
Only the first five entries are shown.}
\label{XMMcross}
\end{table*}

We found a total of 75 X-ray sources that were detected by either \Chandra or \XMM, but not both. For these sources, we use the detection limit of the observatory in which the source was not observed to set a lower-limit on the long-term source variability. X-ray transients (XRTs) observed in the Milky Way and other nearby galaxies originate in either LMXBs or HMXBs, but in all cases the peak X-ray luminosity exceeds the quiescent X-ray luminosity by more than an order of magnitude. We therefore define XRT candidates in NGC~300 as those sources whose maximum luminosity is at least an order of magnitude above the detection limit for the exposure in which they were not observed. The 25 sources meeting this criteria are listed in Table~\ref{transient}.

\begin{table*}[ht]
\centering
\caption{Candidate Transient X-ray Sources from \Chandra and \XMM}
\begin{tabular}{ccccc}
\hline \hline
Source No.$^a$ & R.A. (J2000) & Decl. (J2000) & $\eta^b$ & \fluxratio$^b$  \\
(1)         		& (2)          	& (3)       & (4) & (5)   \\
\hline
CXO-4	& 13.564907 &    -37.554563 & 16.11 & 21.76 \\
CXO-12 	& 13.605871 &    -37.546400 & 11.52 & 15.84 \\
CXO-24     & 13.641366 &    -37.796388 & 15.37 & 24.81 \\
CXO-29     & 13.657195 &    -37.542294 & 9.94 & 12.50 \\
CXO-33     & 13.669315 &    -37.534468 &  32.23 & 42.53 \\
CXO-44     & 13.689520 &    -37.542735 & 12.91 & 18.54 \\
CXO-52     & 13.707457 &    -37.515847 & 12.24 & 17.26 \\
CXO-73     & 13.770201 &    -37.695464 & 13.49 & 20.47 \\
CXO-76     & 13.779925 &    -37.504917 & 8.33 & 10.63 \\
CXO-87     & 13.825446 &    -37.553972 & 12.92 & 18.80 \\
CXO-89     & 13.837805 &    -37.488834 & 41.25 & 68.58 \\
XMM-2  & 13.795250 & -37.809556 &  13.80 & 403.83 \\
XMM-3  & 13.709708 & -37.647083 &  21.06 & 594.39 \\
XMM-4  & 13.795458 & -37.648417 & 11.15 & 85.26 \\  
XMM-5  & 13.669417 & -37.680250 &  5.39 & 57.76 \\  
XMM-8  & 13.835292 & -37.802833  &  3.86 & 35.67 \\  
XMM-10  & 13.835292 & -37.643944 &  6.35 & 60.56 \\  
XMM-16  & 13.706458 & -37.666861  &  5.55 & 51.68 \\  
XMM-18  & 13.700292 & -37.782611  &  5.00 & 43.24 \\  
XMM-23  & 13.583208 & -37.652472  &  4.15 & 30.07 \\  
XMM-25  & 13.926000 & -37.673250 &  5.91 & 28.49 \\  
XMM-43  & 13.722000 & -37.722000  &  3.27 & 18.90 \\  
XMM-52  & 13.566292 & -37.566292  &  2.92 & 12.68 \\  
XMM-53  & 13.807708 & -37.807708  &  3.27 & 12.51 \\  
XMM-59  & 13.849958 & -37.756694  &  3.28 & 12.29 \\
\hline \hline
\end{tabular}
\tablecomments{$^a$Sources labeled as `CXO' were detected in our \Chandra catalog only; sources labeled as `XMM' were detected in the \cite{Carpano+05} \XMM catalog only.\newline
$^b$The long-term variability index $\eta$ and the variability factor \fluxratio listed are lower limits. }
\label{transient}
\end{table*}

We additionally searched the \Chandra source catalog for short-term variability in the 0.35-8 keV band (i.e., variability that occurs on timescales less than the exposure time, $\sim$18 hours). We used a Kolmogorov-Smirnov (K-S) test to compare the cumulative photon arrival time distribution for each X-ray source to a uniform count rate model.  The K-S test returns a probability that both distributions were drawn from the same underlying distribution. One one source (18) has a K-S probability indicating rapid variability at the $3\sigma$ level.

\subsection{Hardness Ratios}\label{hrsec}
\begin{figure*}
\centering
\includegraphics[width=0.5\linewidth,clip=true,trim=2cm 12.5cm 2cm 2.5cm]{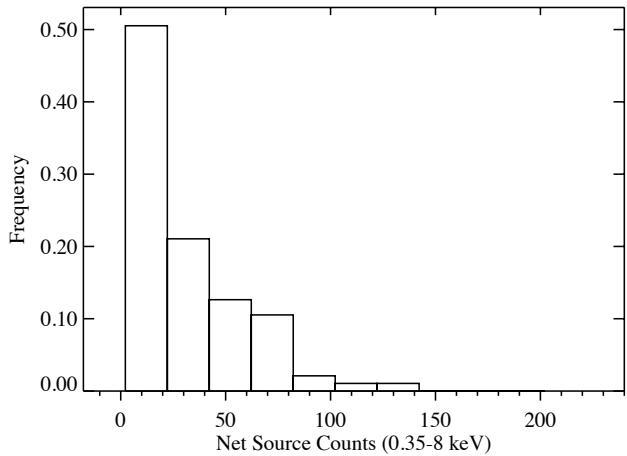}
\caption{Histogram of the net counts observed in the 0.35-8 keV band for our NGC~300 X-ray catalog, excluding source 80 (NGC~300 X-1). We find the majority of our sources have less than 50 counts, making spectral modeling for these sources unreliable.}
\label{count_hist}
\end{figure*}

The X-ray spectrum of a source can provide a key diagnostic for separating different sources. For example, XRBs and AGN are typically well described with power laws with photon indices \PL=1.5-2.0, while soft X-ray sources (such as SNRs) show spectra consistent with a thermal plasma of temperatures of $\sim0.3-0.8$ keV \citep{Long+10}. Still other sources, such as low luminosity AGN, may exhibit an X-ray spectrum consistent with bremsstrahlung emission \citep{Yi+98}. However, in order to constrain the fit parameters of these various models to believable values, a sufficiently high signal-to-noise is required. Even in the best case scenario, low signal-to-noise data (less than $\sim$50 counts) can only have a power law photon index constrained to $\sim$50\% accuracy. The majority of the NGC~300 X-ray sources in our catalog fall into this low-count category (see Figure~\ref{count_hist}).

To assess the X-ray spectral shape for low-count sources, a different approach must be used. Hardness ratios (HRs), also referred to as X-ray colors, are computed for various energy bands to measure the fraction of soft ($S$, 0.35-1.1 keV), medium ($M$, 1.1-2.6 keV), and hard ($H$, 2.6-8.0 keV) photons emitted by a source. A source comprised of a thermally emitted X-rays, such as a SNR, will show characteristically soft X-ray colors, while an obscured background AGN (having had most of its softer X-rays absorbed) will exhibit a significantly harder X-ray color.

We evaluate two HRs for each source using source and background counts determined by \AEx in the bands listed above. We define a ``soft'' color,

\begin{equation}
HR1 = \frac{M-S}{H+M+S},
\end{equation}

\noindent and a ``hard'' color,

\begin{equation}
HR2 = \frac{H-M}{H+M+S}.
\end{equation}

\noindent We use the Bayesian Estimation of Hardness Ratios \citep[{\it BEHR}, ][]{Park+06} in an approach described by \cite{Tullmann+11} \citep[see also ][]{Prestwich+09} to determine the hardness ratios of the NGC~300 X-ray sources. The {\it BEHR} code has the advantage of requiring source and background counts to be non-negative (net counts for low-count sources can be negative due to fluctuations in the source and background estimates, resulting in unphysical HRs); however, it cannot directly compute HRs in the form we require. We therefore use the {\it BEHR} code to generate 50,000 samples of the probability distribution of source fluxes for each energy band. To do this, we use as input to {\it BEHR} the following quantities from the \AEx output for each source: source and background counts, the \AEx ``backscale'' parameter (to account for the ratio of the source and background extraction areas and efficiencies), and a factor converting from counts to flux (i.e., the exposure time multiplied by the mean of the ARF over the extraction region in the appropriate energy band). We then run the {\it BEHR} code twice, once for $S$ and $M$ and again for $S$ and $H$.

We combine the results from our two {\it BEHR} runs to obtain 50,000 values of $HR1$ and $HR2$ for each source. Because the difference of two Poisson distributions does not follow Poisson statistics, we do not use the error estimates directly provided by these {\it BEHR} runs. Instead, the HR value is defined as the mean of the distribution, and the credible interval is evaluated based on the 68.2\% equal-tail estimates (i.e., 0.682/2 of the samples have values below the lower limit, and 0.682/2 of the samples have values above the upper limit). Our HRs for each source are listed in Table~\ref{HRtable}.

\begin{table*}[ht]
\centering
\caption{X-ray Source Hardness Ratios}
\begin{tabular}{cccc}
\hline \hline
Source No. & HR1 & HR2 & Class \\
(1) & (2) & (3) & (4) \\
\hline
1 &  0.72 $\pm$  0.25 & -0.71 $\pm$  0.32 & ABS \\ 
2 &  0.19 $\pm$  0.41 & -0.09 $\pm$  0.51 & XRB \\ 
3 &  0.36 $\pm$  0.13 & -0.56 $\pm$  0.12 & SOFT \\ 
4 &  0.50 $\pm$  0.15 & -0.47 $\pm$  0.21 & ABS \\ 
5 & -0.07 $\pm$  0.46 & -0.15 $\pm$  0.35 & XRB \\ 
\hline \hline
\end{tabular}
\tablecomments{Only the first five entries are shown.}
\label{HRtable}
\end{table*}

\begin{figure*}
\centering
\includegraphics[width=0.7\linewidth,clip=true,trim=2cm 12.5cm 2cm 2.5cm]{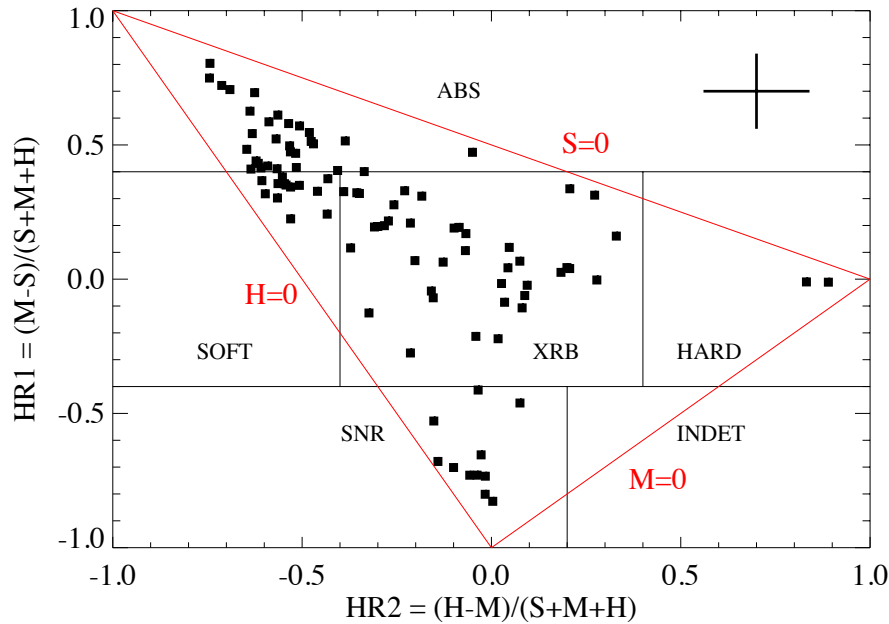}
\caption{X-ray color-color diagrams of all X-ray sources detected in NGC~300. The thick cross in the upper-right corner shows the typical size of the errors. The diagram has been broken into preliminary source identification regions, described in Table~\ref{color_class}. The regions are labeled as: `ABS' for absorbed sources, `XRB' for X-ray binaries, `SNR' for supernova remnants, 'SOFT' for indeterminate soft sources, `HARD' for indeterminate hard sources, and `INDET' for sources with an indeterminate spectral shape (although we do not observe any sources in this category). The red lines indicate the zero-count limits in the hard (`H'), medium (`M'), and soft (`S') bands.}
\label{HR}
\end{figure*}

To aid in the interpretation of our HR calculations, we employ an X-ray color-color classification scheme similar to that used in \cite{Kilgard+05}. Due to the differences in energy bands considered in this work, we modify the classification scheme slightly. We simulated spectra for thermal sources of varying temperature (0.1-2 keV) and power law sources (with \PL=0.4-3) with different levels of absorption and calculated HR1 and HR2 for each source to inform our X-ray color cuts. We define six categories: XRBs (which likely also contain a significant fraction of background AGN), SNRs, absorbed sources (`ABS'), indeterminate soft sources (`SOFT'), indeterminate hard sources (`HARD'), and sources with an indeterminate spectral type (labeled `INDET', although we do not observe any sources with hardness ratios fitting this criteria). Indeterminate sources are those whose X-ray spectra are ambiguous; for example, `HARD' sources may have exceptionally flat power law spectra, suffer extreme absorption, have multiple spectral components that cannot be distinguished due to our low number of source counts, or some combination of these effects. The color-color classification scheme is summarized in Table~\ref{color_class}, and a plot of the HRs for the NGC~300 X-ray source population is shown in Figure~\ref{HR}. 

\begin{table*}[ht]
\centering
\caption{Hardness Ratio Source Classification Scheme}
\begin{tabular}{ccc}
\hline \hline
Classification & Definition & \# Sources \\
(1) & (2) & (3) \\
\hline
X-ray binary (`XRB')				& $-0.4 < HR2 < 0.4$, $-0.4 < HR1 < 0.4$	& 39 \\
Absorbed source (`ABS')			& $HR1 > 0.4$					& 30 \\
Supernova remnant (`SNR')		& $HR2 < 0.4$, $HR1 < -0.4$		& 11 \\
Indeterminate hard source (`HARD')	& $HR2 > 0.4$, $-0.4 < HR1 < 0.4$	& 2 \\
Indeterminate soft source (`SOFT')	& $HR2 < -0.4$, $-0.4 < HR1 < 0.4$	& 13 \\
\hline \hline
\label{color_class}
\end{tabular}
\end{table*}

We find most of the NGC~300 sources fall within the `XRB' category (41\%), and an additional $\sim$12\% of sources are categorized as SNRs. This is not surprising - X-ray emission from non-active spiral galaxy with recent star formation is expected to be dominated by XRBs and bright SNRs. Nearly one-third of the NGC~300 X-ray sources show evidence for absorption beyond the Galactic column. Two sources exhibit unusually hard X-ray colors, and are likely background AGN whose soft X-ray emission has suffered a high degree of absorption. Thirteen sources show unusually soft X-ray colors. These may be very soft XRBs, sources with unusual absorption, or foreground objects not associated with NGC~300.

\subsection{X-ray Spectral Analysis} \label{spectra}
As discussed above, the X-ray spectrum of a source depends on the physical X-ray production mechanism. The value of the spectral fit parameters can reveal further clues to the nature of the X-ray emission. For example, XRBs containing a NS primary will show a hard power law with \PL$\sim1-1.5$, while BH primaries produce a significantly softer spectrum with \PL$\sim2-2.5$.

\AEx utilizes the \texttt{CIAO} tools \texttt{dmextract}, \texttt{mkacisrmf}, and \texttt{mkarf} when extracting spectra and generating response products. Due to the low number of counts found for most of the sources in our catalog, we fit spectral models to the unbinned spectra and use $C$-statistics in lieu of traditional \chisq statistics. The only X-ray source to have more than 200 net counts in the 0.35-8 keV band is source 80, the HMXB NGC~300 X-1. Detailed spectral fitting and timing analysis was carried out for this source in \cite{Binder+11b}.

The ungrouped spectra for 53 of the 95 sources were automatically fit in \AEx with one of four models: a power-law model, an \texttt{APEC} model \citep{Smith+01}, or either a power-law or thermal model with additional absorption. The \texttt{tbabs} model \citep{Wilms+00} was used to model the Galactic absorption \citep[\nH=4.09$\times10^{20}$ cm$^{-2}$; ][]{Kalberla+05}. To determine the best-fit model, we first rejected those models which predicted unreasonably high photon indices (\PL$>4$) or plasma temperatures ($kT>6$ keV). To assign a goodness-of-fit comparison for each of the models using $C$-statistics, we used \texttt{XSPEC}'s \texttt{goodness} command to perform 5,000 Monte Carlo calculations of the goodness-of-fit for each model. The command simulated spectra based on the model and computed the fraction of simulations with a fit statistic less than that found for the data -- if the observed spectrum was produced by that model, the \texttt{goodness} should be around 50\%. We therefore chose the best-fit model as the one for which the \texttt{goodness} command returned closest to 50\%. In some cases, the \texttt{goodness} command returned values near 50\% for multiple models, or did not return a reasonable value for any model. In these cases, we checked each spectrum for indications for emission features. In cases where evidence of thermal emission was present, the thermal model (with fixed \nH) was adopted; for all other sources, the power law with fixed \nH was chosen. The best-fit model and parameters for sources in NGC~300 with greater than 50 net counts are listed in Table~\ref{spectral_fits}.

\begin{table*}[ht]
\centering
\caption{Best-Fit Spectral Models for Sources With More Than 50 Counts}
\begin{tabular}{ccccccccccc}
\hline \hline
No. & Best-fit$^a$  & \nH$^b$ &	\PL &	$kT$ & 	Norm$^c$ & $dof^d$ & $C$ & $L_X$ (0.5-2.0 keV)$^e$ & $L_X$ (0.5-8.0 keV)$^f$	& ``goodness$^g$'' \\
       & Model            & (cm$^{-2}$) &       & 	(keV) &			&                  &                     & \multicolumn{2}{c}{(10$^{36}$ erg s$^{-1}$)}       	& (\%)         \\
 (1) & (2)		     & (3)  & (4)		&  (5)		& (6)		& (7)			& (8)			& (9) & (10)	\\
\hline
3 & pow &		  ...  				&  1.7$\pm0.3$ 		&  ...  				&  3.12$\times10^{-6}$ & 1016 &  576.19 &  3.3$\pm1.1$ 			&  8.5$^{+ 3.3}_{- 2.1}$ 	& 45.38	\\
11 & pow &	  ...  				&  1.5$\pm0.4$ 		&  ...  				&  2.39$\times10^{-6}$ & 1016 &  625.87 &  2.6$^\pm0.7$ 		&  8.0$^{+ 2.2}_{- 1.8}$ 	& 82.80	\\
23 & pow &	  0.7$^{+0.6}_{-0.4}$	&  2.2$^{+0.9}_{-0.7}$ 	&  ...  				&  1.10$\times10^{-5}$ & 1016 &  464.43 &  2.7$^{+1.1}_{-0.9}$ 	& 10.3$^{+ 4.2}_{- 3.3}$ 	& 32.70	\\
24 & apec &	  ... 	 			&  ...  				&  4.8$^{+20.7}_{- 2.7}$ 	&  1.22$\times10^{-5}$ & 1016 &  750.98 &  3.4$^{+ 0.9}_{- 3.1}$ 	&  7.9$^{+ 1.8}_{- 3.8}$ 	& 98.38	\\
25 & pow &	  $<$0.4			&  2.3$^{+1.4}_{-0.7}$  	&  ... 					&  4.42$\times10^{-6}$ & 1016 &  385.29 &  3.2$^{+1.9}_{-1.0}$ 	&  6.2$^{+ 3.8}_{- 1.9}$ 	& 55.28	\\
28 & pow &	  $<$0.4  			&  1.4$\pm0.2$ 		&  ...  				&  2.17$\times10^{-6}$ & 1016 &  438.60 &  2.4$\pm0.7$ 			&  8.2$^{+ 2.7}_{- 2.0}$ 	& 44.54	\\
33 & pow & 	0.2$^{+0.5}_{-0.2}$ 	&  1.8$^{+ 0.9}_{- 0.7}$ 	&  ...  				&  6.99$\times10^{-6}$ & 1015 &  679.83 &  7.4$^{+ 2.6}_{- 7.0}$ 	& 16.7$^{+ 3.3}_{-10.5}$ 	& 55.04	\\
35 & pow & 	1.1$^{+1.0}_{-0.5}$ 	&  1.3$^{+ 0.7}_{- 0.4}$ 	&  ...  				&  6.66$\times10^{-6}$ & 1015 &  706.75 &  7.4$^{+ 1.0}_{- 2.8}$ 	& 27.5$^{+ 6.1}_{-10.9}$ 	& 41.04	\\
39 & pow & 	 ...  				&  1.5$\pm0.4$ 		&  ...  				&  2.29$\times10^{-6}$ & 1016 &  590.38 &  2.5$^{+ 0.7}_{- 0.5}$ 	&  7.7$^{+ 2.5}_{- 1.8}$ 	& 64.26	\\
43 & apec & 	 ... 				&  ...  				&  0.4$\pm0.1$ 		&  1.92$\times10^{-5}$ & 1016 &  456.08 &  6.8$^{+ 1.2}_{- 1.7}$ 	&  6.8$^{+ 1.2}_{- 1.4}$ 	& 66.84	\\
50 & pow &  	 ...  				&  1.6$\pm0.3$ 		&  ...  				&  4.12$\times10^{-6}$ & 1016 &  631.44 &  4.4$^{+ 0.8}_{- 0.9}$ 	& 12.3$^{+ 3.0}_{- 2.1}$ 	& 91.80	\\
52 & pow & 	 0.9$^{+0.9}_{-0.6}$ 	&  2.5$^{+1.7}_{-1.1}$ 	&  ...  				&  8.22$\times10^{-6}$ & 1016 &  390.55 &  1.6$^{+1.1}_{-0.7}$ 	&  5.7$^{+ 3.9}_{- 2.5}$ 	& 55.88	\\
53 & pow &	 $<$0.3  			&  2.1$^{+1.0}_{-0.4}$ 	&  ... 					&  3.31$\times10^{-6}$ & 1016 &  380.30 &  3.5$^{+ 1.7}_{- 0.7}$ 	&  6.7$^{+ 3.2}_{- 1.3}$ 	& 52.93	\\
87 & apec & 	 $<0.2$  			&  ...  				&  2.6$^{+ 1.5}_{- 0.8}$ 	&  7.31$\times10^{-6}$ & 1016 &  371.56 &  2.7$^{+ 1.6}_{- 0.8}$ 	&  5.0$^{+ 2.9}_{- 1.5}$ 	& 72.86	\\
89 & pow & 	0.3$^{+0.4}_{-0.3}$ 	&  2.4$^{+ 0.6}_{- 0.5}$ 	&  ...  				&  1.62$\times10^{-5}$ & 1015 &  564.66 & 17.3$^{+ 4.0}_{-12.0}$ 	& 27.7$^{+ 7.0}_{-12.7}$ 	& 71.14	\\
95 & apec &	 ...  				&  ...  				&  0.6$^{+ 0.8}_{- 0.3}$ 	&  1.79$\times10^{-6}$ & 1016 &  541.01 &  0.7$^{+ 0.3}_{- 0.6}$ 	&  0.7$^{+ 0.5}_{- 0.6}$ 	& 98.80	\\
\hline \hline
\label{spectral_fits}
\end{tabular}
\tablecomments{\newline
$^a$ The best-fit model is either a \texttt{powerlaw} or \texttt{vapec} model, depending on which one has the lower \Cdof. \newline
$^b$ Intrinsic source absorption (in units of 10$^{22}$ cm$^{-2}$), if beyond the Galactic column was required.\newline
$^c$ Normalization constant of the fit, in units of photons keV$^{-1}$ cm$^{-2}$ s$^{-1}$ and cm$^{-5}$ for the power law and thermal plasma models, respectively. \newline
$^d$ Degrees of freedom. \newline
$^{e,f}$ Unabsorbed X-ray luminosities, assuming a distance to NGC~300 of 2.0 Mpc \citep{Dalcanton+09}. \newline
$^g$ Results of the \texttt{XSPEC} ``goodness'' command, run using 5000 realizations.
}
\end{table*}

\subsection{Radial Source Distribution}\label{radial_src}
We assign an inclination-corrected galactocentric distance to each X-ray sources assuming the following galaxy parameters for NGC~300, as in \cite{Gogarten+10}: $\alpha_0$ = 13\mbox{\ensuremath{.\!\!^{\circ}}}722833, $\delta_0$ = -37\mbox{\ensuremath{.\!\!^{\circ}}}684389 (galaxy center), $i=42^{\circ}$, (inclination), and $\theta=111^{\circ}$ \citep[position angle][]{Kim+04C}. The maximum radius we find for an X-ray source in our \Chandra field is $r=8.06$ kpc.

We divide the X-ray sources into radial bins based on their inclination-corrected distance from the center of the galaxy. The bins are spaced 0.9 kpc apart in radius so that we can directly compare the X-ray radial distribution to the stellar population analysis presented in \cite{Gogarten+10}. We use the cumulative \lognlogs distribution of \cite{Cappelluti+09} to estimate the expected contamination by background AGN, which should be nearly flat with galactocentric radius.

\begin{figure*}
\centering
\includegraphics[width=0.65\linewidth,clip=true,trim=2cm 12.5cm 2cm 3cm]{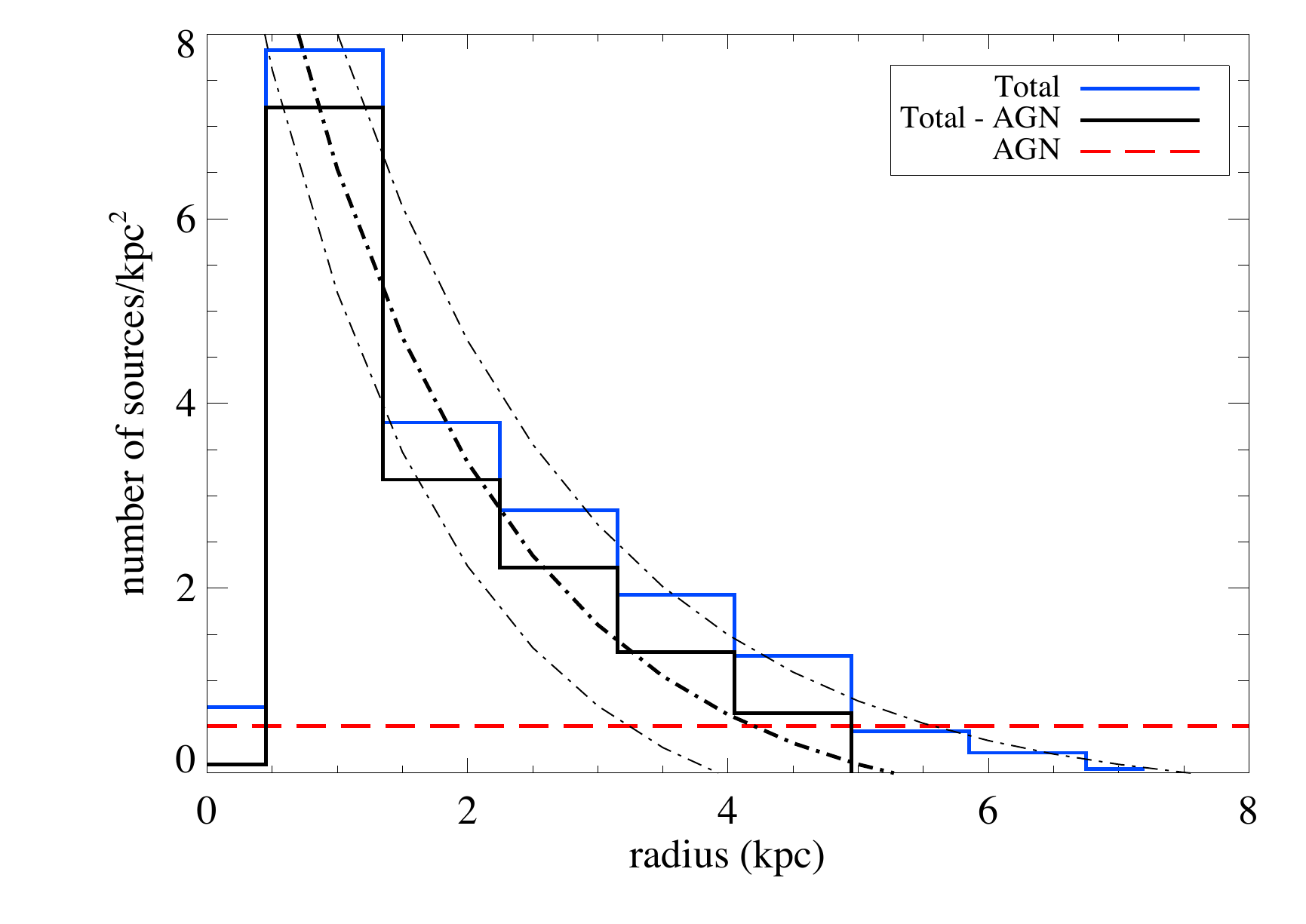}
\caption{The galactocentric source density profile of NGC~300 in the 0.35-8.0 keV band. The blue histogram shows the total number of sources per square kiloparsec. The red dashed line shows the anticipated number of background AGNs, while the black histogram shows the expected number of sources after correcting for background AGN.The thick dot-dashed line shows our exponential fit to the AGN-subtracted radial source distribution, with the lighter dot-dashed lines indicating the 1$\sigma$ uncertainty.}
\label{radialdist}
\end{figure*}

The radial X-ray source distribution is shown in Figure~\ref{radialdist}, and clearly shows that the number of X-ray sources per square kiloparsec decreases with galactocentric radius.  Since NGC~300 does not possess a substantial bulge, we fit the background-subtracted radial X-ray source distribution to a simple exponential profile, and find the best-fit relation to be

\begin{equation}
\mathcal{N}_X(r) = (12.8\pm3.0)e^{-r/(1.7\pm0.2)} - (0.57\pm0.85),
\end{equation}

\noindent where $\mathcal{N}_X(r)$ is the number of X-ray sources per kpc$^2$ expected at a given galactocentric radius, $r$ (in kpc). The exponential scale length is 1.7$\pm$0.2 kpc, similar to the 1.3$\pm$0.1 kpc scale length found by \cite{Gogarten+10} and in excellent agreement with estimates derived from the $K$-band surface brightness \citep{Munoz+07}. We additionally find the X-ray scale length to be consistent with multiwavelength measurements. The radial source distribution suggests that the X-ray sources in NGC~300 are the evolutionary end-points of a parent population of disk stars. 

For an exponential disk, we expect 90\% of stars to be inside of 4 kpc (6\amm9), suggesting that the majority of sources beyond this radius are not associated with the stellar disk of NGC~300. This suggests that 37 X-ray sources beyond 4 kpc are likely to be AGN (i.e., 40\% of the sample). This interpretation is consistent with the estimated number of background and hard sources identified in our hardness ratio analysis.

We use the source classifications determined by our hardness ratio analysis to investigate the differences in galactocentric radii among the different types of sources. The results are summarized in Table~\ref{mean_dist}. We find both the preliminary SNR and XRB candidates to have median radii less than 4 kpc, suggesting that many of these are indeed associated with NGC~300. However, XRB-classified sources span a wide range of galactocentric radii (out to 6.7 kpc). Given that background AGN are anticipated to dominate the X-ray sources at greater than $\sim$4 kpc, the `XRB' class likely suffers from AGN contamination. `SOFT' sources additionally span a wide range in radii, with a median distance of 4 kpc, suggesting that at least some of these sources are likely to be associated with NGC~300. Absorbed and indeterminate hard sources have median galactocentric radii greater than 4 kpc, implying that many of these sources are background or foreground objects.

\begin{table*}[ht]
\centering
\caption{Galactocentric Radii By Hardness Ratio IDs}
\begin{tabular}{cccc}
\hline \hline
Source classification & $r_{\rm min}$ (kpc) & $r_{\rm max}$ (kpc) & $r_{\rm av}$ (kpc) \\
(1) & (2) & (3) & (4) \\
\hline
SNR	& 0.4 & 6.0 & 2.5 \\
XRB & 1.3 & 6.7 & 3.4 \\
ABS & 1.2 & 8.1 & 4.1 \\
SOFT & 2.1 & 6.2 & 4.0 \\
HARD & 2.8 & 6.9 & 4.9 \\
\hline \hline
\end{tabular}
\label{mean_dist}
\end{table*}

\section{Source Classification}\label{classifications}
\subsection{Supernova Remnants}
In Section~\ref{hrsec}, we found that 11 X-ray sources in our catalog had hardness ratios consistent with those of SNRs. Of these, only one source (\#43) had a sufficient number of counts for spectral modeling to be performed. Source 43 was best-fit by a thermal plasma model with a temperature $kT$=0.4 keV, consistent with SNRs observed in M33 \citep{Long+10}. Taking this as our ``standard SNR'' model, we estimate the unabsorbed 0.5-2 keV luminosities for the remaining 10 SNR candidates and find a range in luminosities of $\sim5\times10^{35}$ lum to $\sim6\times10^{36}$ \lum, also consistent with observations of M33 SNRs.

\begin{figure*}
\centering
\begin{tabular}{ccc}
Source 7 & Source 15 & Sources 32, 34  \\
\includegraphics[width=0.27\linewidth,clip=true,trim=1cm 4cm 1cm 4cm]{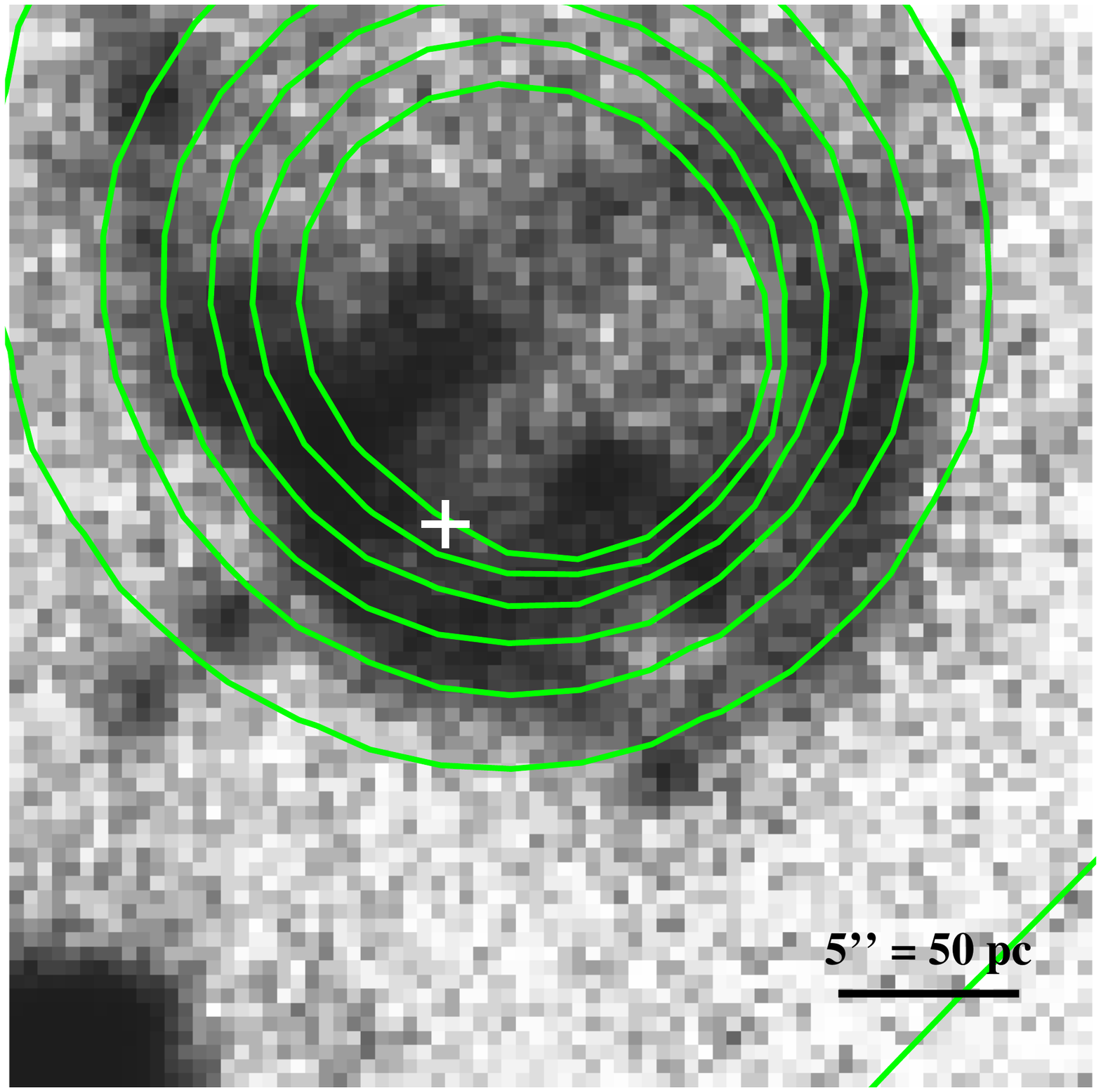} &
\includegraphics[width=0.27\linewidth,clip=true,trim=1cm 4cm 1cm 4cm]{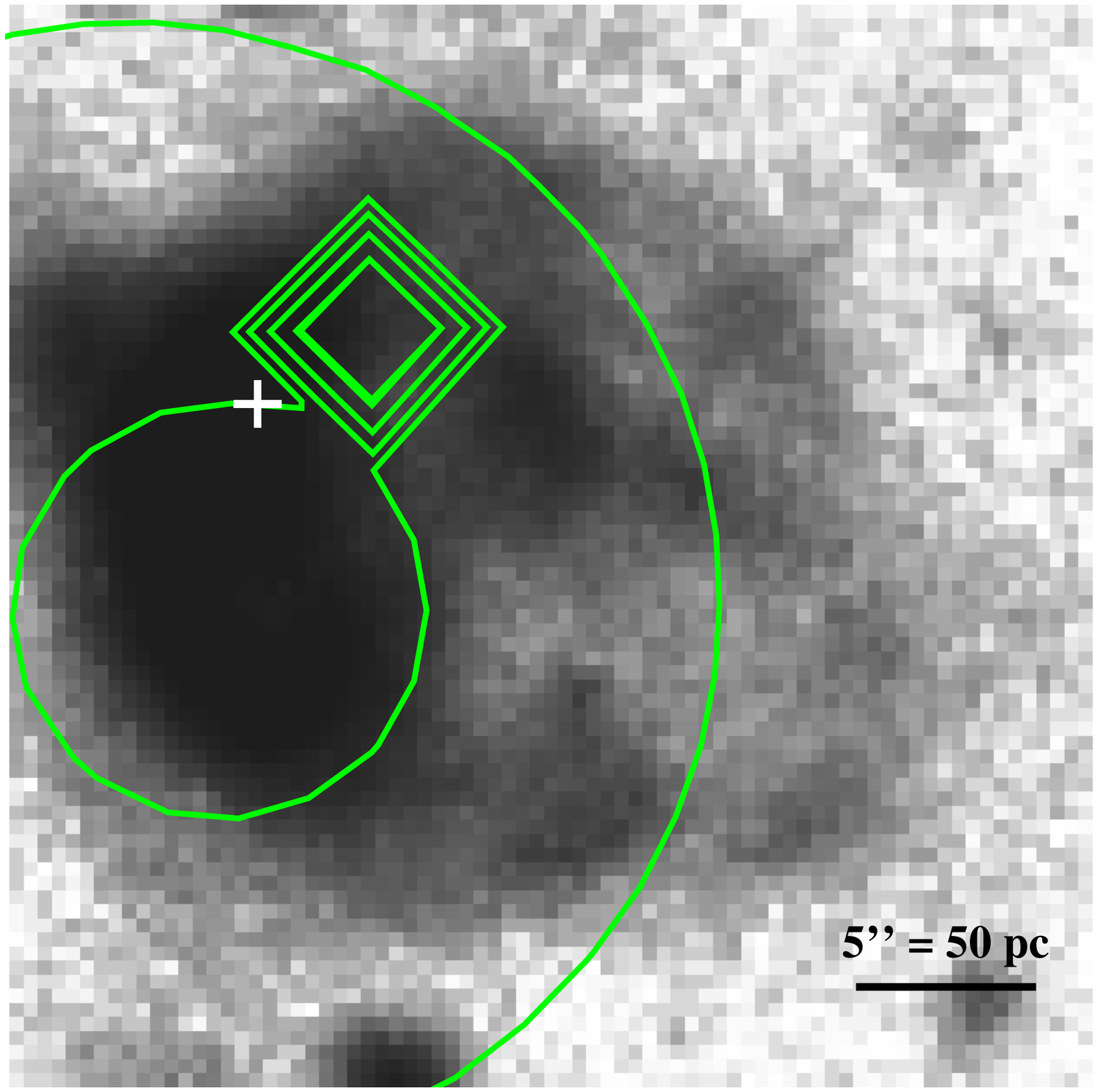} &
\includegraphics[width=0.27\linewidth,clip=true,trim=1cm 4cm 1cm 4cm]{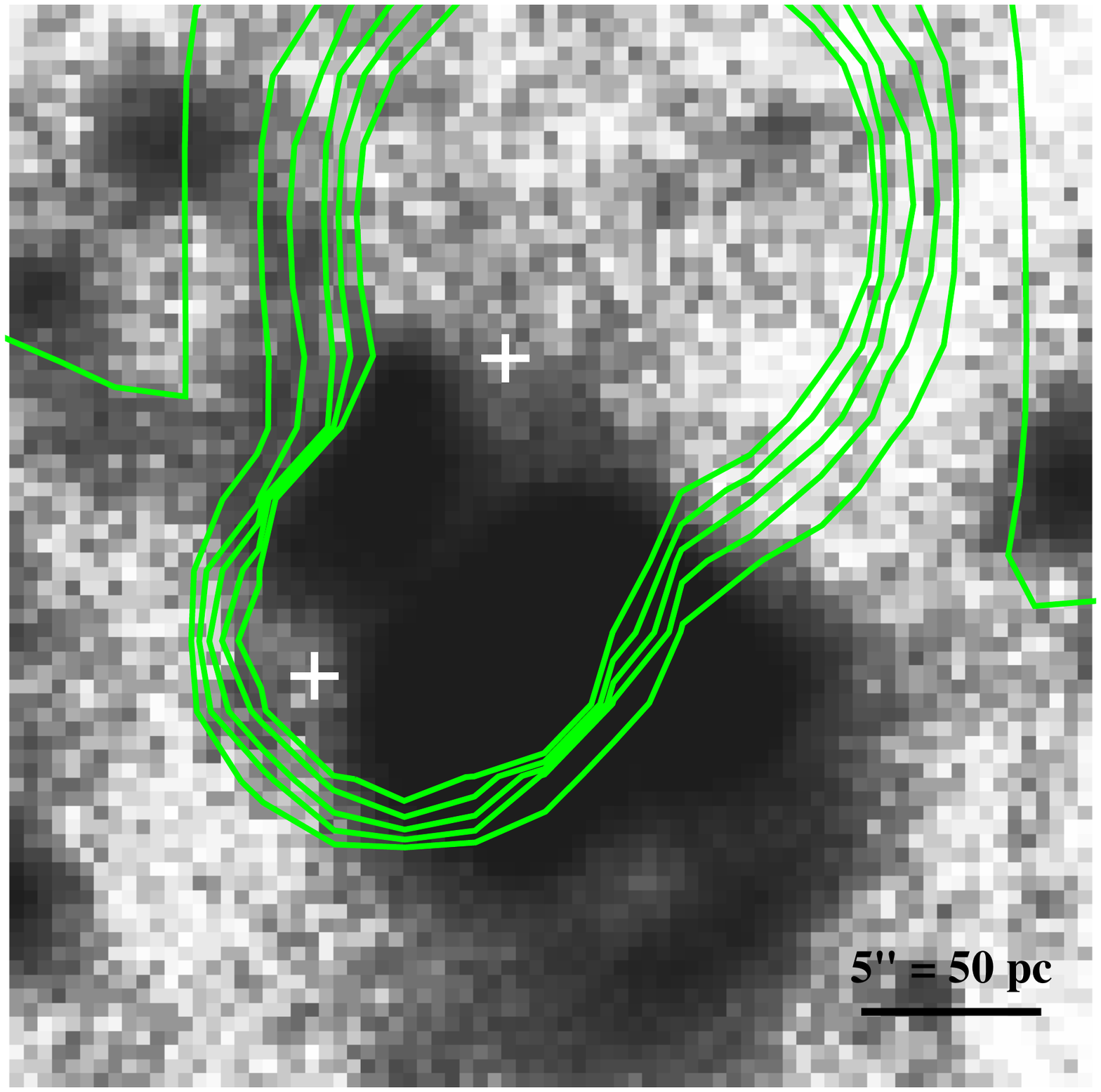} \\
Source 41 & Source 43 & Source 54 \\
\includegraphics[width=0.27\linewidth,clip=true,trim=1cm 4cm 1cm 4cm]{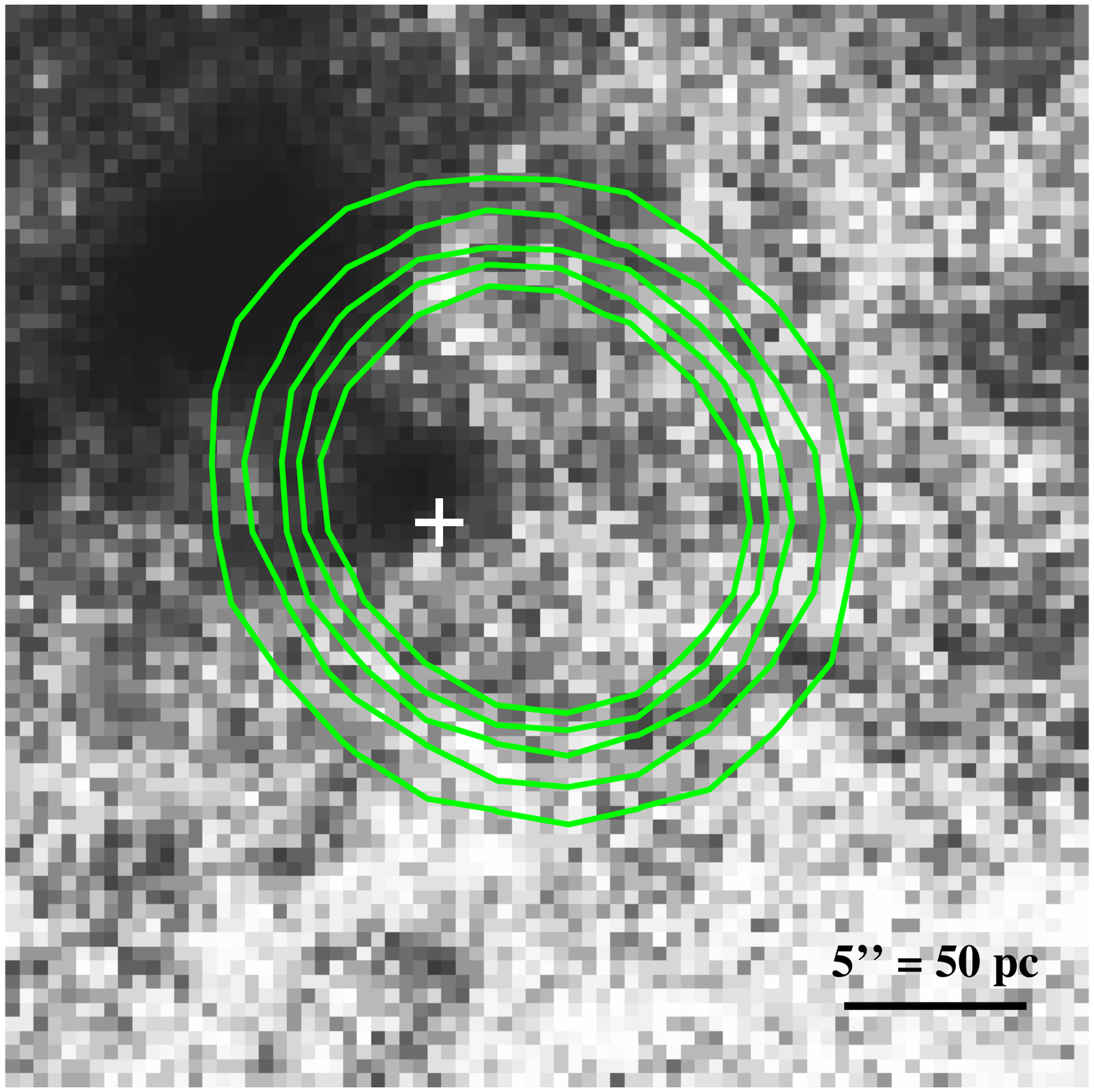} &
\includegraphics[width=0.27\linewidth,clip=true,trim=1cm 4cm 1cm 4cm]{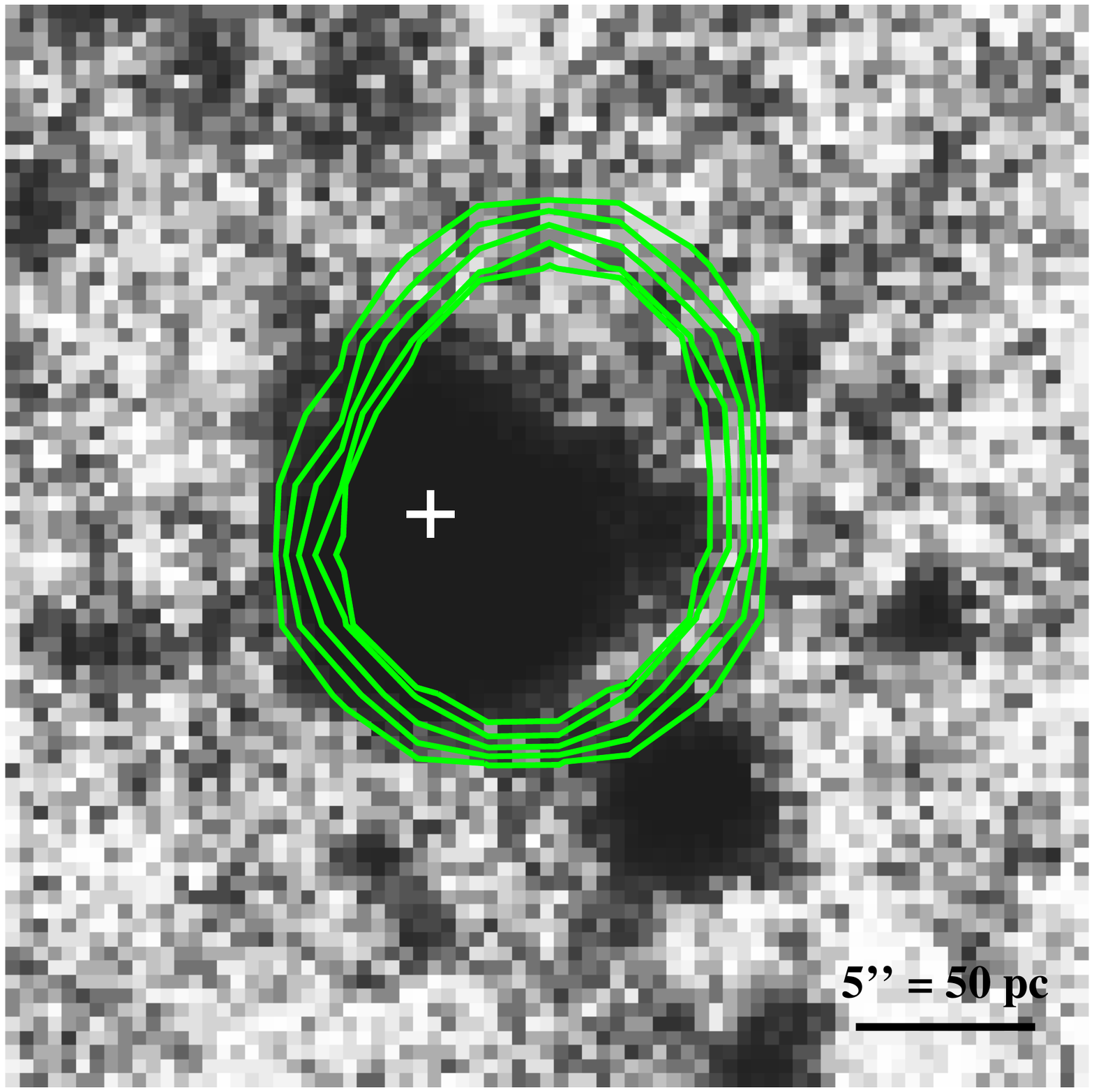} &
\includegraphics[width=0.27\linewidth,clip=true,trim=1cm 4cm 1cm 4cm]{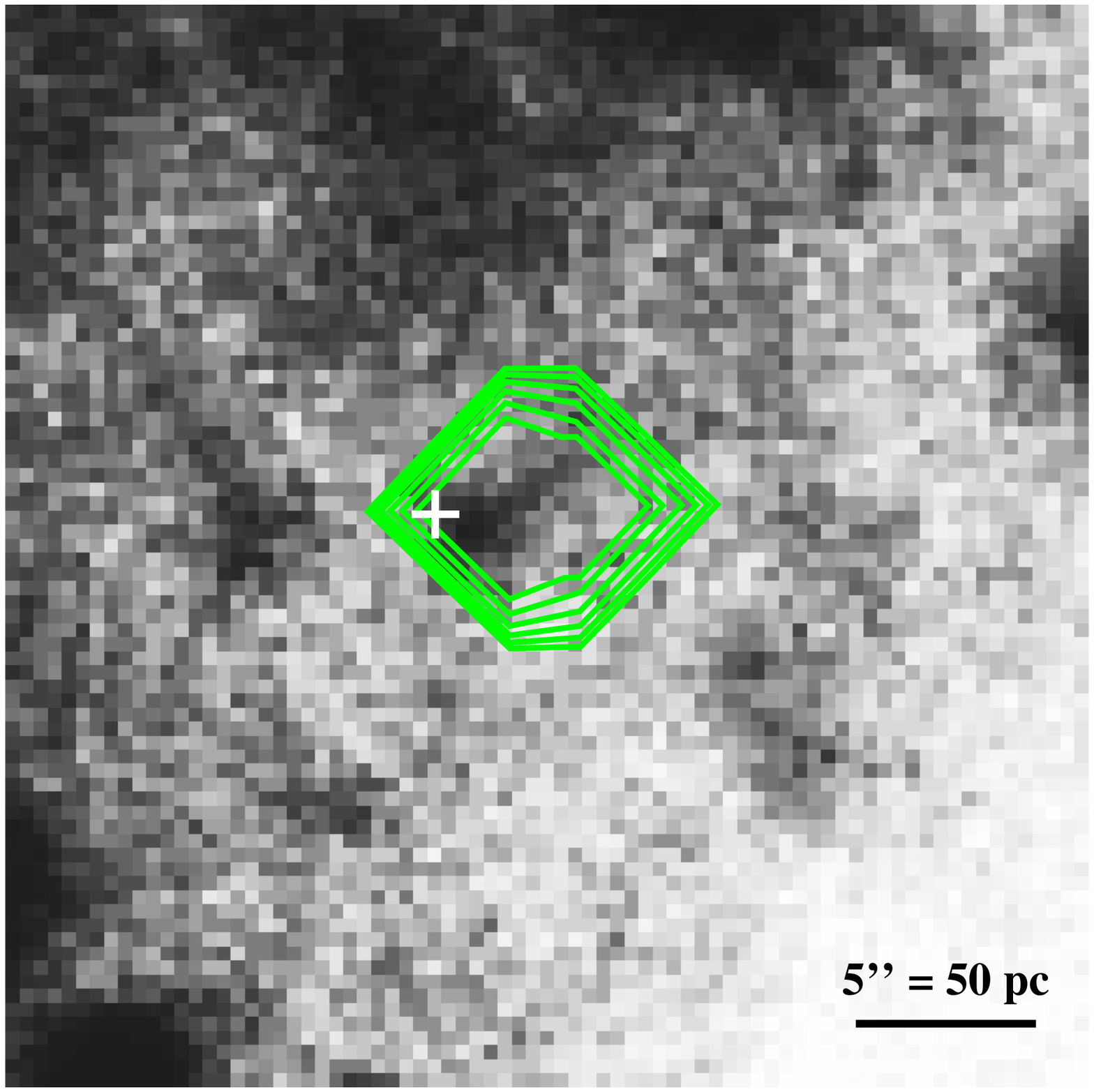} \\
Source 77 & Source 81 & Source 85 \\
\includegraphics[width=0.27\linewidth,clip=true,trim=1cm 4cm 1cm 4cm]{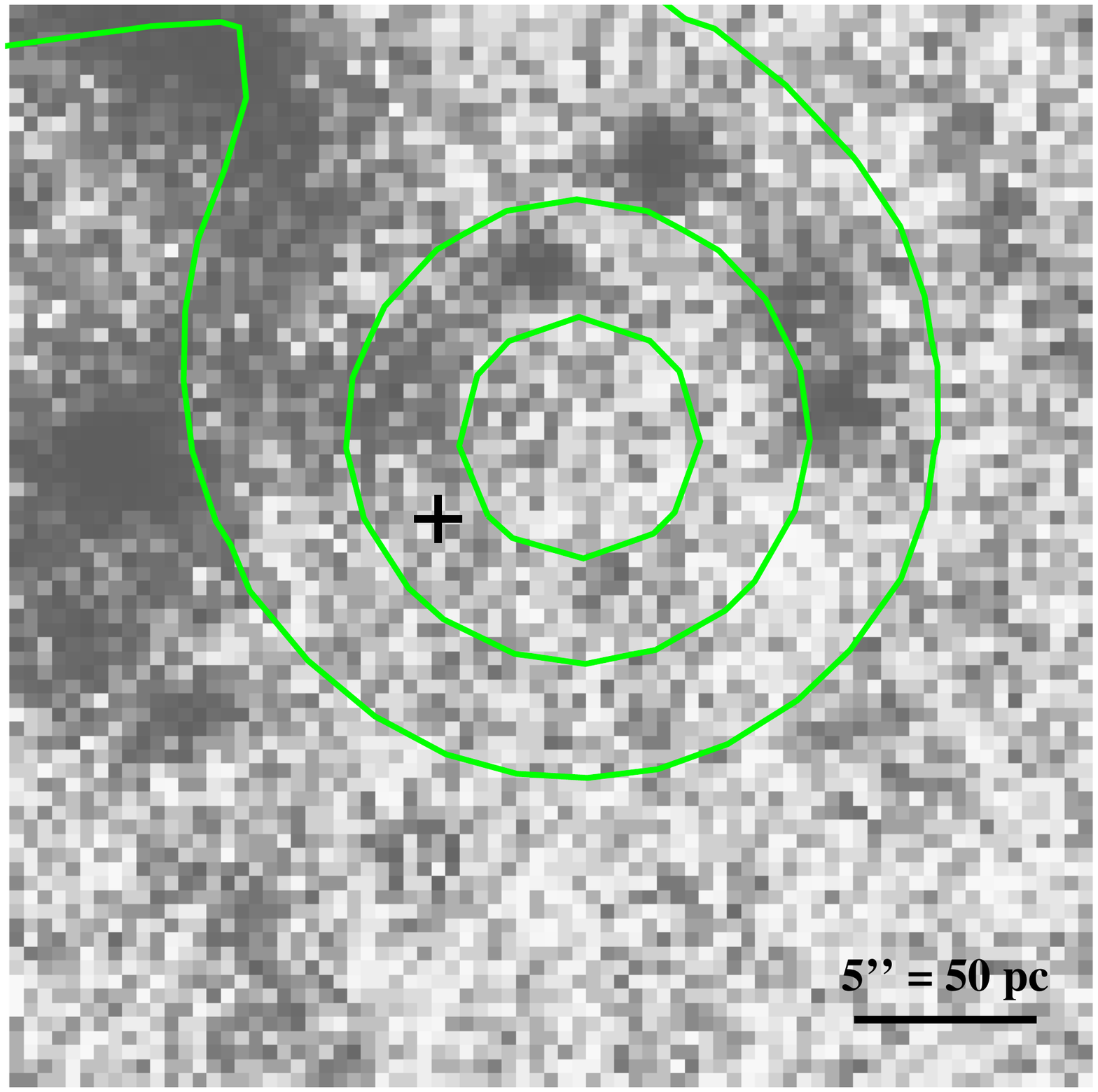} &
\includegraphics[width=0.27\linewidth,clip=true,trim=1cm 4cm 1cm 4cm]{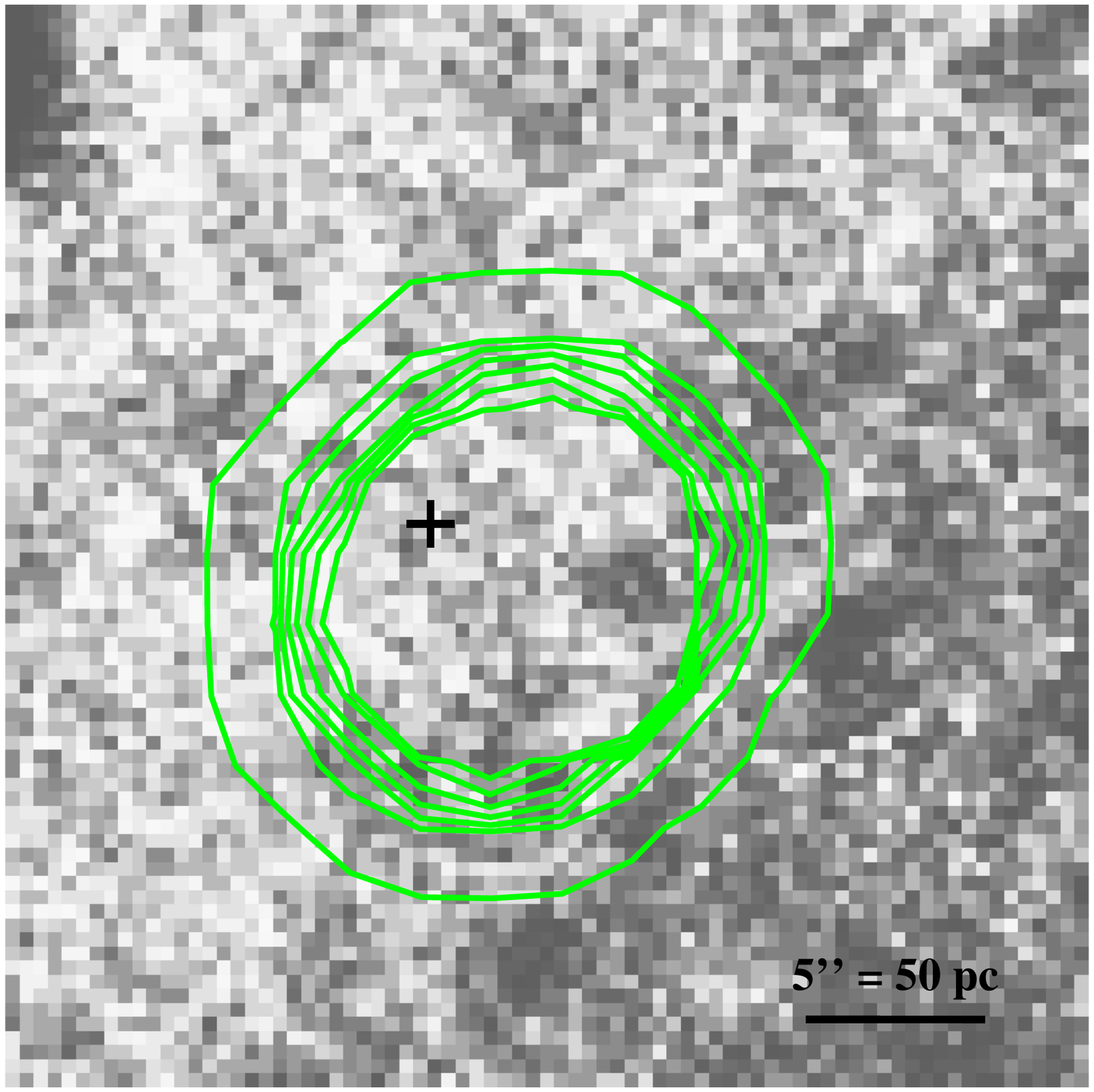} &
\includegraphics[width=0.27\linewidth,clip=true,trim=1cm 4cm 1cm 4cm]{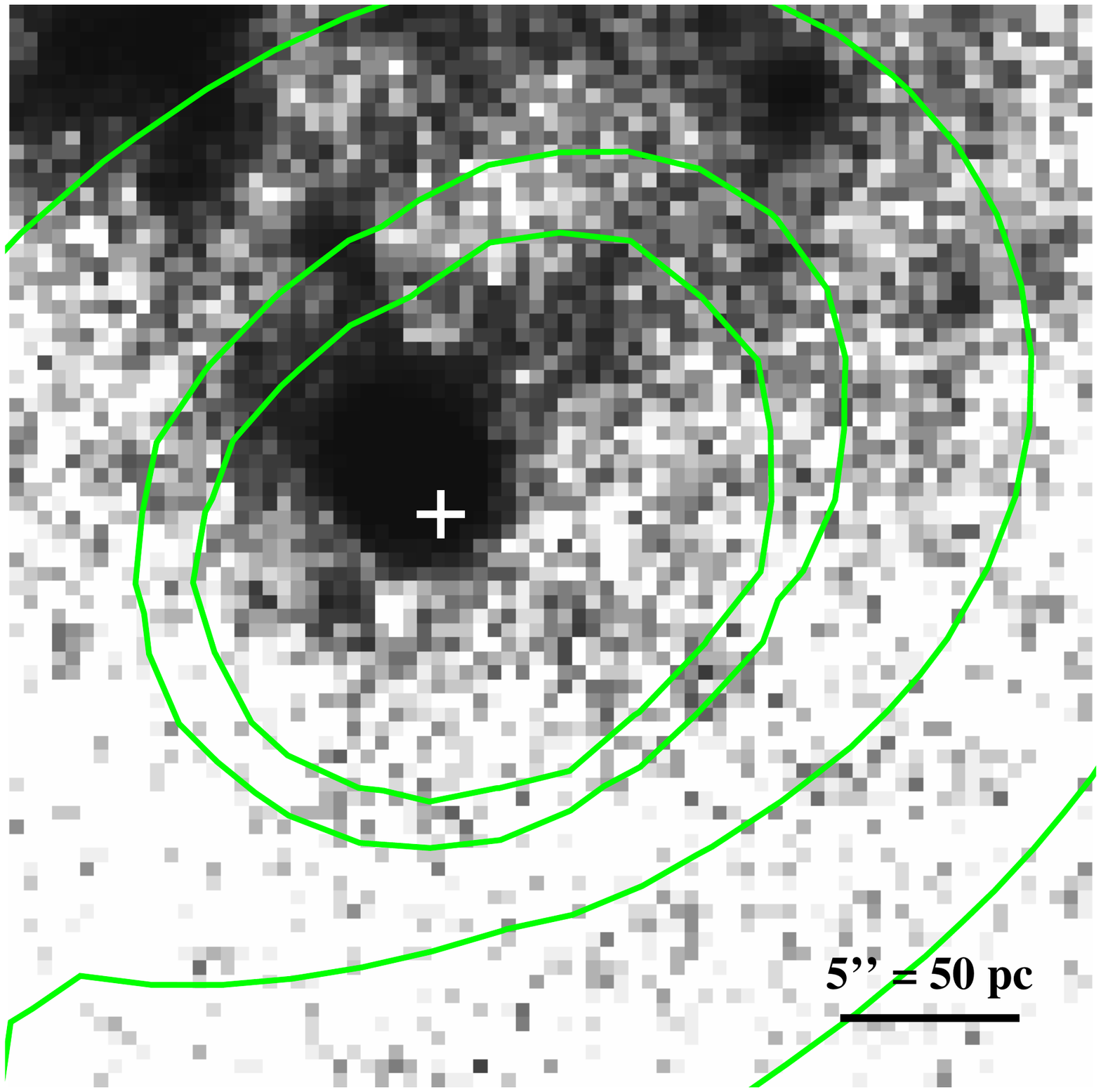} \\
\end{tabular}
\caption{H$\alpha$ emission from each of our SNR candidates (except source 46, for which H$\alpha$ imaging is not available). Each postage stamp is 0\amm5$\times$0\amm5, centered on the apparent H$\alpha$ emission. Crosses indicate the \Chandra X-ray source position (determined by the 0.35-8 keV emission). The green contours show the 0.35-1 keV X-ray emission.}
\label{SNR_Halpha}
\end{figure*}

To see if any of our SNR candidates have been previously identified, we cross-correlated our 11 sources with the multiwavelength NGC~300 SNR catalogs of \cite{Blair+97} and \cite{Pannuti+00} and the catalog of NGC~300 \ion{H}{2} regions of \cite{Deharveng+88}. We find 7 of our 11 sources are located within 3\as5 of a SNR or \ion{H}{2} region, and two of our X-ray sources are coincident with optically confirmed SNRs. Table~\ref{SNRtable} summarizes our SNR candidates. Figure~\ref{SNR_Halpha} shows 0\amm5$\times$0\amm5 postage stamp images of the H$\alpha$ emission coincident with each X-ray source, with the soft 0.35-1 keV X-ray contours superimposed.

\begin{table*}[ht]
\centering
\caption{Candidate Supernova Remnants and \ion{H}{2} Regions}
\begin{tabular}{ccccc}
\hline \hline
Source No. & Counterpart & Separation & 0.5-2 keV unabs. & Notes \\
		  &			&			& luminosity (\lum) & \\
(1) & (2) & (3) & (4) & (5) \\
\hline
7 & SNR & 2\as6 & 6.9$\times10^{35}$ & \cite{Blair+97} \\
15 & \ion{H}{2} & 3\as2 & 6.4$\times10^{35}$ & \cite{Deharveng+88} \\
32 & & & 1.1$\times10^{36}$ & \\
\multirow{2}{*}{34} & SNR	& 3\as3 &  \multirow{2}{*}{1.5$\times10^{36}$} & \cite{Blair+97}, opt. identified \\
     & \ion{H}{2} & 0\as6 & & \cite{Deharveng+88} \\
41 & & & 1.5$\times10^{36}$ \\
43 & SNR & 2\as8 & 5.6$\times10^{36}$ & \cite{Pannuti+00} \\
46 & & & 6.3$\times10^{35}$ & \\
54 & SNR & 3\as0 & 1.8$\times10^{36}$ & \cite{Pannuti+00} \\
77 & & & 7.5$\times10^{35}$ & \\
81 & \ion{H}{2} & 3\as1 & 4.7$\times10^{35}$ & \cite{Deharveng+88} \\
\multirow{2}{*}{85} & SNR & 0\as3 & \multirow{2}{*}{2.5$\times10^{36}$} & \cite{Blair+97}, opt. identified \\
      & \ion{H}{2} & 3\as4 & & \cite{Deharveng+88} \\
\hline \hline
\label{SNRtable}
\end{tabular}
\end{table*}

\subsection{Optical Counterparts}
We searched for optical counterpart candidates for each X-ray source that was contained within one or more overlapping \HST field. For many of the X-ray sources, we found multiple optical counterpart candidates within the \Chandra X-ray error circle (typically on the order of $\sim$0\as5); in these cases, the optical counterpart candidates are labeled `a,' `b,' `c,' etc. in order of increasing distance away from the X-ray source location. We kept only those optical sources that met the quality cuts described in \cite{Williams+09} and \cite{Gogarten+10} (i.e., sources with sufficient signal-to-noise, not flagged as unusable, meeting predetermined crowding and sharpness thresholds, etc.).

For each source, we compute a logarithmic X-ray-to-optical flux ratio $log(f_X/f_V)$, where $f_X$ is the 2-10 keV X-ray flux and and the $f_V$ is the optical flux measured in either $F555W$ or $F606W$, whichever is available. Although the $F606W$ has a slightly larger bandpass than the $F555W$ filter, we estimate that for the majority of stars accessible in our \HST exposures the affect on $log(f_X/f_V)$ will be small, $\sim0.1$.

Table~\ref{optical_counterparts} lists the optical and X-ray properties of all the X-ray source candidate optical counterparts. Figure~\ref{opticalIDs} shows RGB-rendered 5\asn$\times$ 5\asn finding charts for each of the 32 X-ray sources with overlapping \HST fields. The X-ray source position is shown by a white cross, with a white circle indicating the extent of the \Chandra error circle. Red lines are used to indicate the likely optical counterpart; in cases of multiple optical counterparts, the optical sources are labeled. 

\begin{table*}[ht]
\centering
\caption{Optical Counterparts to \Chandra X-ray Sources}
\renewcommand{\tabcolsep}{3pt}
\begin{tabular}{ccccccccc}
\hline \hline
Source	& $d^a$		& $F435W$	& $F475W$	& $F555W$	& $F606W$	& $F814W$	& $f_X^b$	& $log(f_X/f_V)^c$	\\
Name	& (arcsec)		&			&			&			&			&			& ($10^{-15}$ \flux)		&			\\
(1)		& (2)			& (3)			& (4)			& (5)			& (6)			& (7)			& (8)			& (9)			\\
\hline
4 (galaxy) & 1\as46		& 21.9		&		& 20.3			&			& 18.5		& \multirow{2}{*}{12.0} & -0.24		 \\
4 (point source) & 0\as18	& 24.7		& 		& 23.8			&			& 23.7		&			& 1.17 \\
\\
5a		& 0\as22		& 			& 26.99$\pm$0.10 &			& 26.39$\pm$0.06 & 25.47$\pm$0.08 & \multirow{4}{*}{2.1}	& 1.44 		\\
5b		& 0\as27		& 25.45$\pm$0.05 & 25.51$\pm$0.03 & 25.47$\pm$0.05 & 25.60$\pm$0.03 & 25.53$\pm$0.07, 25.86$\pm$0.08 & & 1.07 (1.12)	\\
5c		& 0\as62		& 			& 26.86$\pm$0.09 &			& 26.17$\pm$0.05 & 25.42$\pm$0.05 &		& 1.35		\\
5d		& 0\as70		&			& 26.24$\pm$0.05 &			& 26.14$\pm$0.05 & 26.52$\pm$0.18 &		& 1.34		\\
 \\
7a		& 0\as39		& 26.91$\pm$0.14 & 		& 26.65$\pm$0.12 &			& 26.55$\pm$0.14 & \multirow{3}{*}{1.6}	& 1.42		\\
7b		& 0\as42		& 27.13$\pm$0.18 &			& 26.93$\pm$0.17 &			& 26.13$\pm$0.10 & & 1.53		\\
7c		& 0\as55		& 24.04$\pm$0.02 &			& 24.13$\pm$0.06 &			& 24.25$\pm$0.03 & & 0.41		\\
 \\
8		& 0\as04		& 19.7			&		& 18.5			&		& 16.6			& 3.9		& -1.44		\\
 \\
9a		& 0\as34		& 27.02$\pm$0.15 & 26.27$\pm$0.05 & 25.48$\pm$0.06 & 25.09$\pm$0.02 & 23.98$\pm$0.04, 24.04$\pm$0.02 & \multirow{2}{*}{2.7}	& 1.19 (1.04)	\\
9b		& 0\as39		& 25.98$\pm$0.07 & 25.43$\pm$0.03 & 24.75$\pm$0.03 & 24.40$\pm$0.01 & 23.51$\pm$0.04, 23.53$\pm$0.02 & & 0.90 (0.76)	\\
\hline \hline
\label{optical_counterparts}
\end{tabular}
\tablecomments{
$^a$The positional offset from the \Chandra X-ray source location.\newline
$^b$X-ray flux is unabsorbed in the 0.35-8 keV energy band.\newline
$^c$$f_V$ is calculated from either the $F55W$ or $F606W$ magnitude, depending on which is available for a given observation.\newline
Only the first five X-ray sources with optical counterpart candidates are listed.}
\end{table*}

\begin{figure*}
\centering
\begin{tabular}{cccc}
Source 3 & Source 4 & Source 5 & Source 7 \\
\includegraphics[width=0.21\linewidth,clip=true,trim=2.7cm 5.2cm 2.7cm 5.2cm]{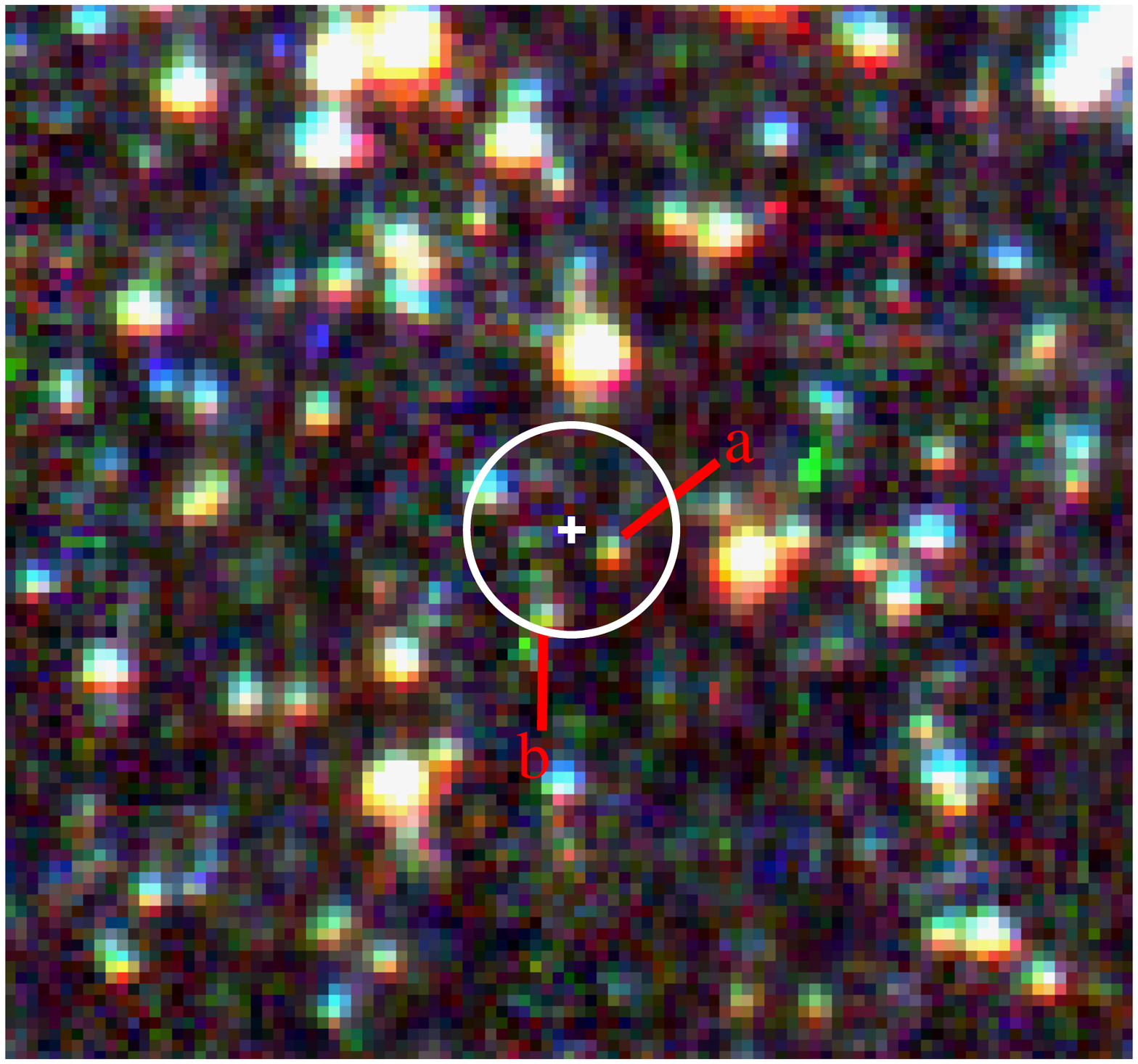} &
\includegraphics[width=0.21\linewidth,clip=true,trim=2cm 0cm 2cm 2cm]{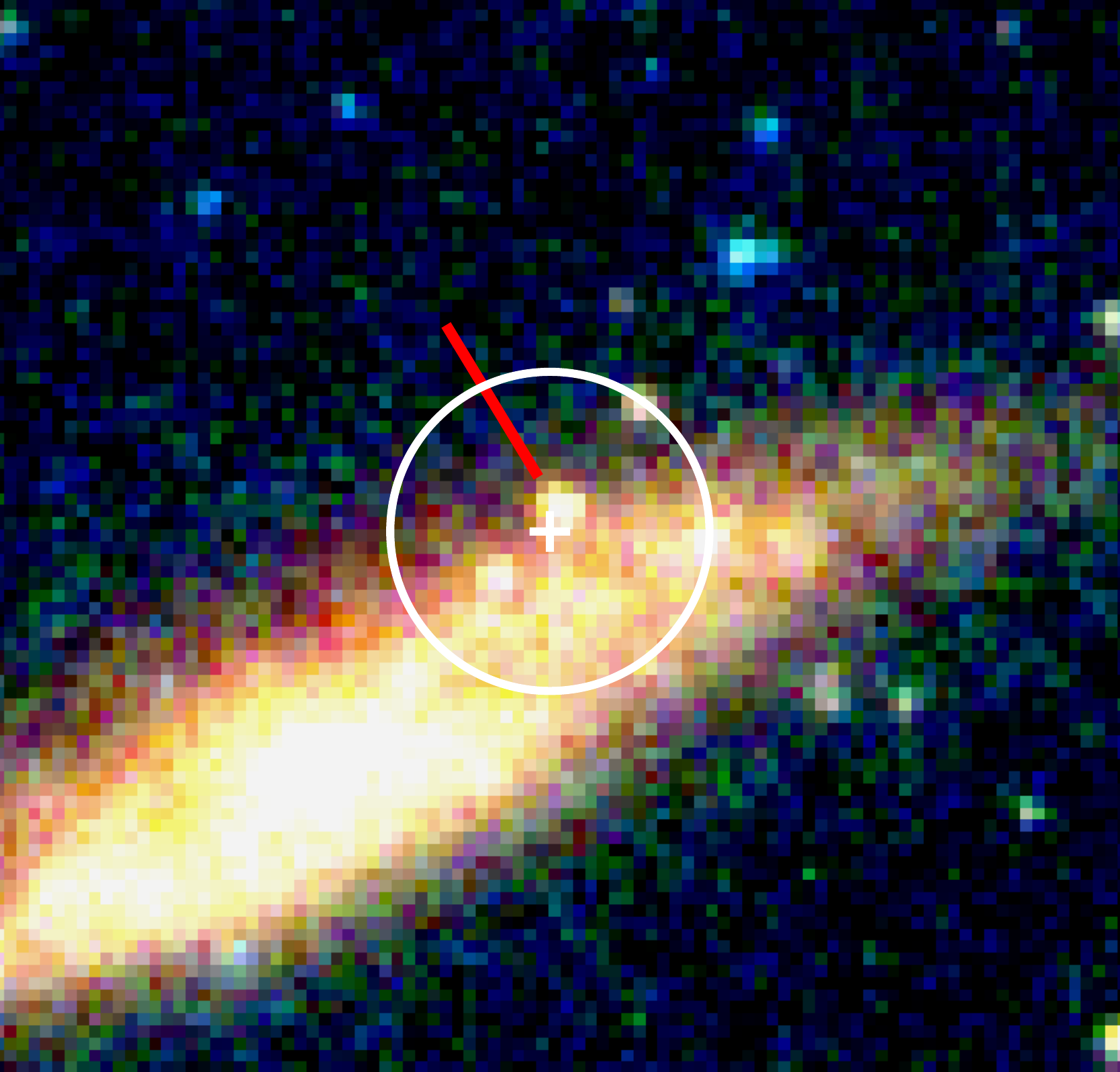} &
\includegraphics[width=0.21\linewidth,clip=true,trim=2.7cm 5.2cm 2.7cm 5.2cm]{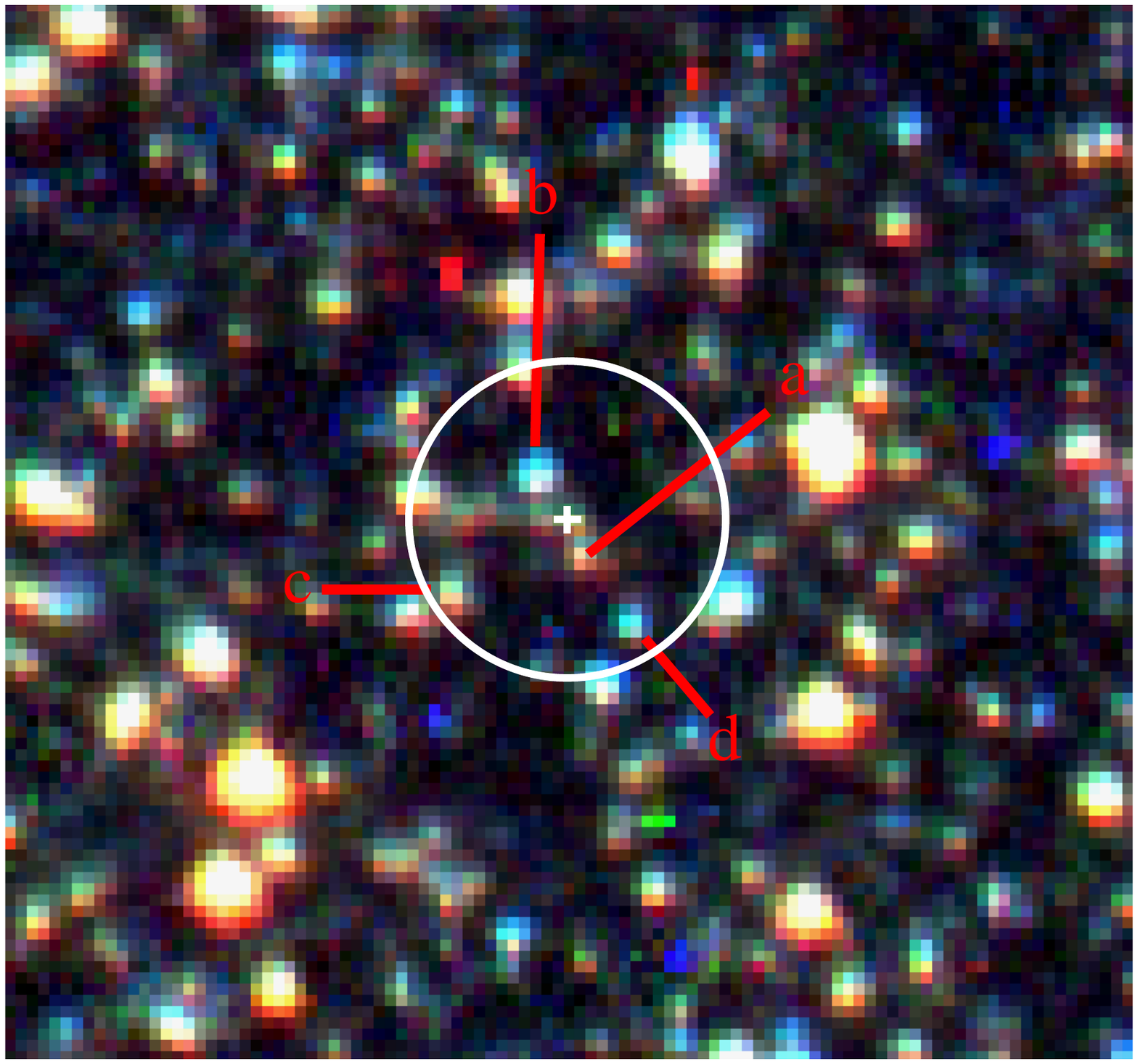} &
\includegraphics[width=0.21\linewidth,clip=true,trim=2.7cm 5.2cm 2.7cm 5.2cm]{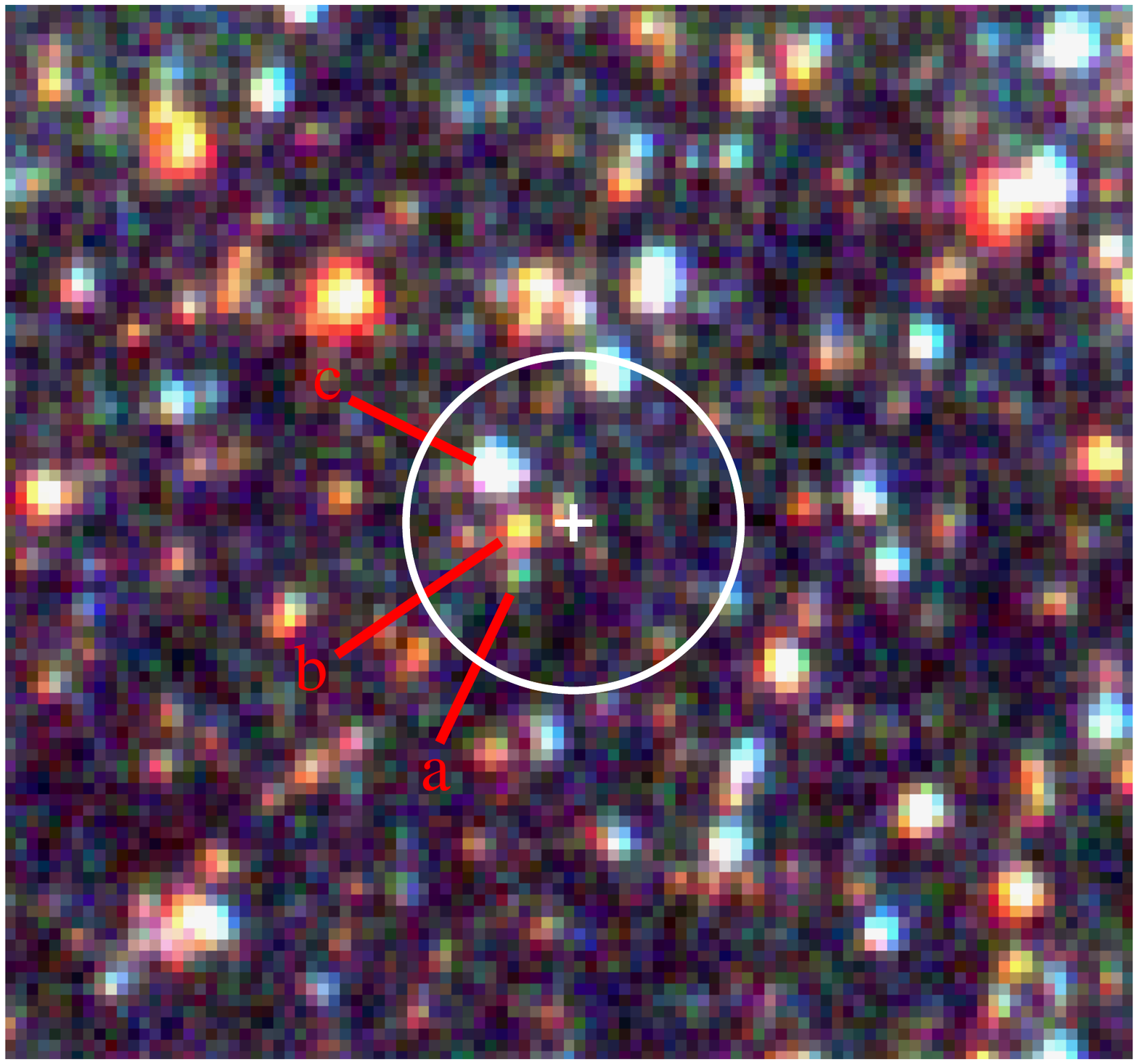} \\
Source 8 & Source 9 &  Source 10 & Source 11 \\
\includegraphics[width=0.21\linewidth,clip=true,trim=2.7cm 5.2cm 2.7cm 5.2cm]{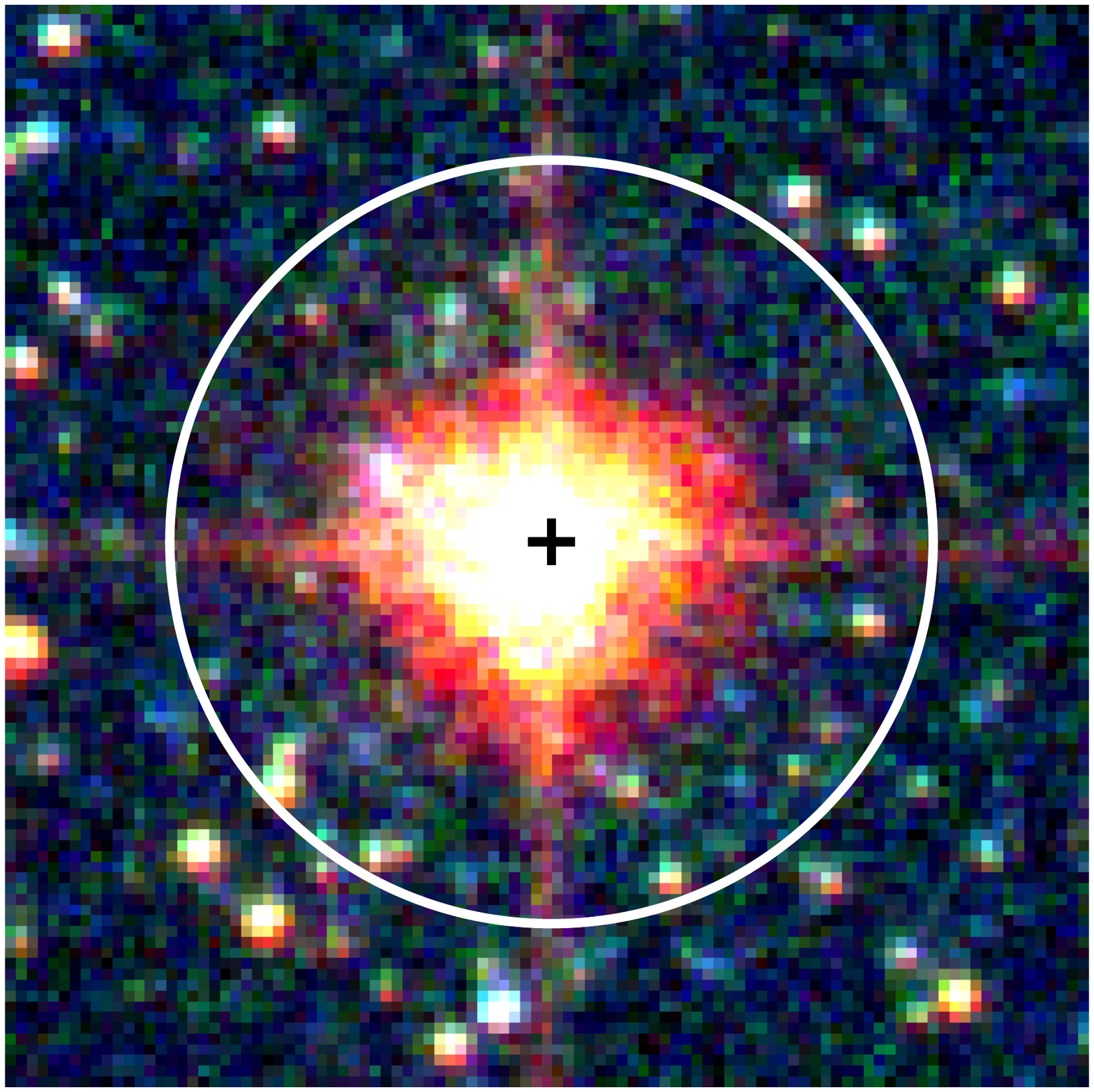} &
\includegraphics[width=0.21\linewidth,clip=true,trim=2.7cm 5.2cm 2.7cm 5.2cm]{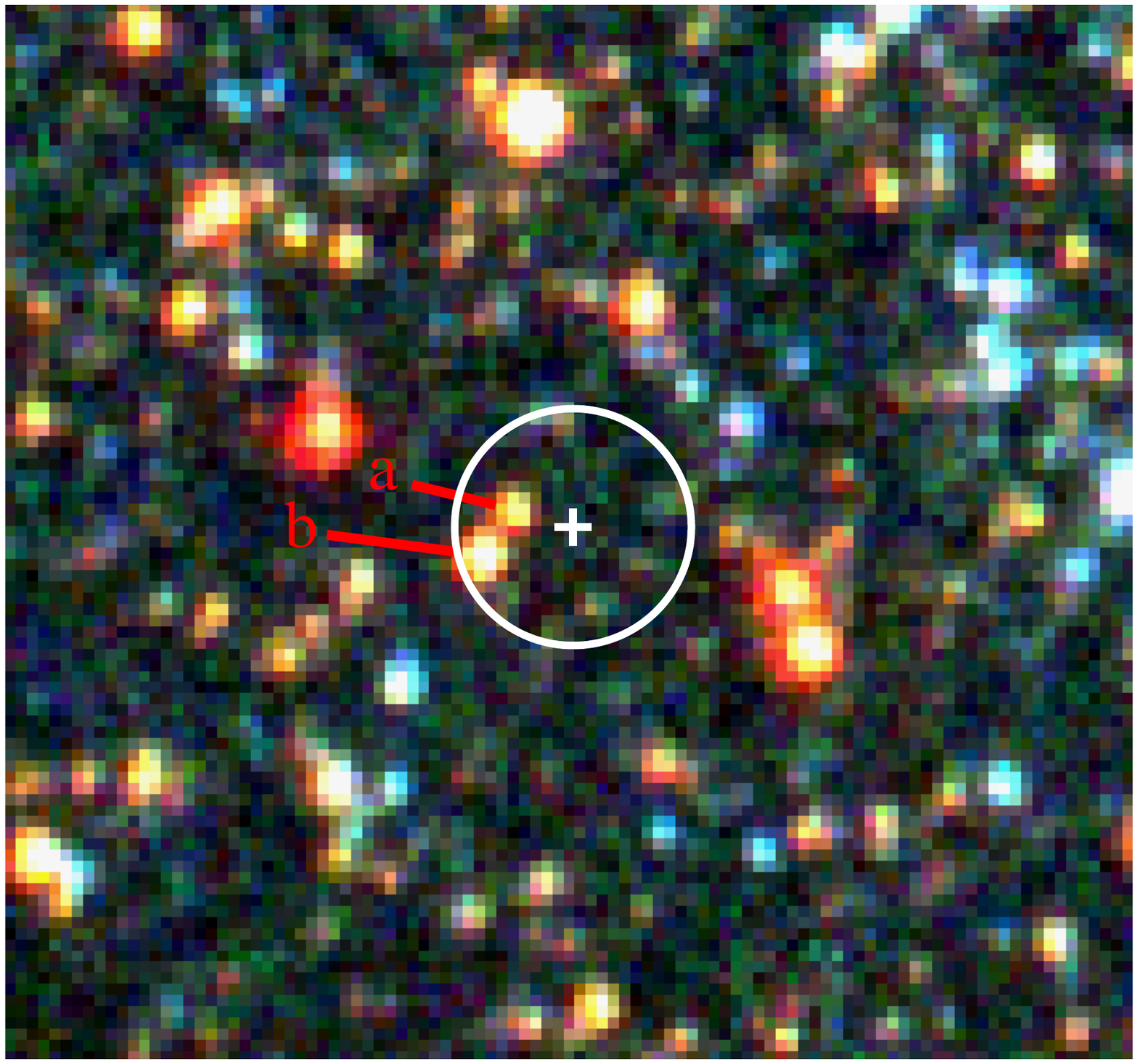} &
\includegraphics[width=0.22\linewidth,clip=true,trim=2cm 1.7cm 2cm 1.7cm]{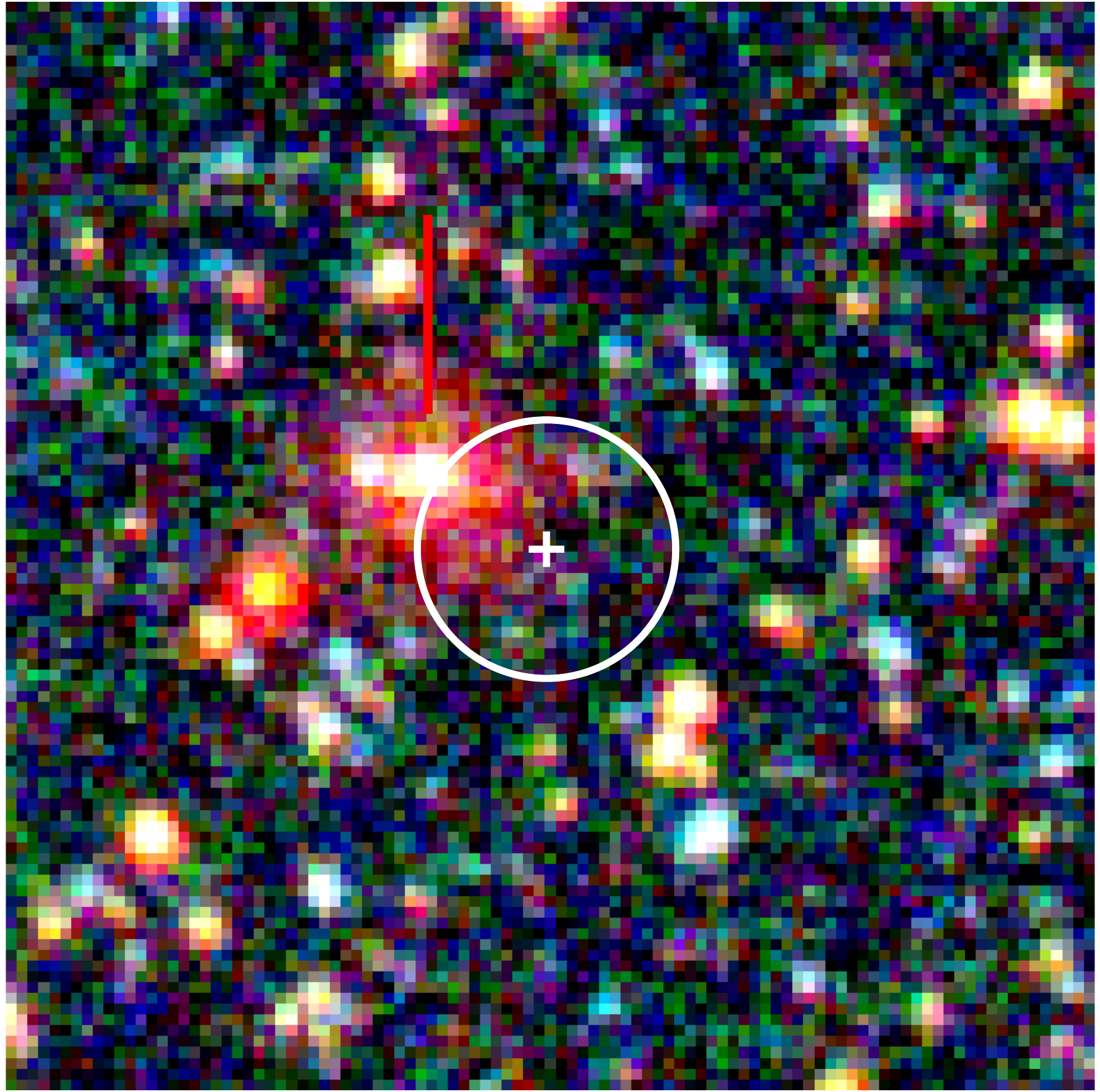} &
\includegraphics[width=0.21\linewidth,clip=true,trim=2.7cm 5.2cm 2.7cm 5.2cm]{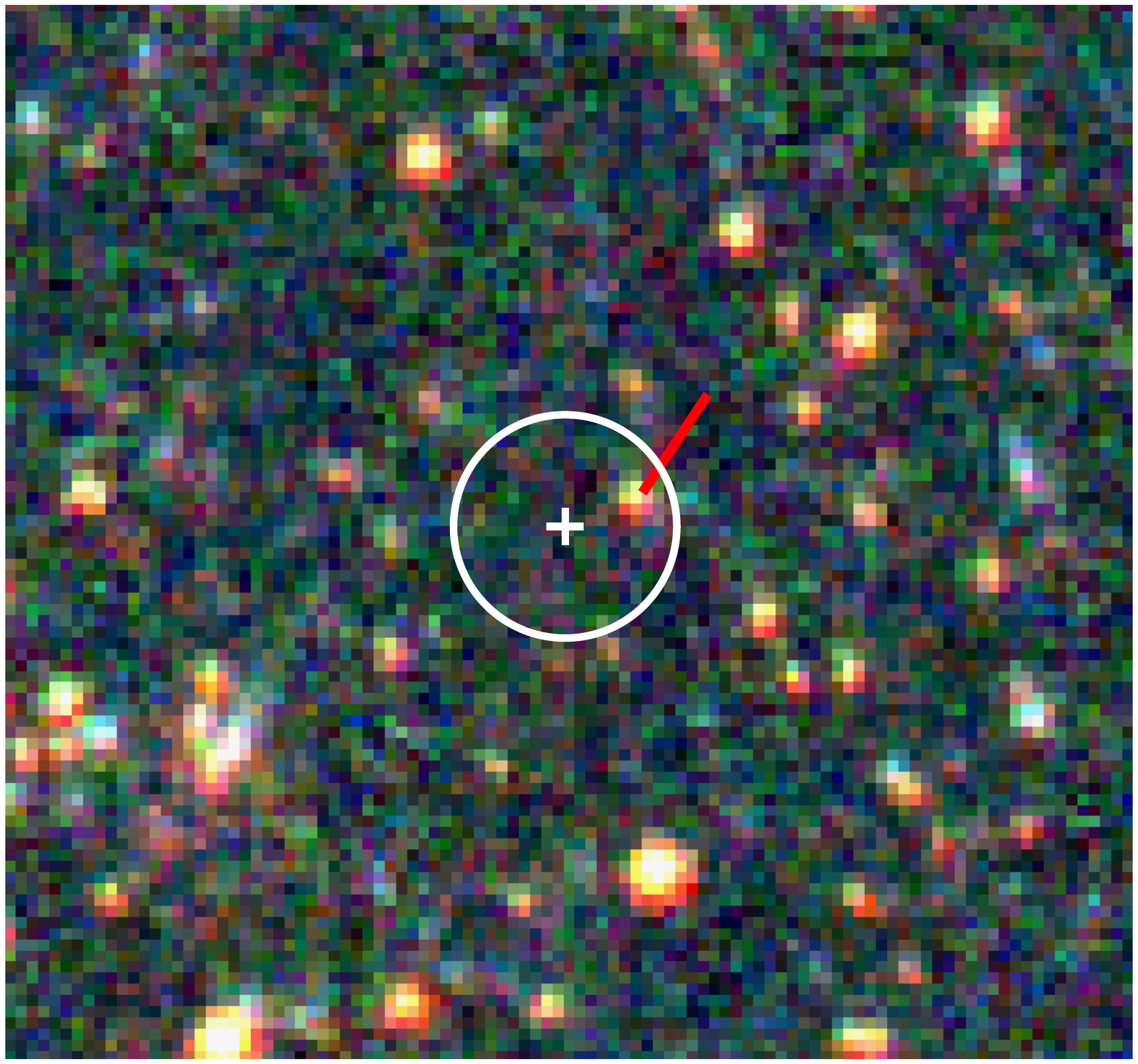} \\
Source 12 & Source 13 & Source 14 & Source 17 \\
\includegraphics[width=0.21\linewidth,clip=true,trim=2.7cm 5.2cm 2.7cm 5.2cm]{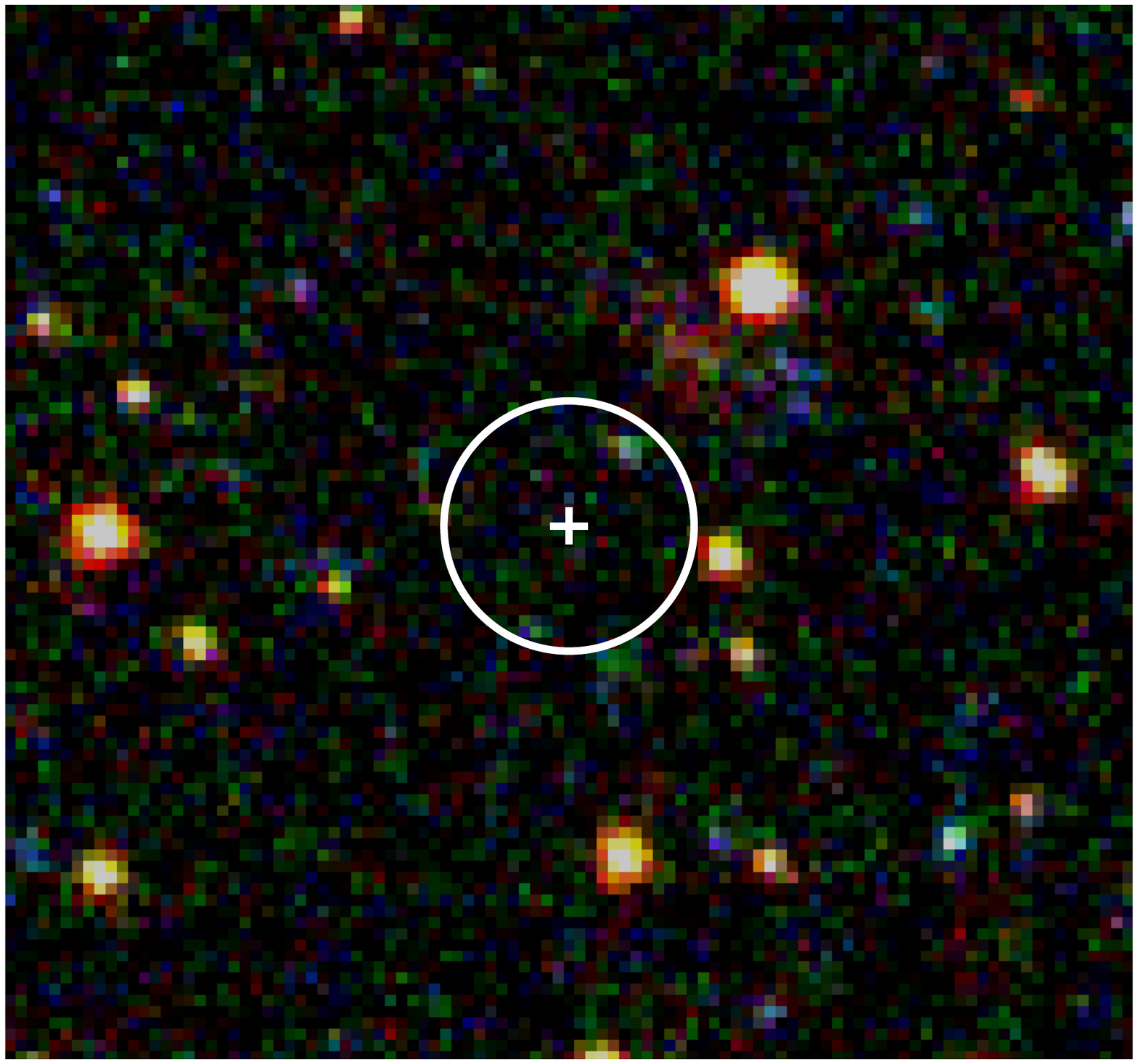} &
\includegraphics[width=0.21\linewidth,clip=true,trim=2.7cm 5.2cm 2.7cm 5.2cm]{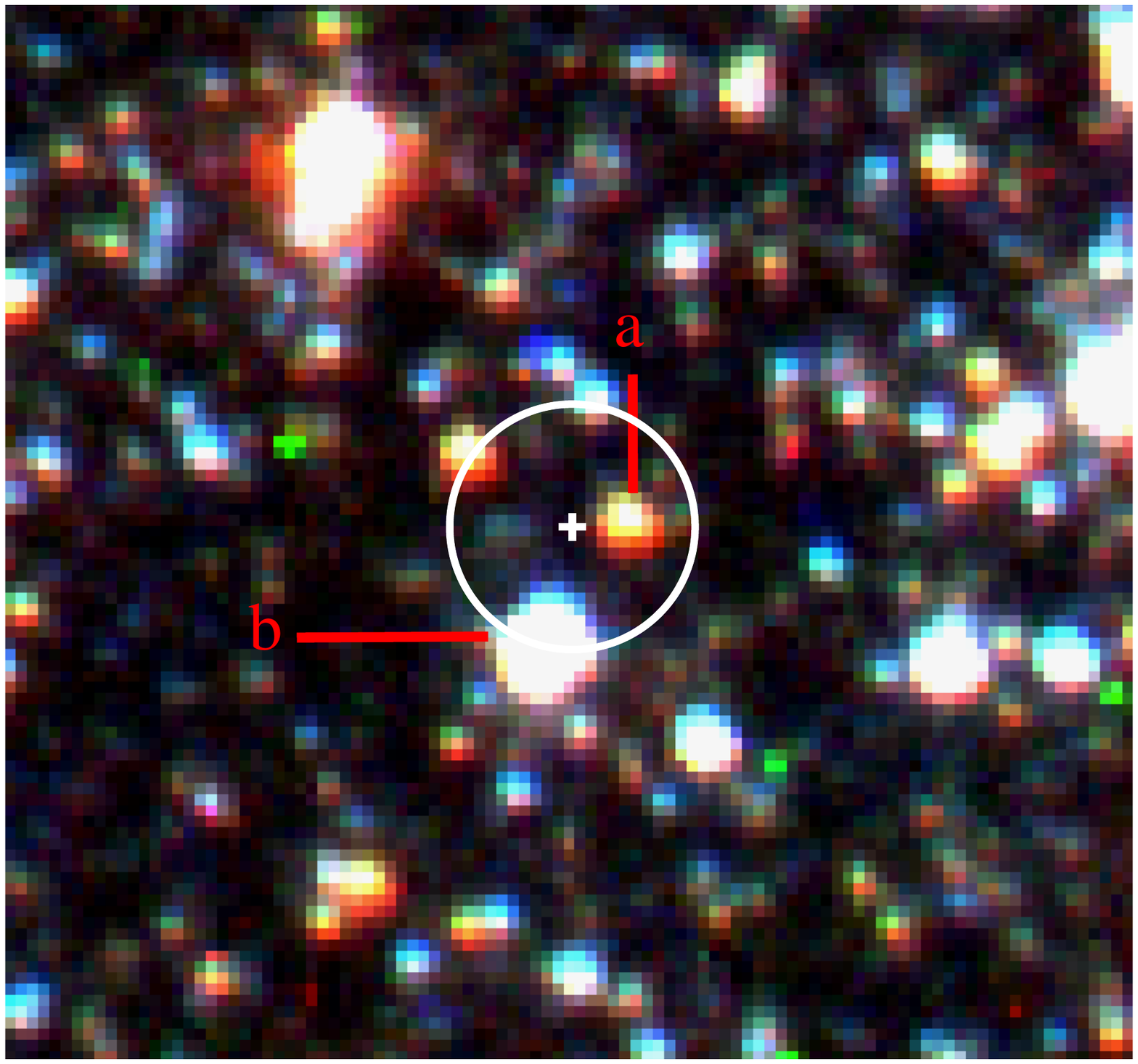} &
\includegraphics[width=0.21\linewidth,clip=true,trim=2.7cm 5.2cm 2.7cm 5.2cm]{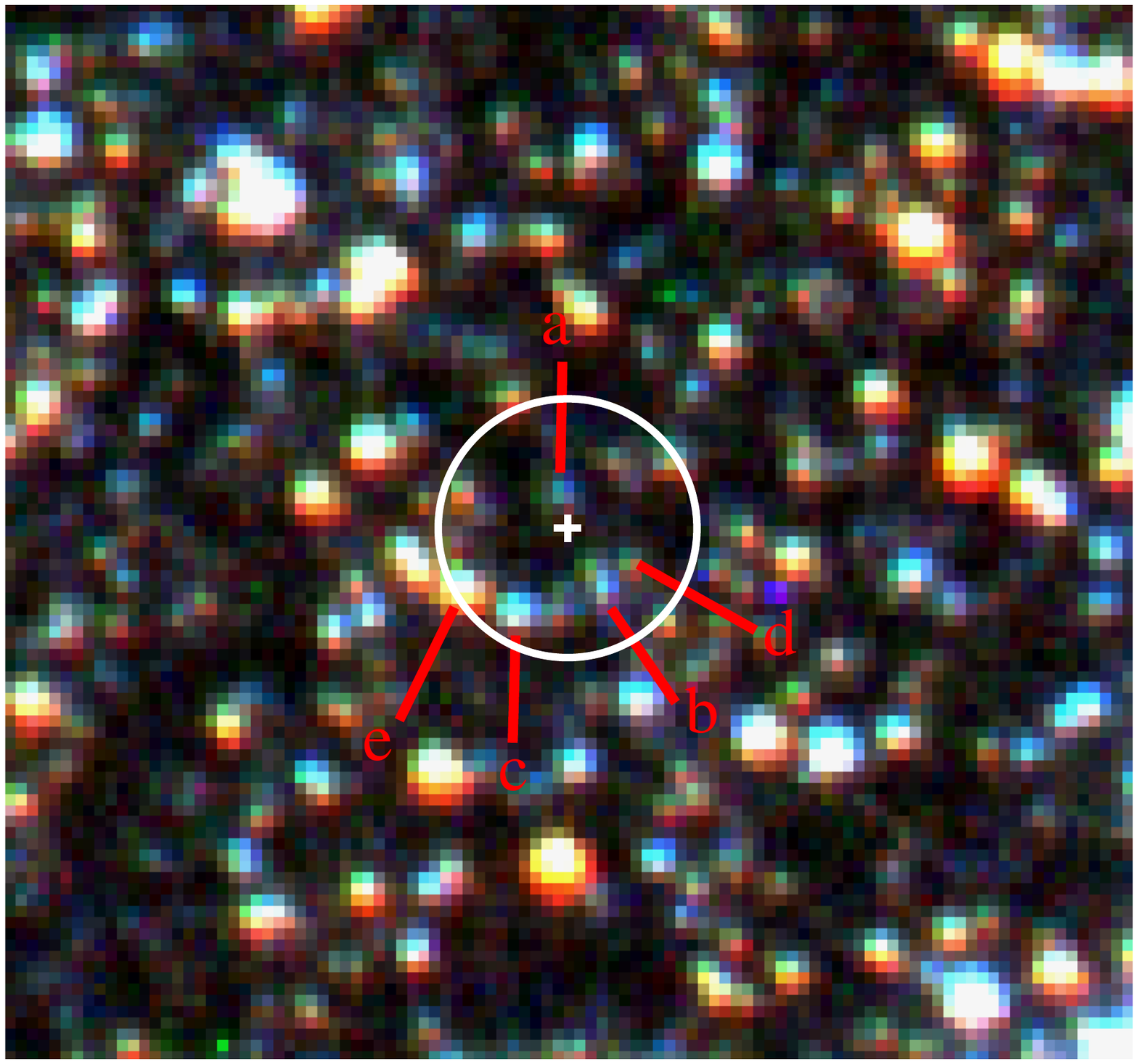} &
\includegraphics[width=0.21\linewidth,clip=true,trim=2.7cm 5.2cm 2.7cm 5.2cm]{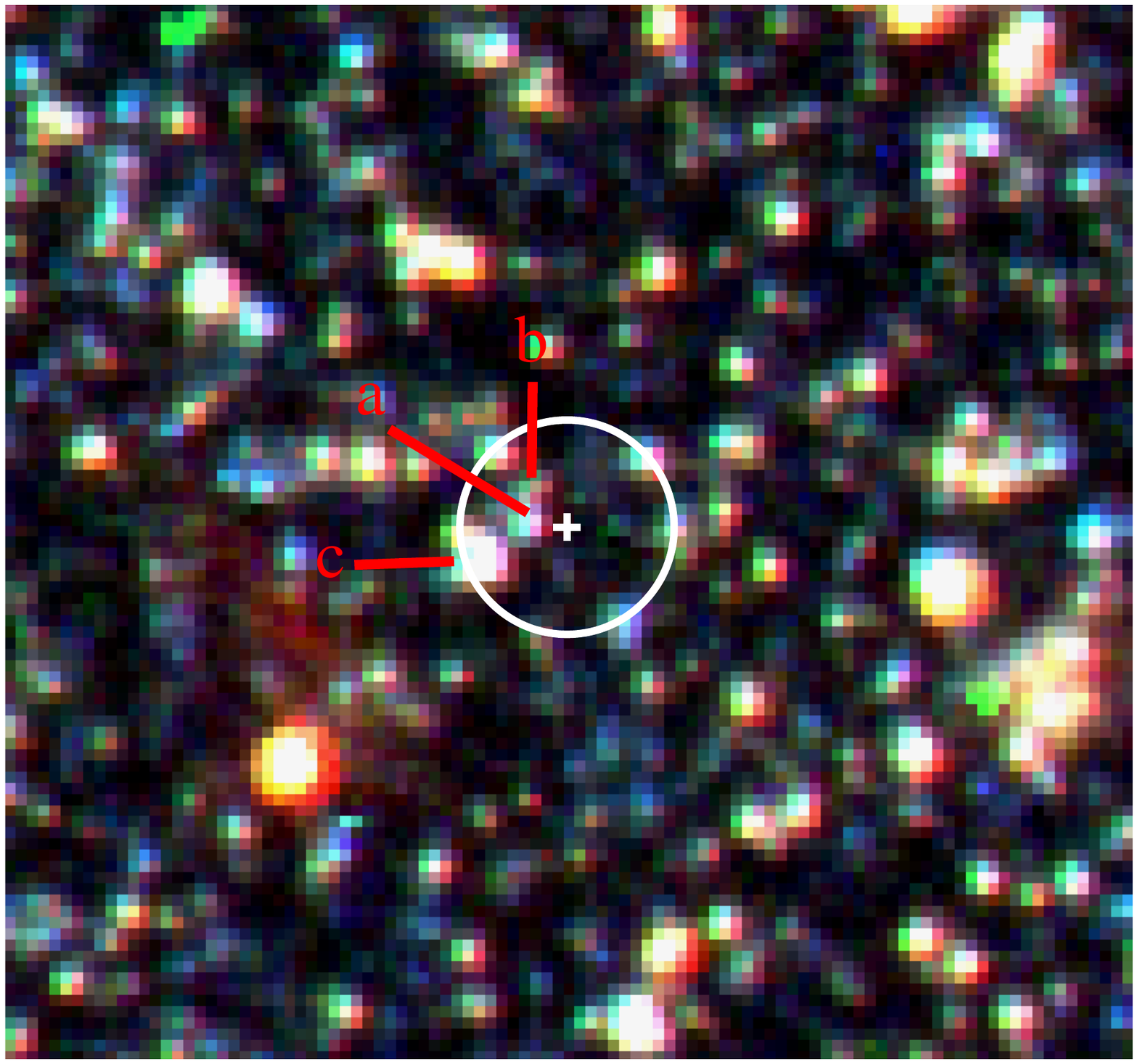} \\
Source 18 & Source 19 & Source 20 & Source 21 \\
\includegraphics[width=0.21\linewidth,clip=true,trim=2.7cm 5.2cm 2.7cm 5.2cm]{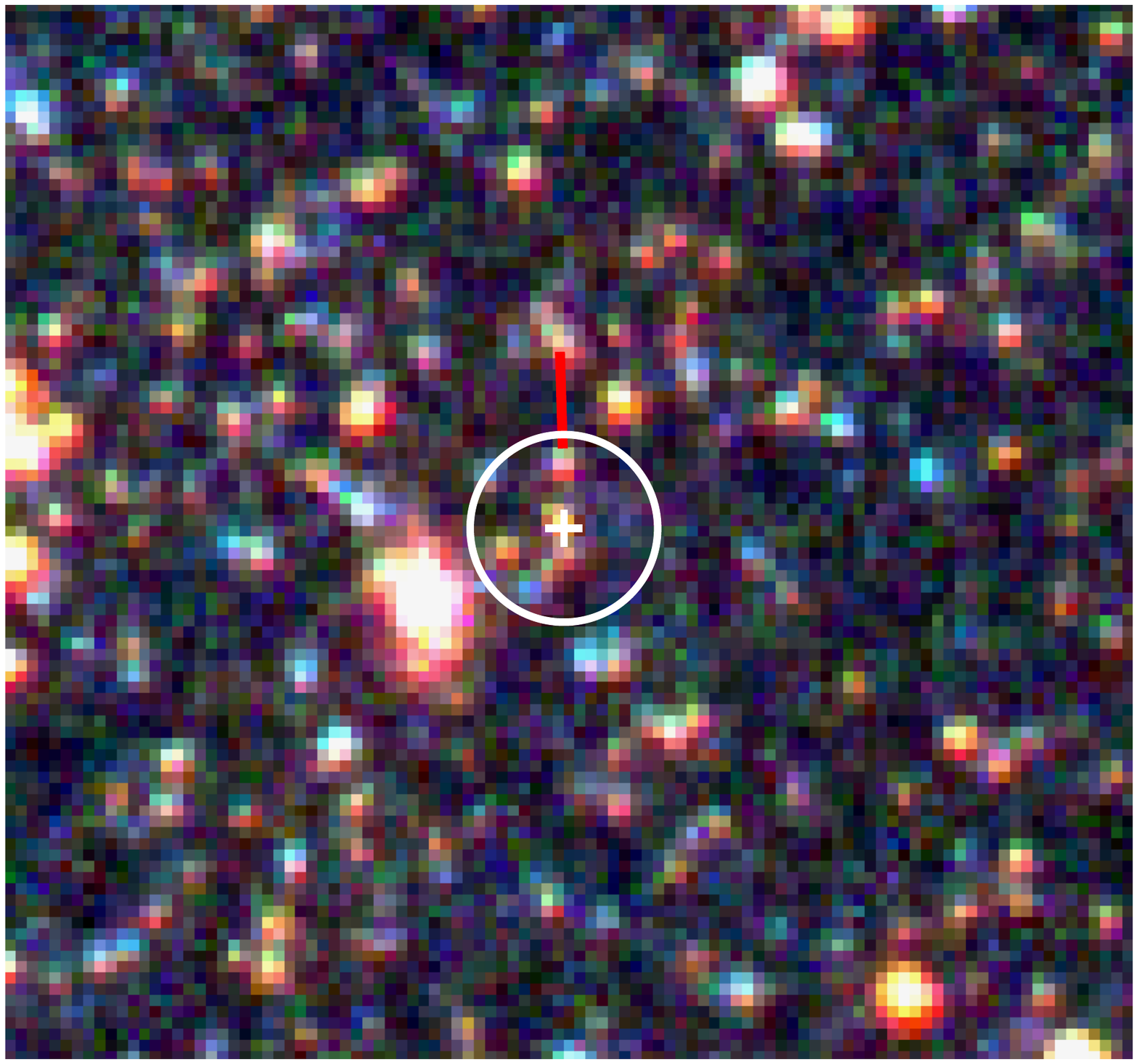} &
\includegraphics[width=0.21\linewidth,clip=true,trim=2.7cm 5.2cm 2.7cm 5.2cm]{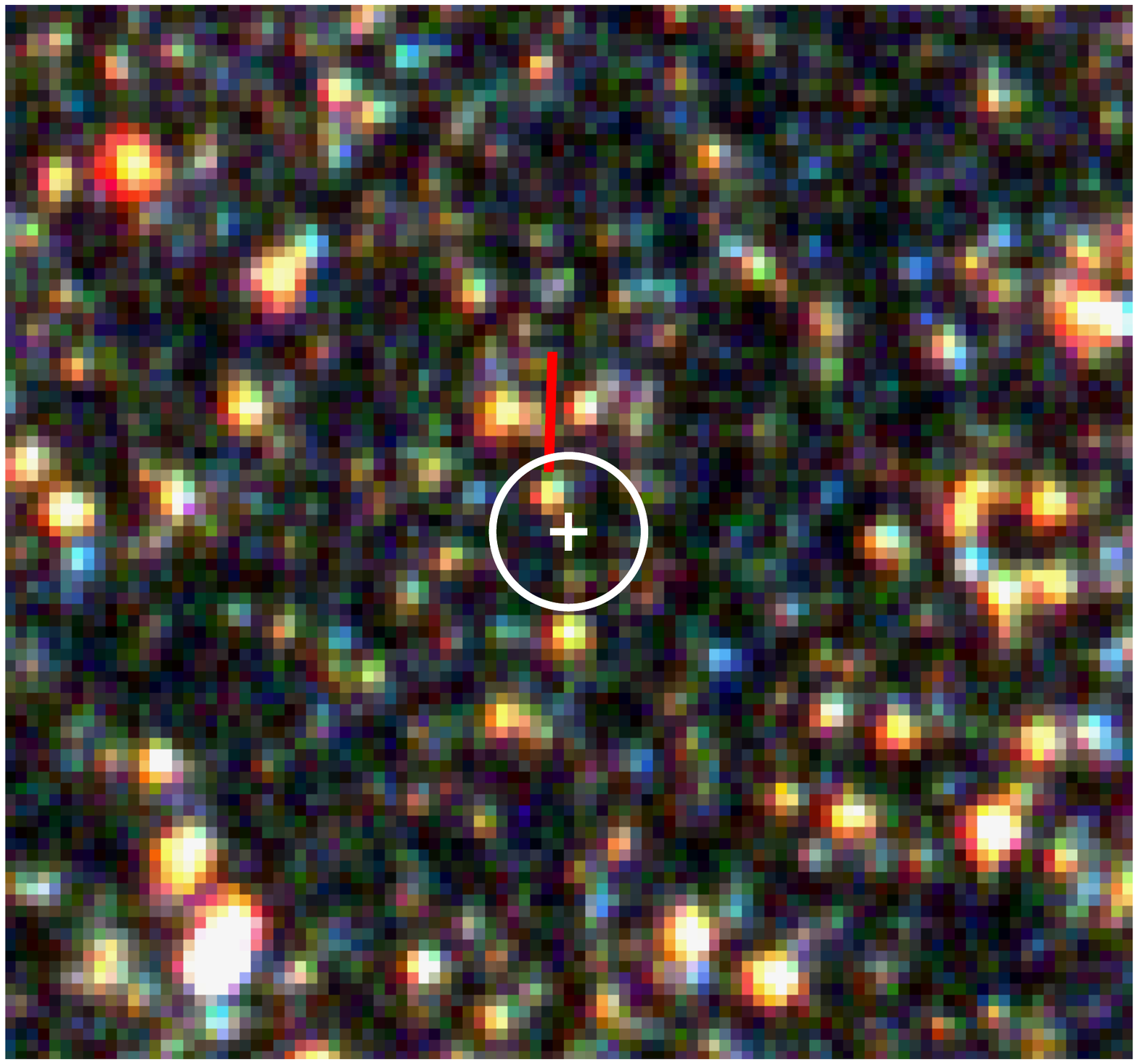} &
\includegraphics[width=0.21\linewidth,clip=true,trim=2.7cm 5.2cm 2.7cm 5.2cm]{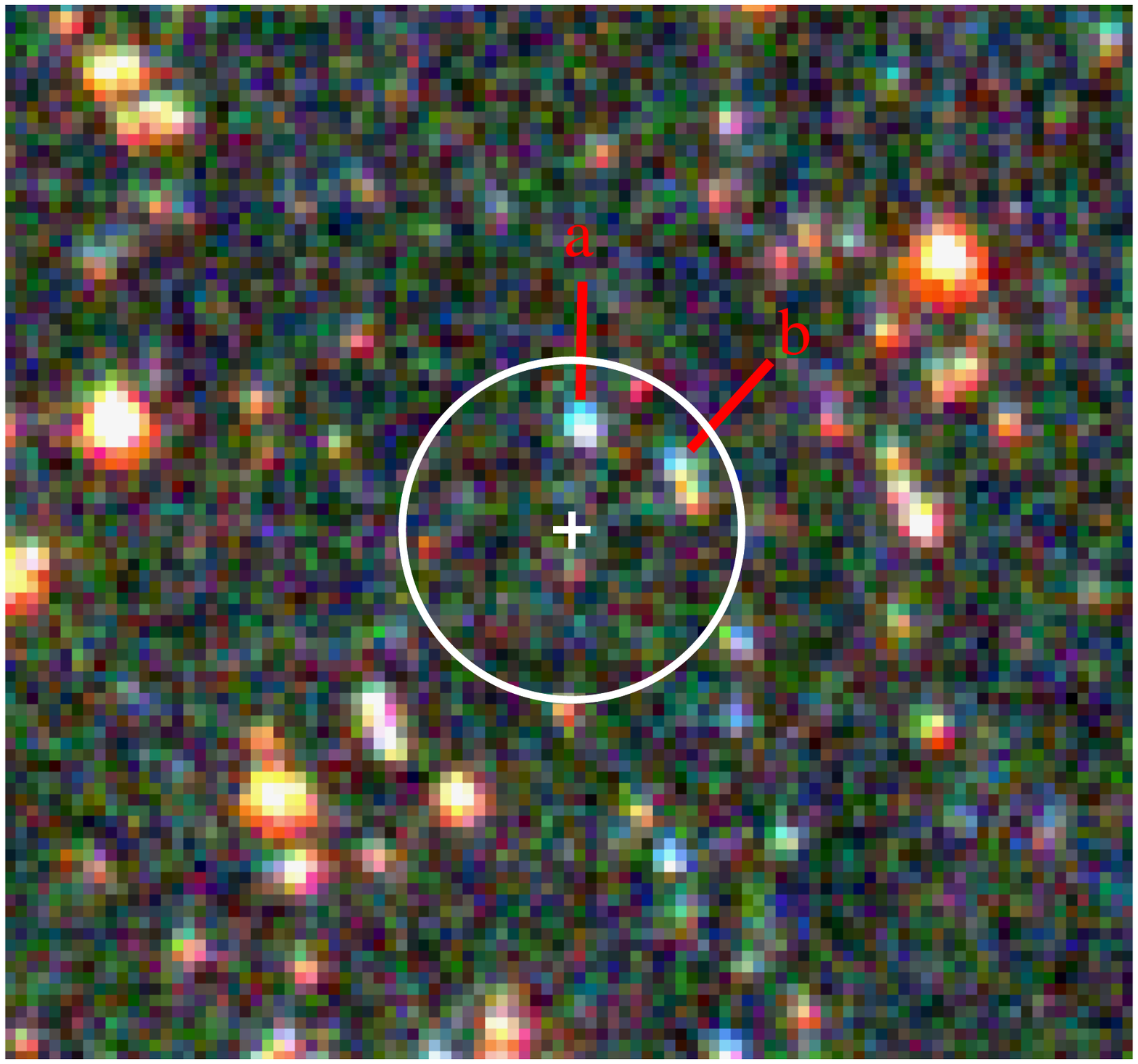} &
\includegraphics[width=0.21\linewidth,clip=true,trim=2.7cm 5.2cm 2.7cm 5.2cm]{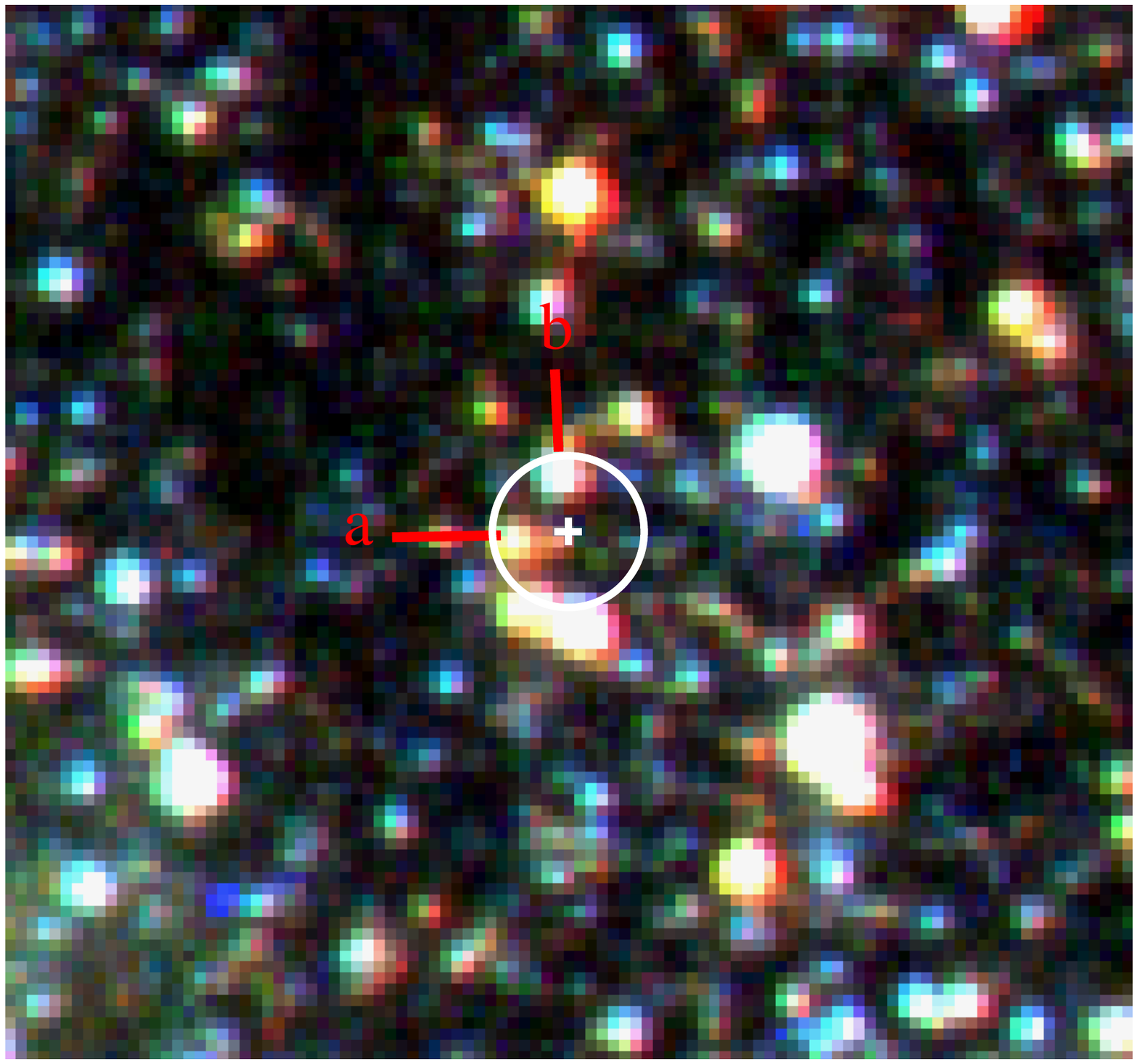} \\
Source 22 & Source 23 & Source 25 & Source 26 \\
\includegraphics[width=0.21\linewidth,clip=true,trim=2.7cm 5.2cm 2.7cm 5.2cm]{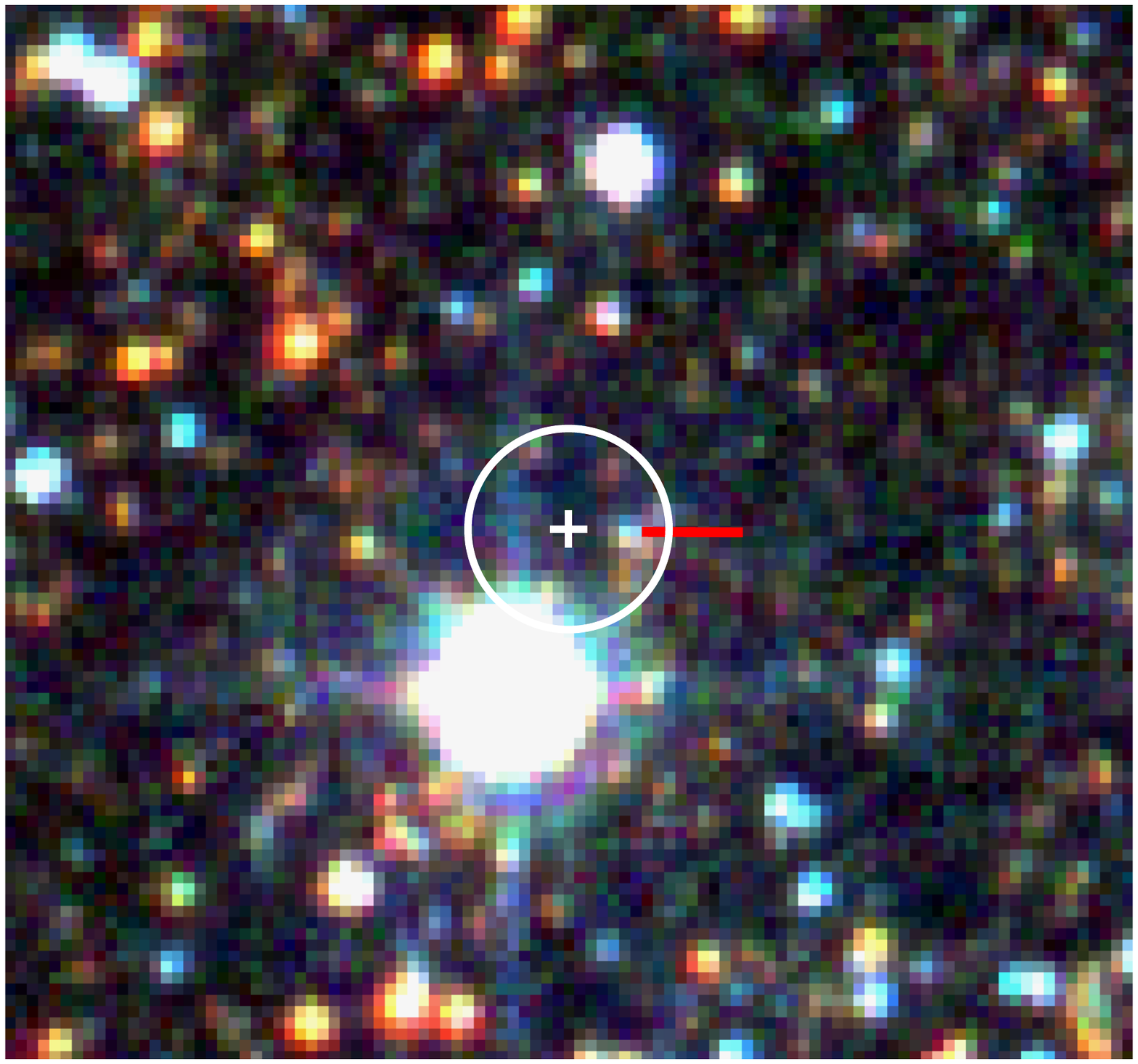} &
\includegraphics[width=0.22\linewidth,clip=true,trim=2cm 5cm 2cm 5cm]{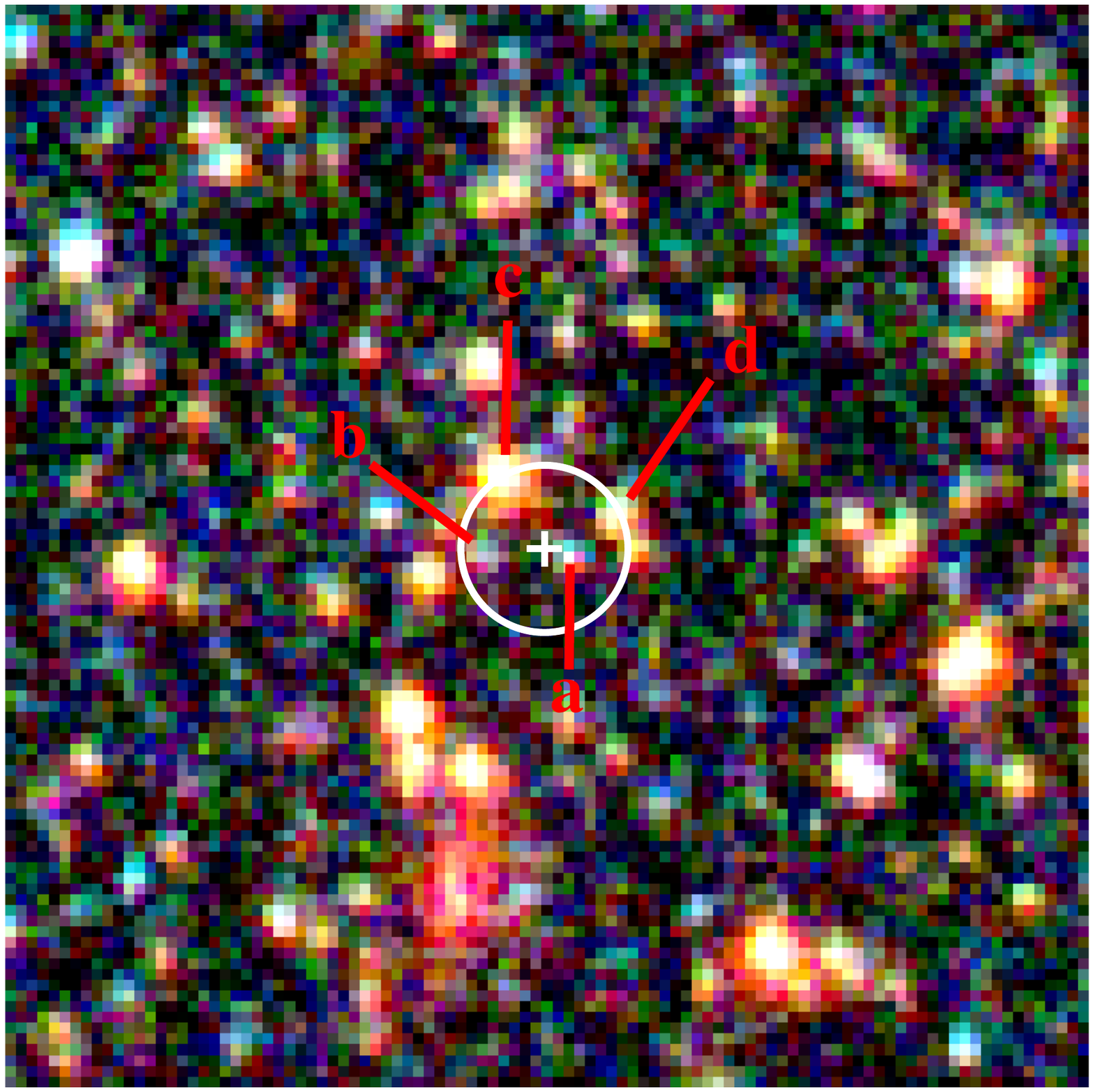} &
\includegraphics[width=0.21\linewidth,clip=true,trim=2.7cm 5.2cm 2.7cm 5.2cm]{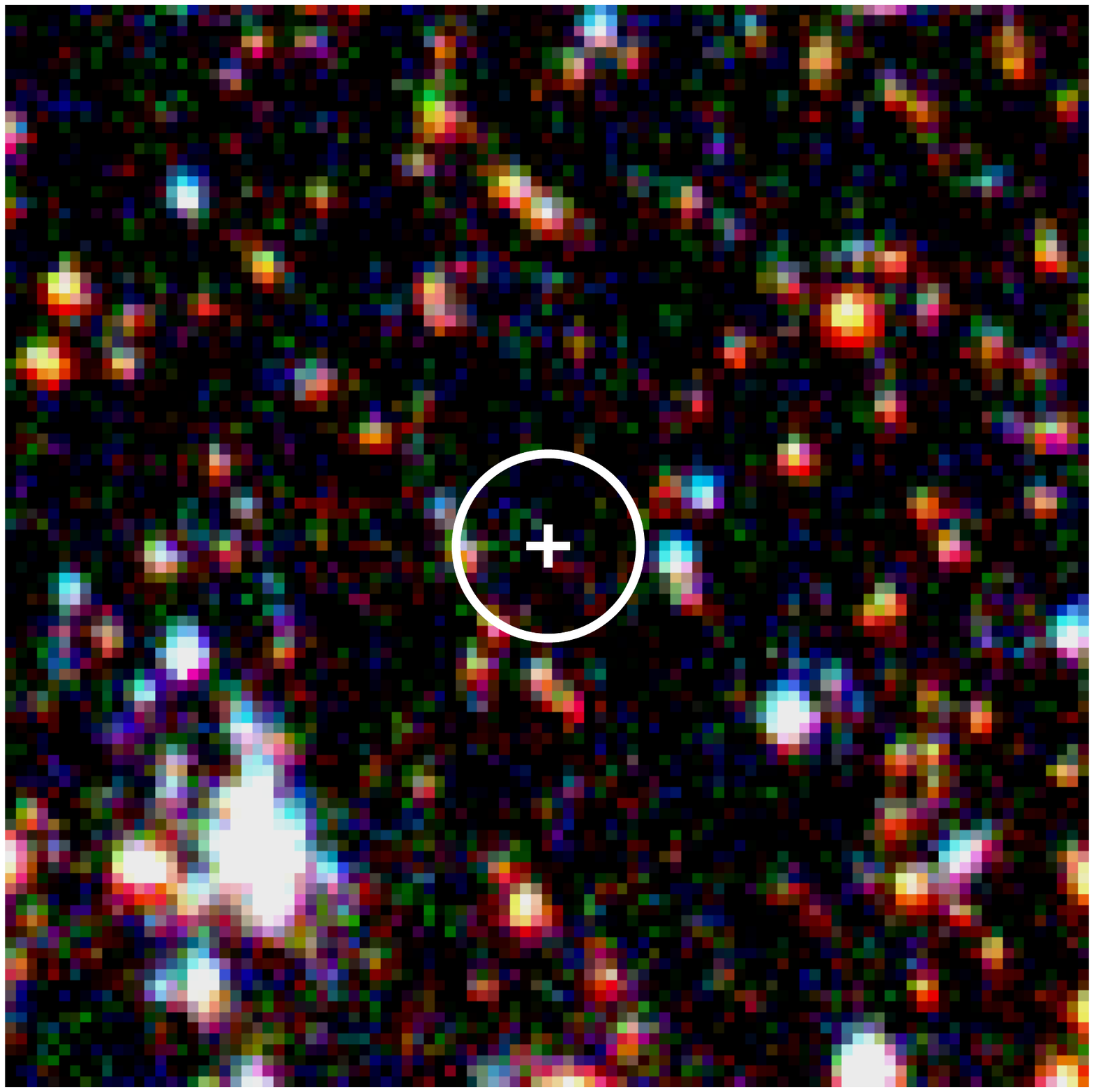} &
\includegraphics[width=0.21\linewidth,clip=true,trim=2.7cm 5.2cm 2.7cm 5.2cm]{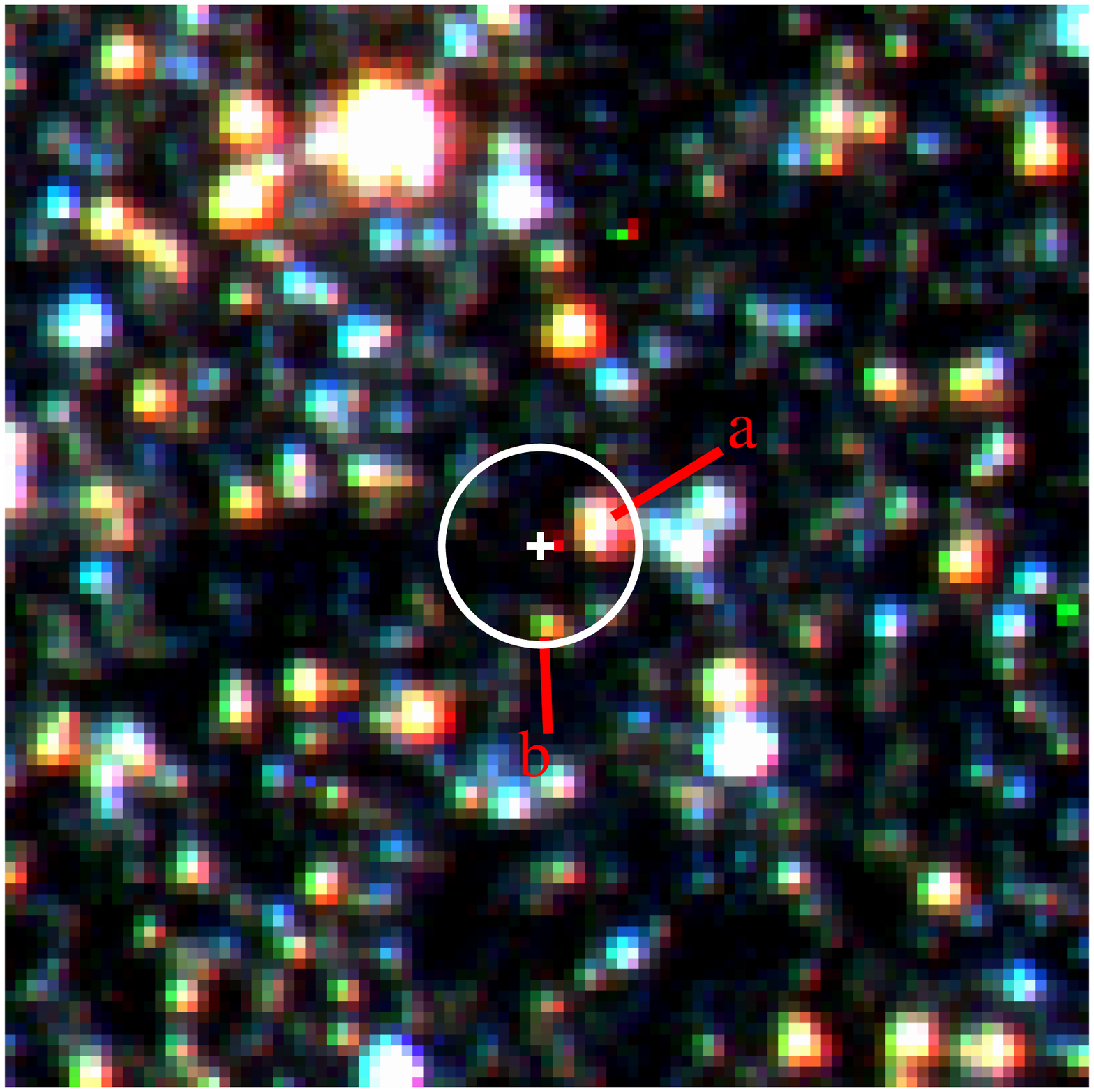} \\
\end{tabular}
\caption{Optical RGB-rendered finding charts at the location of each X-ray source, indicated by a cross. White circles show the radius of the 90\% \Chandra error circle for each source. Candidate optical counterparts are indicated with red lines. X-ray sources with multiple candidate optical counterparts have their optical counterparts labeled `a,' `b,' `c,' etc.}
\label{opticalIDs}
\end{figure*}

\setcounter{figure}{6}
\begin{figure*}
\centering
\begin{tabular}{cccc}
Source 32 & Source 34 & Source 41 & Source 42 \\
\includegraphics[width=0.21\linewidth,clip=true,trim=2.7cm 5.2cm 2.7cm 5.2cm]{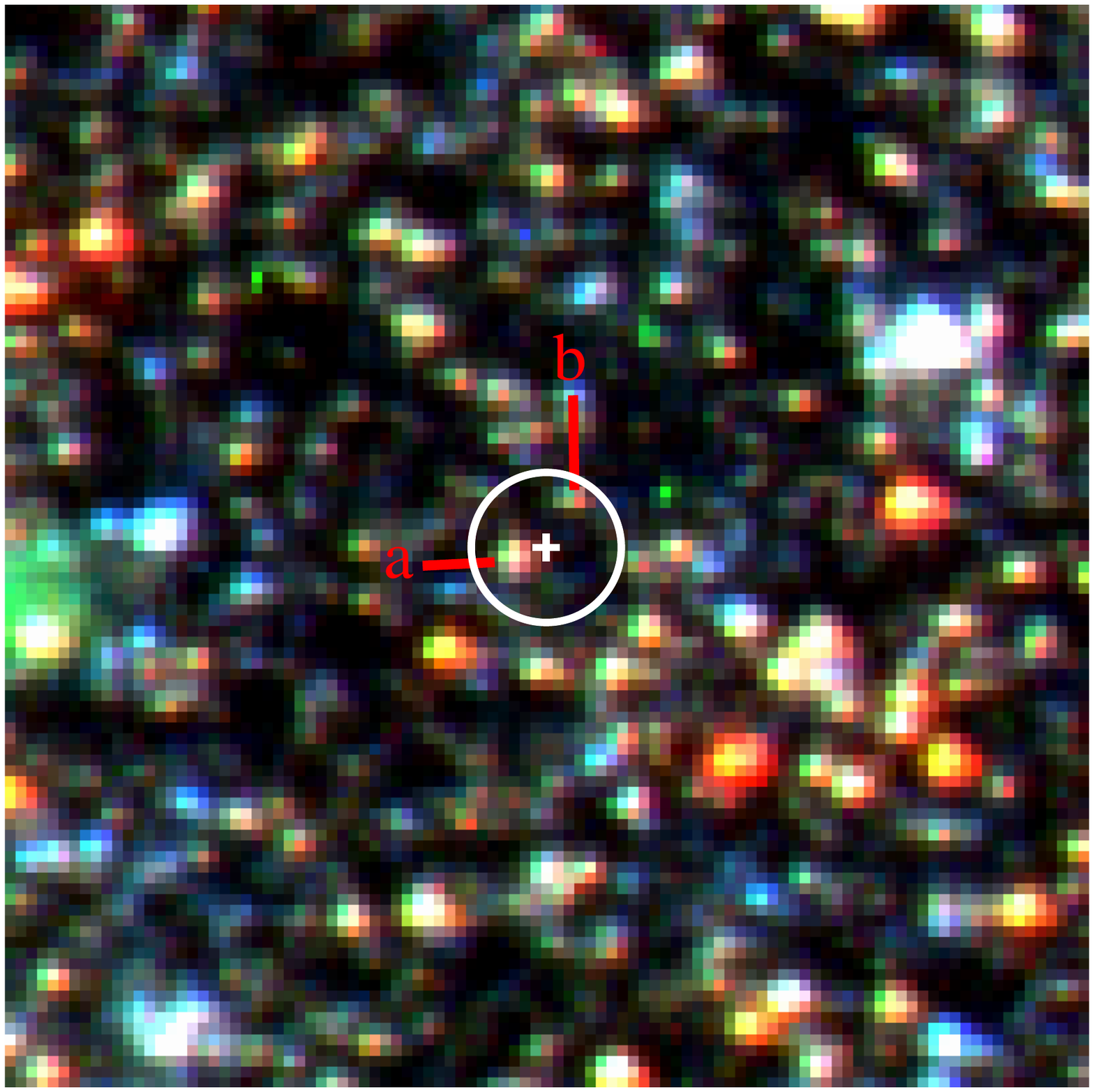} &
\includegraphics[width=0.21\linewidth,clip=true,trim=2.7cm 5.2cm 2.7cm 5.2cm]{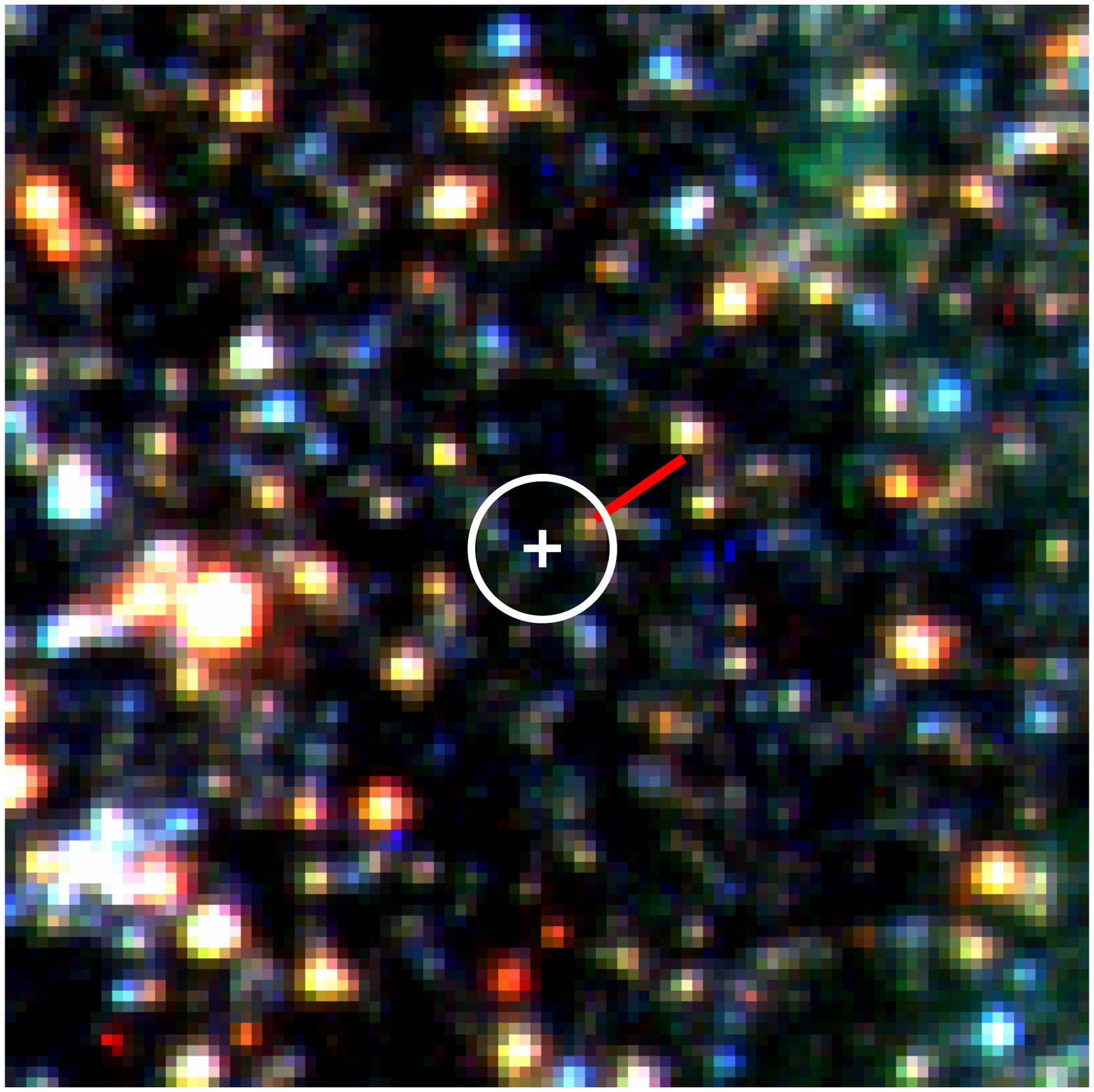} &
\includegraphics[width=0.21\linewidth,clip=true,trim=2.7cm 5.2cm 2.7cm 5.2cm]{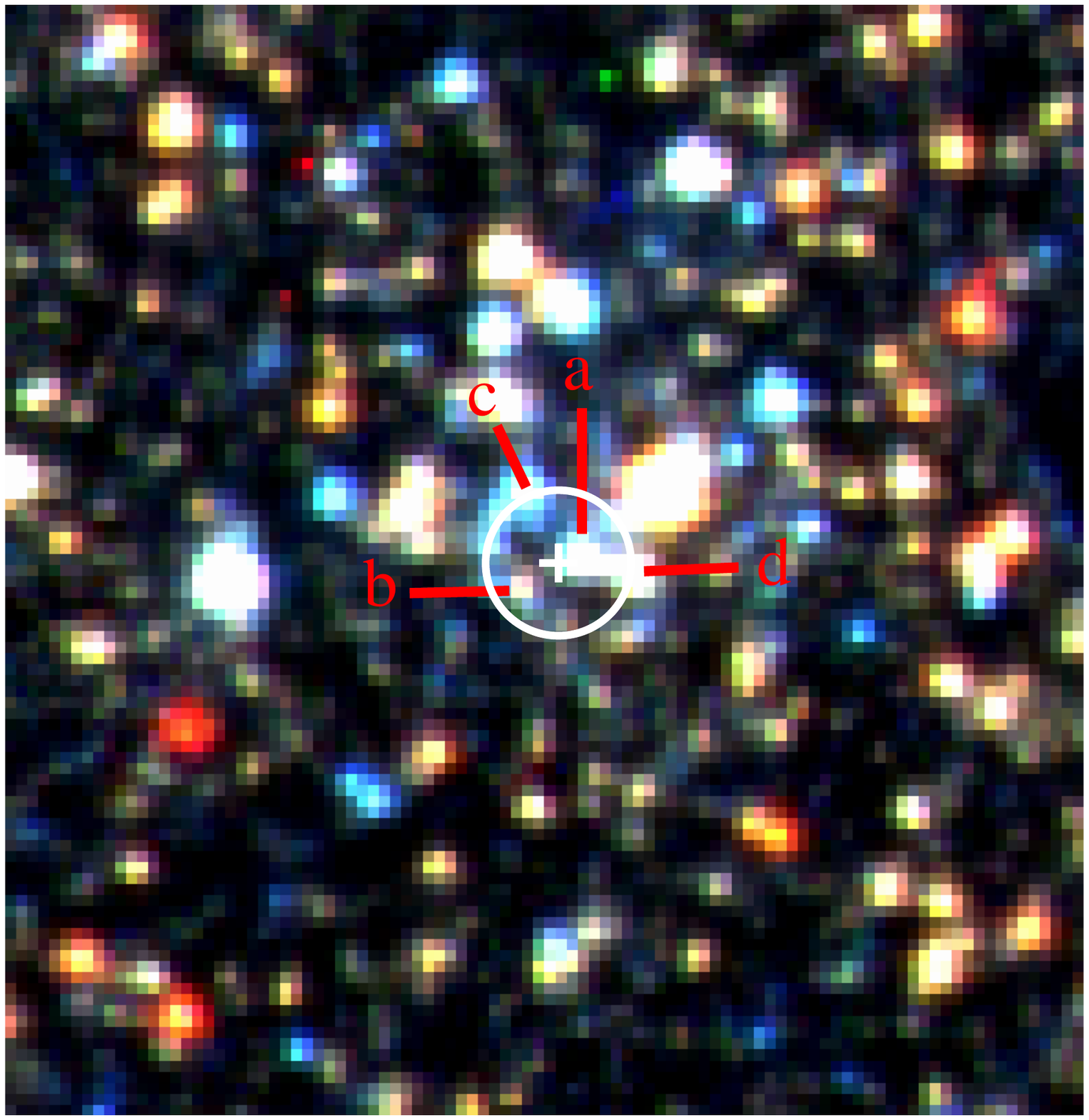} &
\includegraphics[width=0.21\linewidth,clip=true,trim=2.7cm 5.2cm 2.7cm 5.2cm]{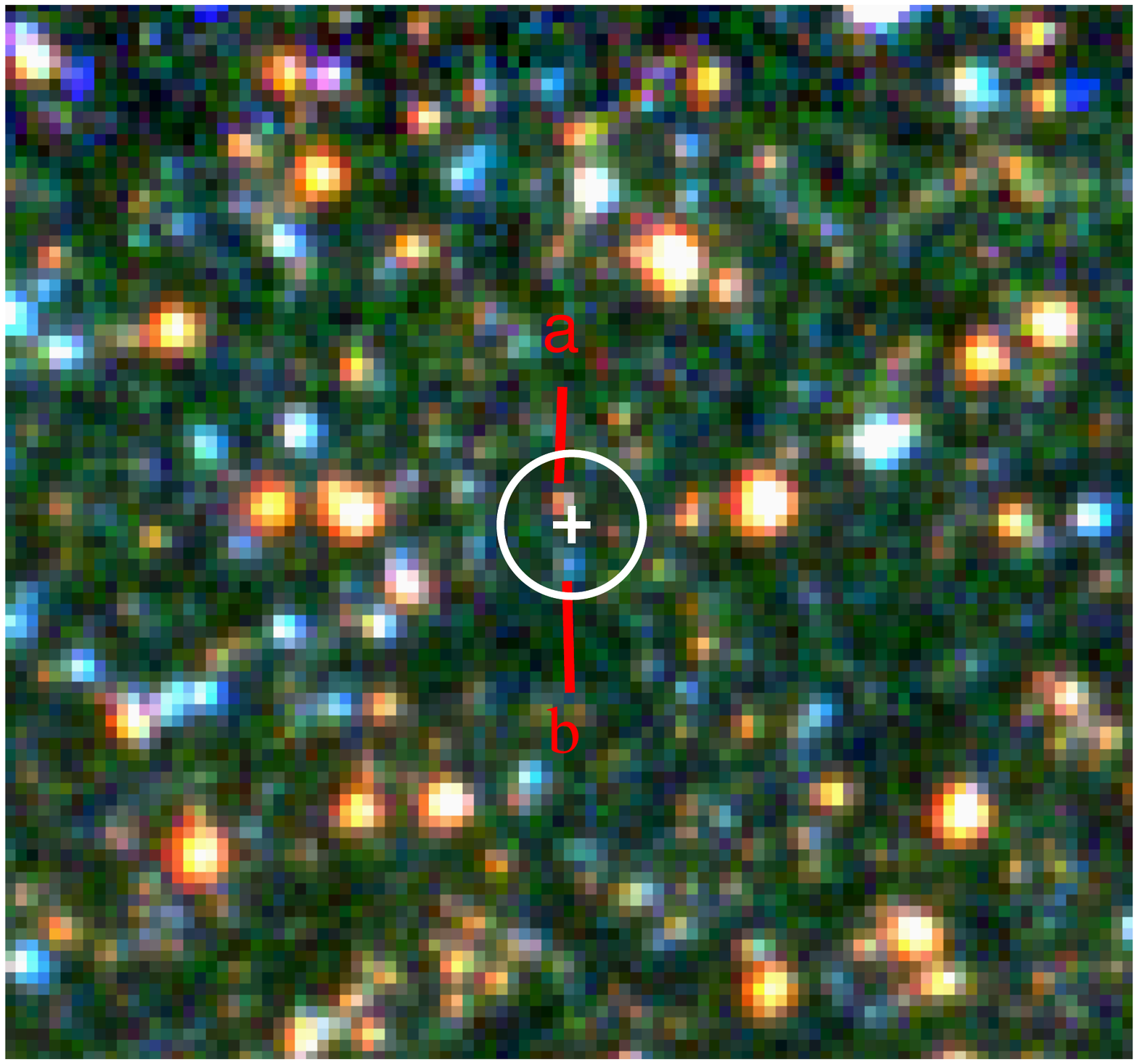} \\
Source 43 & Source 54 & Source 58 & Source 63 \\
\includegraphics[width=0.21\linewidth,clip=true,trim=2.7cm 5.2cm 2.7cm 5.2cm]{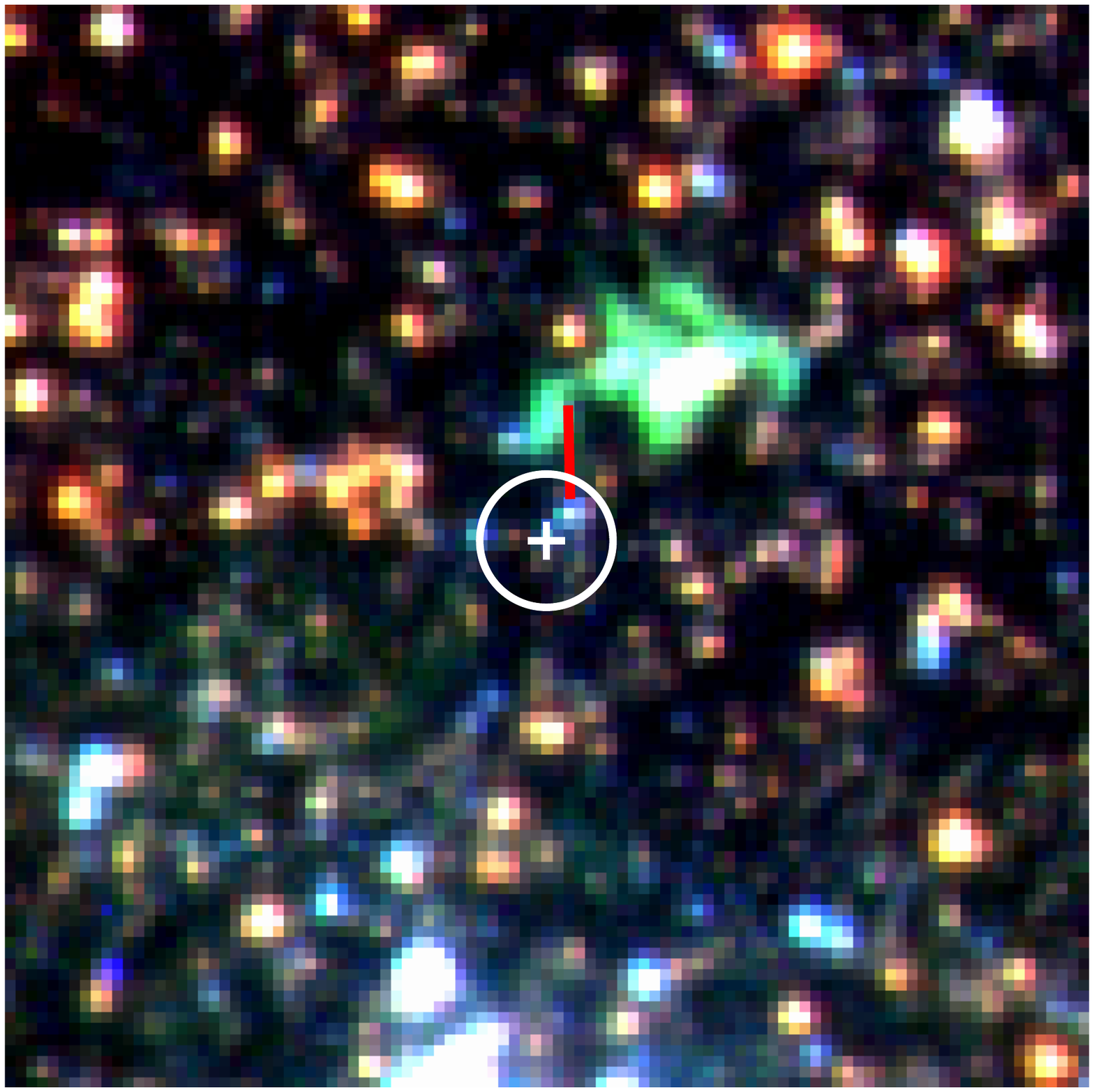} &
\includegraphics[width=0.21\linewidth,clip=true,trim=2.7cm 5.2cm 2.7cm 5.2cm]{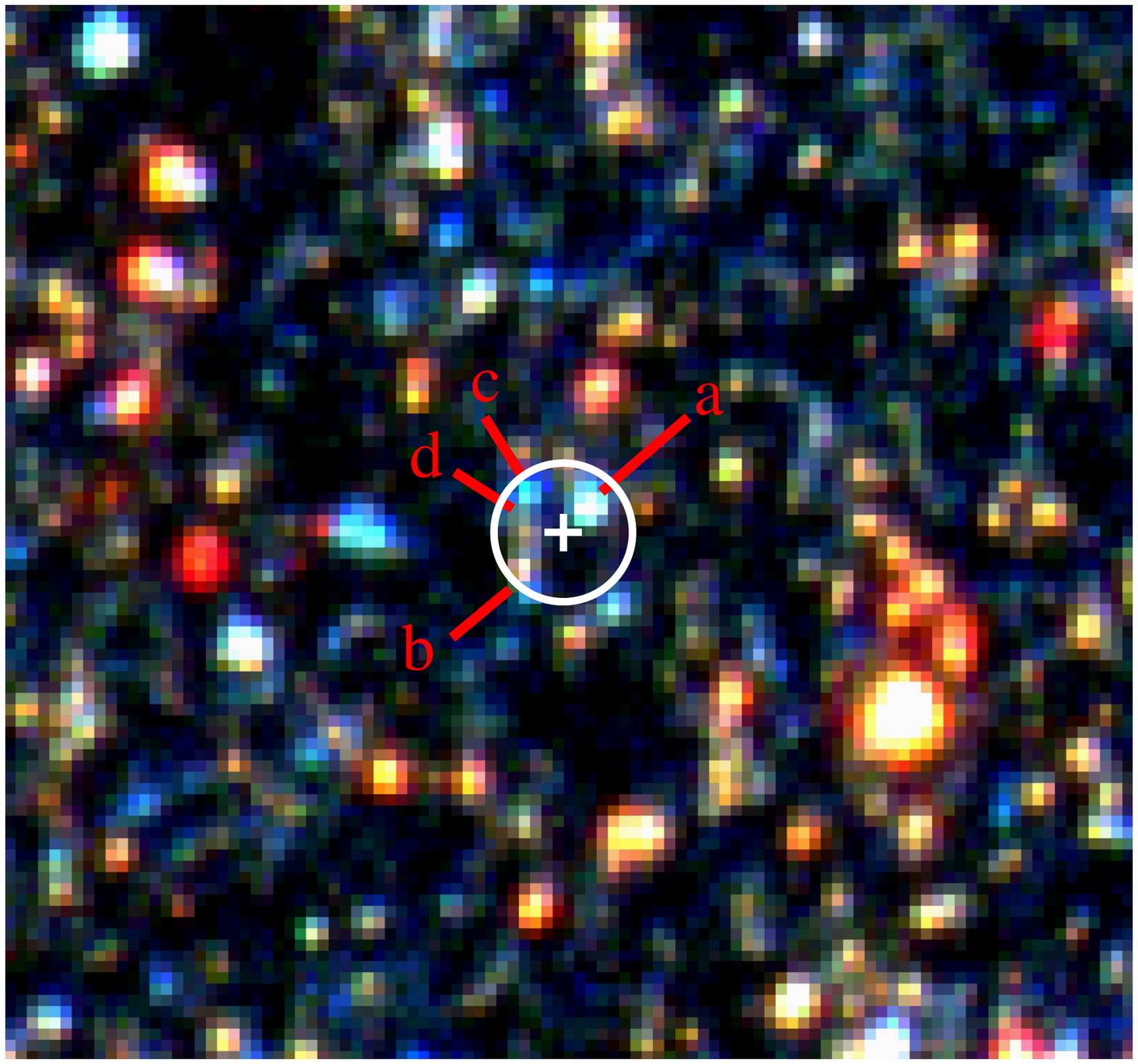} &
\includegraphics[width=0.21\linewidth,clip=true,trim=2.7cm 5.2cm 2.7cm 5.2cm]{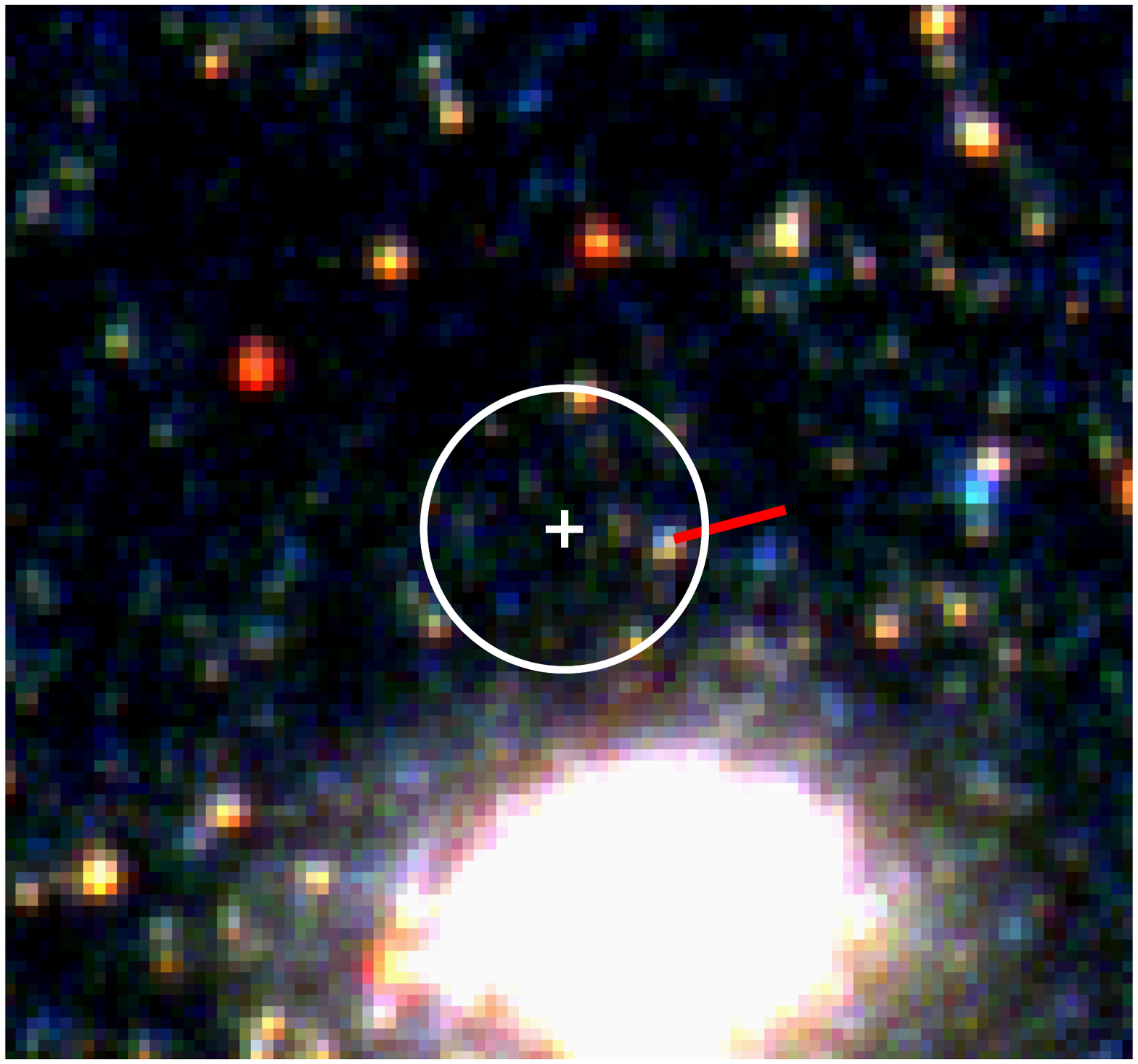} &
\includegraphics[width=0.21\linewidth,clip=true,trim=2.7cm 5.2cm 2.7cm 5.2cm]{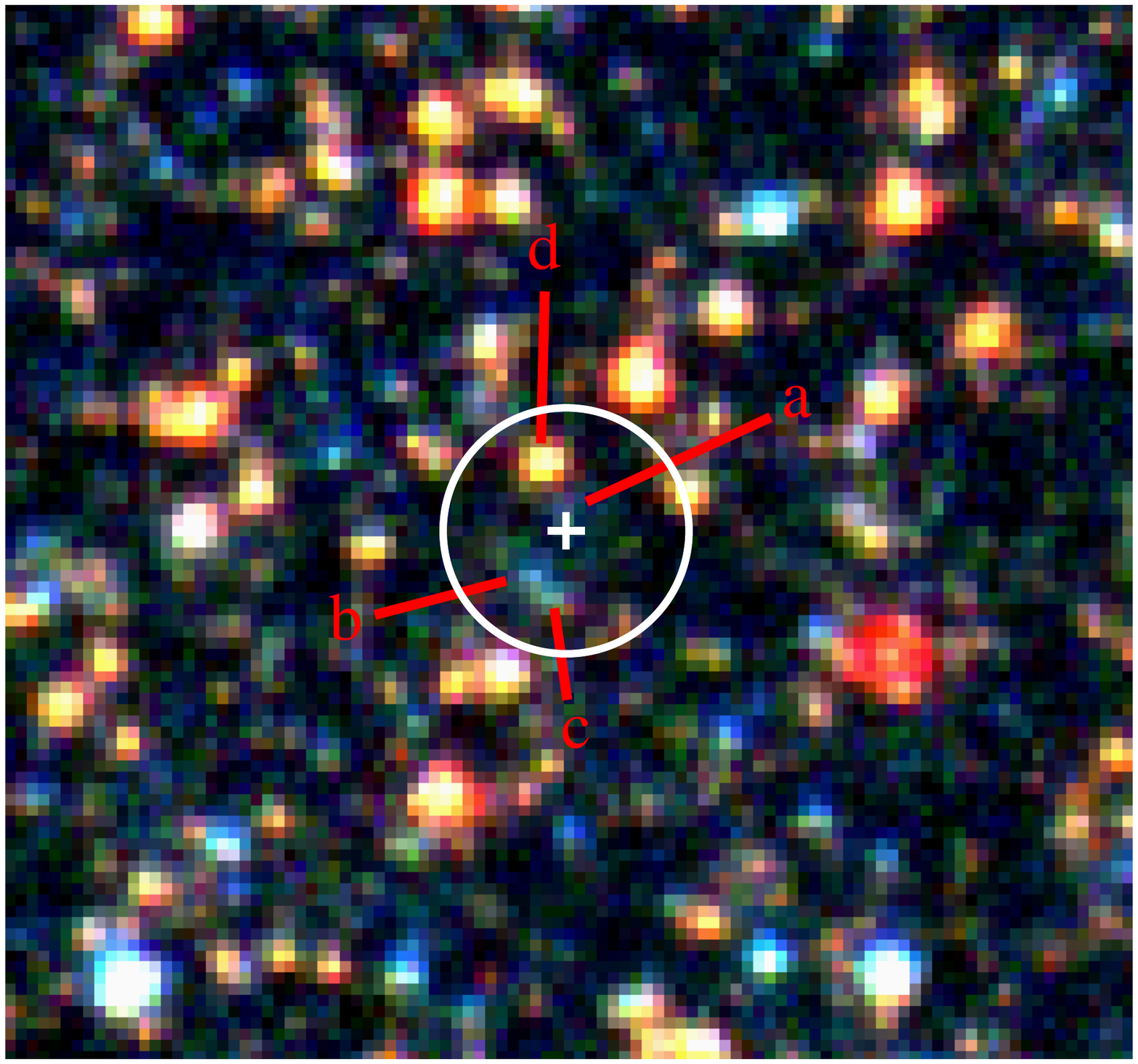} \\
Source 64 & Source 68 & Source 72 & Source 95 \\
\includegraphics[width=0.21\linewidth,clip=true,trim=2.7cm 5.2cm 2.7cm 5.2cm]{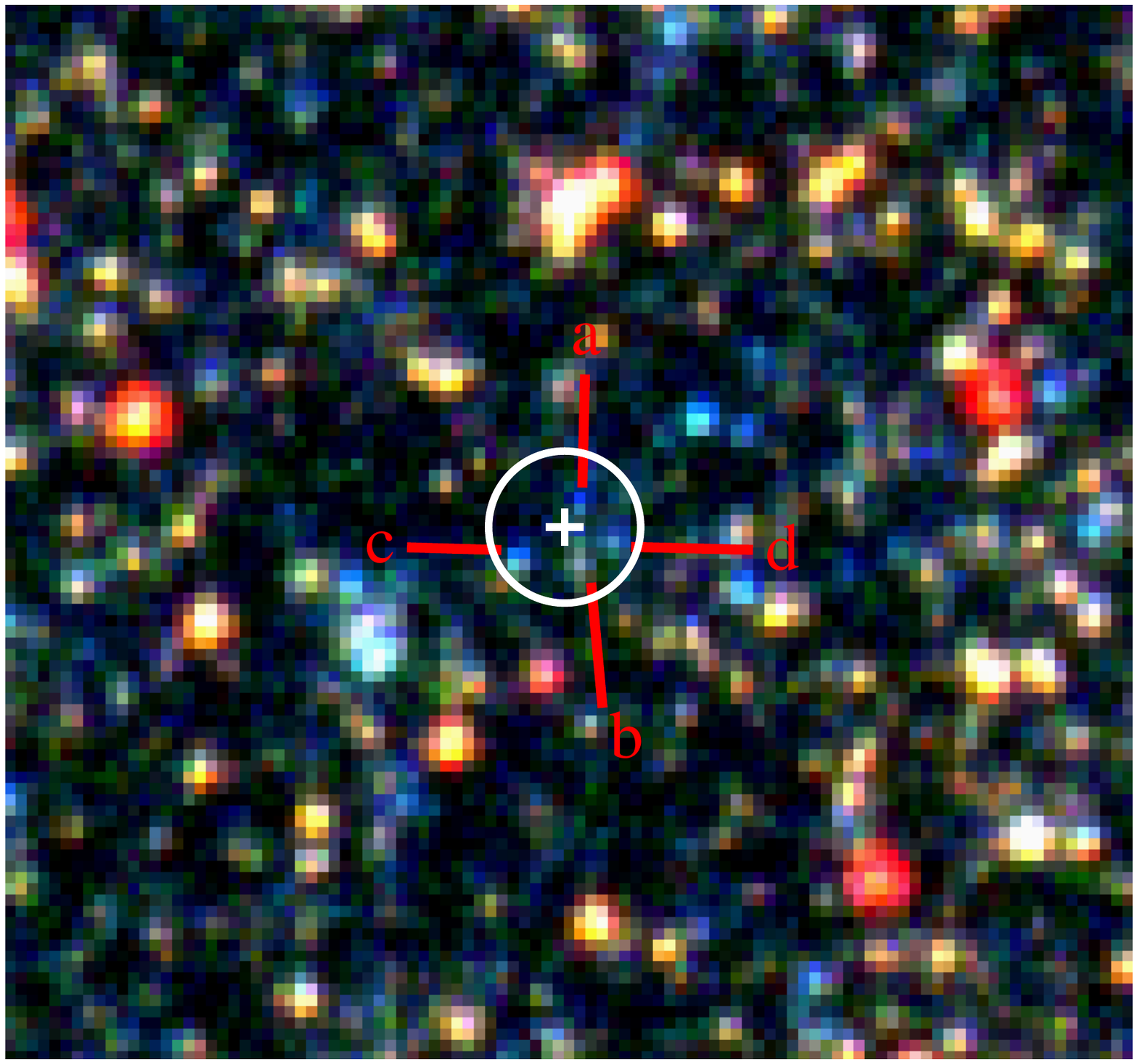} &
\includegraphics[width=0.22\linewidth,clip=true,trim=2cm 5cm 2cm 5cm]{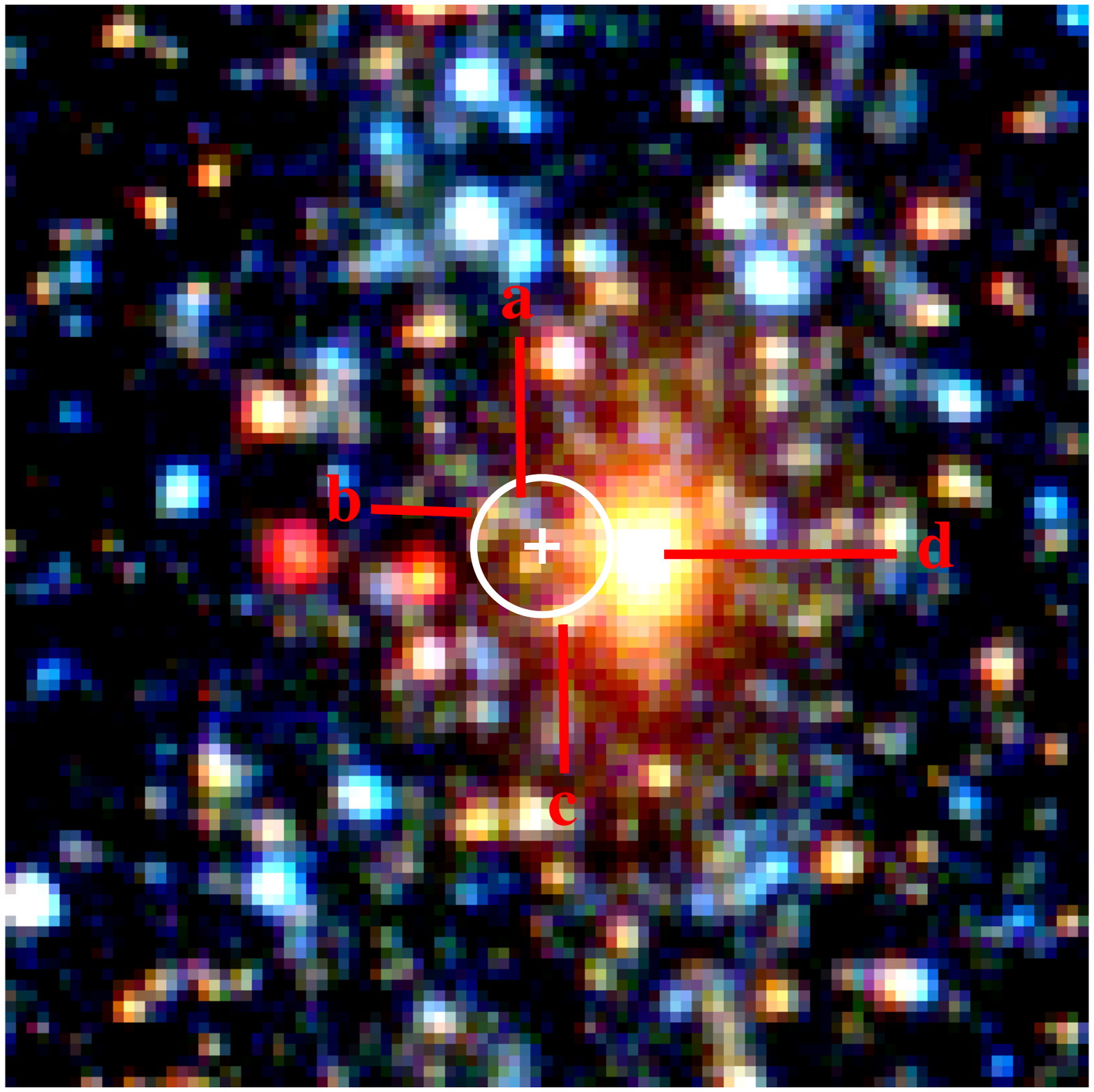} &
\includegraphics[width=0.21\linewidth,clip=true,trim=2.7cm 5.2cm 2.7cm 5.2cm]{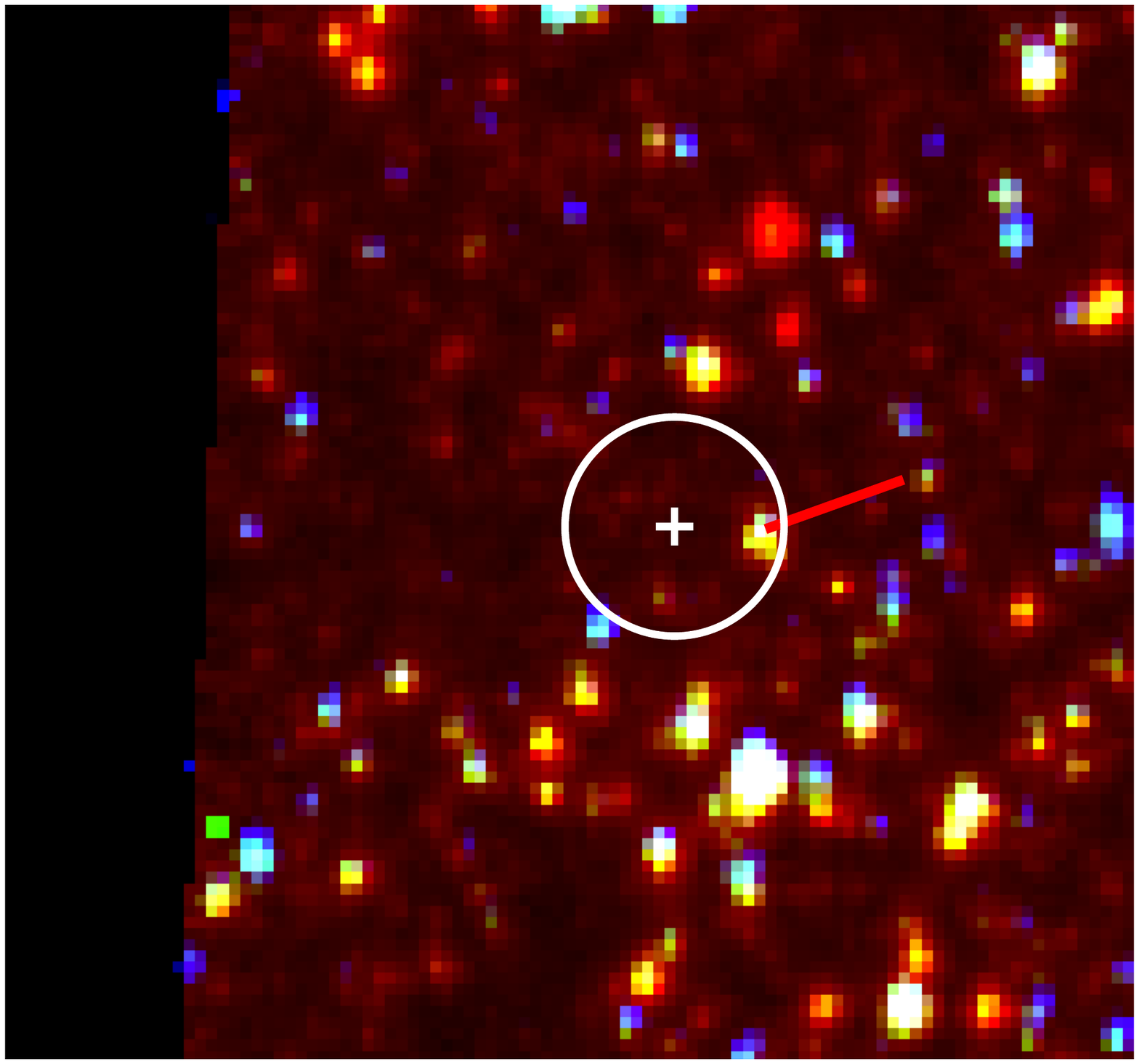} &
\includegraphics[width=0.21\linewidth,clip=true,trim=2.7cm 5.2cm 2.7cm 5.2cm]{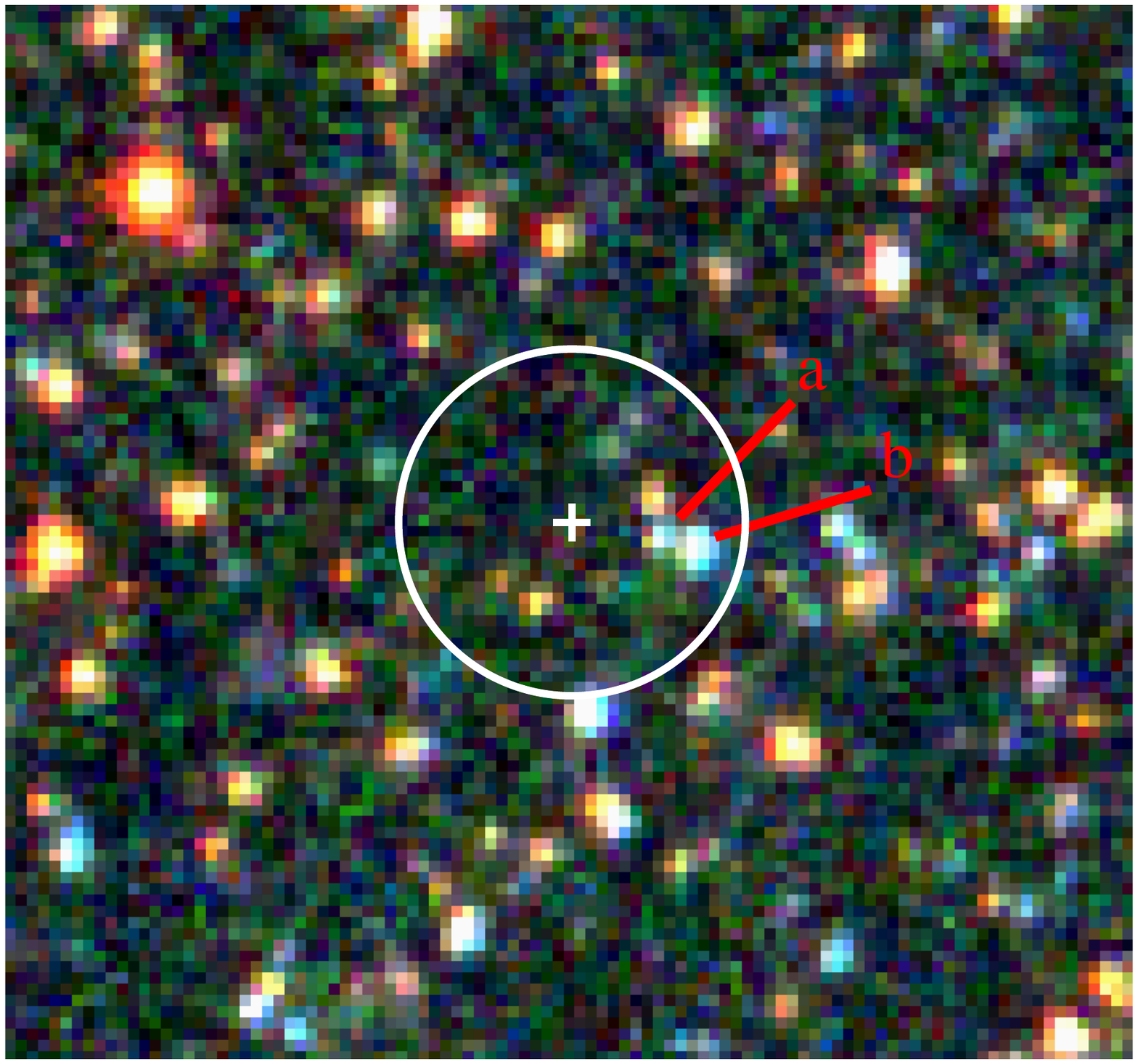} \\
\end{tabular}
\caption{(continued) Optical RGB-rendered finding charts at the location of each X-ray source, indicated by a cross. White circles show the radius of the 90\% \Chandra error circle for each source. Candidate optical counterparts are indicated with red lines. X-ray sources with multiple candidate optical counterparts have their optical counterparts labeled `a,' `b,' `c,' etc.}
\end{figure*}

Two of the X-ray sources (4 and 8) were coincident with a saturated foreground star and the disk of a background galaxy (with one visible point source nearly coincident with the X-ray position), respectively. Since both of these optical sources failed to meet the \HST quality criteria of a stellar object in NGC~300, they were not included in the \HST star catalogs. We used the IRAF task \texttt{imexamine} to estimate the magnitudes of the foreground star, the whole background galaxy, and the point source within the X-ray error circle centered on the disk of the background galaxy. The foreground star and background galaxy magnitudes were found to be consistent with those available in SIMBAD. Multi-band photometric measurements for each of the optical counterpart candidates to the X-ray sources are given in Table~\ref{optical_counterparts}. Source 4 is discussed further in \S\ref{src4}.
 
\subsection{X-ray/Optical Source Classifications}\label{xoptclasstext}
The detection of optical counterpart candidates for each X-ray source with overlapping \HST fields allows some sources to be assigned a source classification based on their multiwavelength properties. However, many X-ray sources are faint or do not have optical coverage, making their physical nature ambiguous. We therefore attempt to combine statistical quantities (i.e., using the radial source distribution, location in spiral arms, inter-arm regions, or background regions) with individual source properties (variability, X-ray spectral shape, and optical colors) to determine the most likely source classification.

We first consider the statistical properties of each X-ray source. For example, an X-ray source with a galactocentric radius less than 4 kpc that is coincident with a spiral arm is likely to be associated with NGC~300, while an X-ray source at a large galactocentric radius that is not obviously associated with NGC~300 is likely to be a background object. As described in Section~\ref{radial_src}, we expect the majority of X-ray sources at radii beyond 4 kpc are not associated with the stellar disk of NGC~300. We then examined ground-based, optical $R$-band and H$\alpha$ images to identify spiral arms and inter-arm regions in NGC~300, and define background regions not obviously associated with NGC~300. All sources with $r>4.5$ kpc ($\theta>$ 7\amm7) were located in `background' regions according to the ground-based imaging, making these sources probable background AGN.

We next consider the individual X-ray properties of each source. Long-term X-ray variability is observed for both AGN and XRBs. However, SNRs and some HMXB systems do not show significant X-ray variability \citep[i.e., NGC~300 X-1; see][and references therein]{Binder+11b}. We therefore do not discount the possibility of an XRB nature for non-variable sources. The data on galactocentric radius, location relative to NGC~300, and X-ray variability is then combined with the preliminary source classifications defined using our hardness ratio analysis (see Section~\ref{hrsec}). For sources without overlapping \HST exposures, we then assign a likely source classification based on our X-ray observations. For example: a source with $r=1$ kpc, located within a spiral arm, no significant long-term X-ray variability, and a soft X-ray spectral shape is classified as a likely SNR.

To determine a likely source classification for X-ray sources with overlapping \HST fields, we generated color-magnitude diagrams (CMDs) of each \HST field and candidate optical counterparts were classified as falling either on the main sequence (MS) or red giant branch (RGB). Additionally, we constructed X-ray color-magnitude diagrams (XCMDs), where the optical color was plotted as a function of $log(f_X/f_V)$. Many authors have utilized optical and X-ray colors in discriminating between XRBs and background AGN \citep[e.g., ][]{Horn+01,Shty+05}; notably, \cite{McGowan+08} found that known pulsars and HMXBs located in the Small Magellanic Cloud occupied a region of the XCMD bounded by $B-V\lesssim0$ and $f_X/f_V\lesssim1$.

\begin{figure*}
\centering
\begin{tabular}{cc}
\includegraphics[width=0.5\linewidth,clip=true,trim=2cm 12.5cm 1cm 1cm]{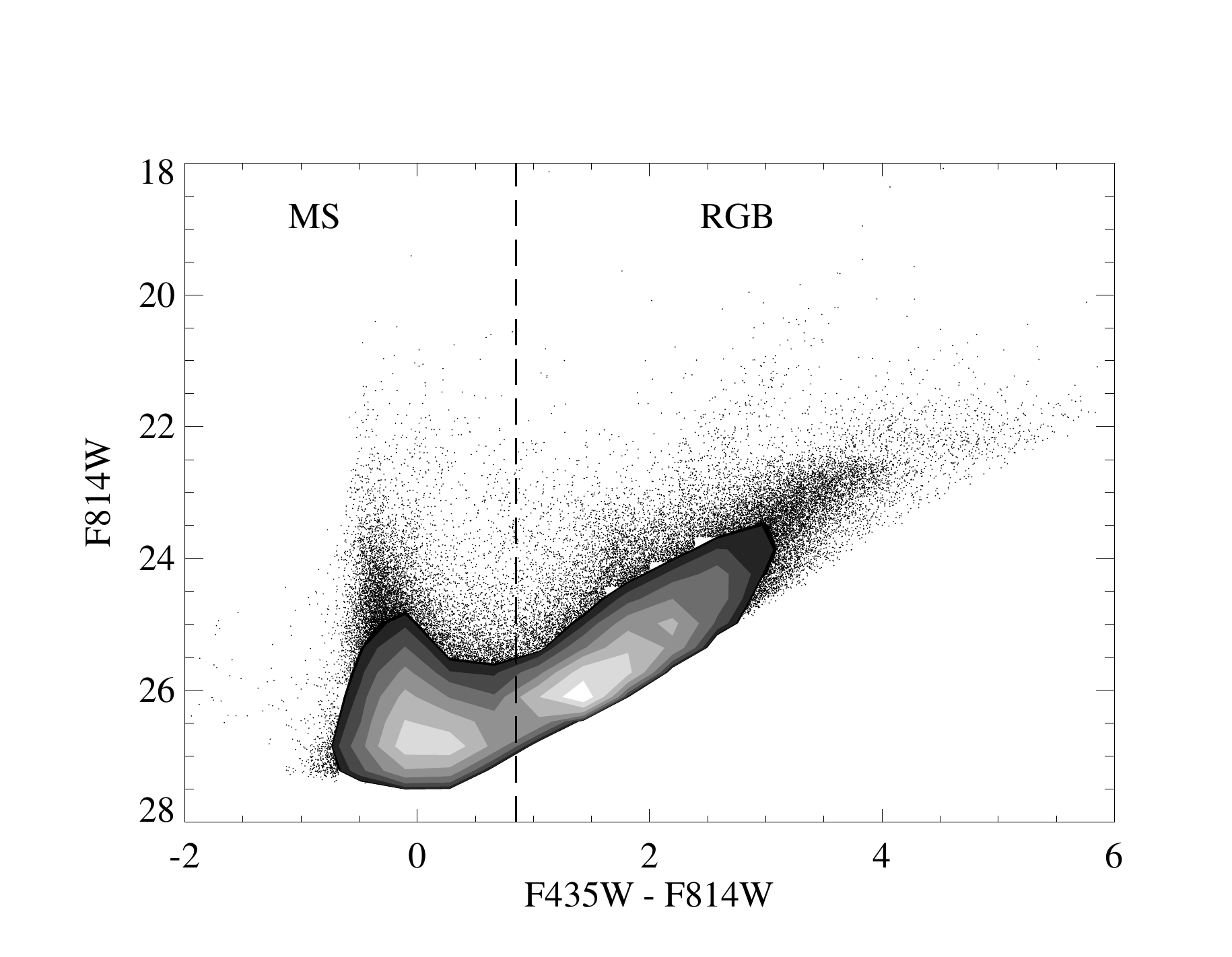} &
\includegraphics[width=0.5\linewidth,clip=true,trim=2cm 12.5cm 1cm 1cm]{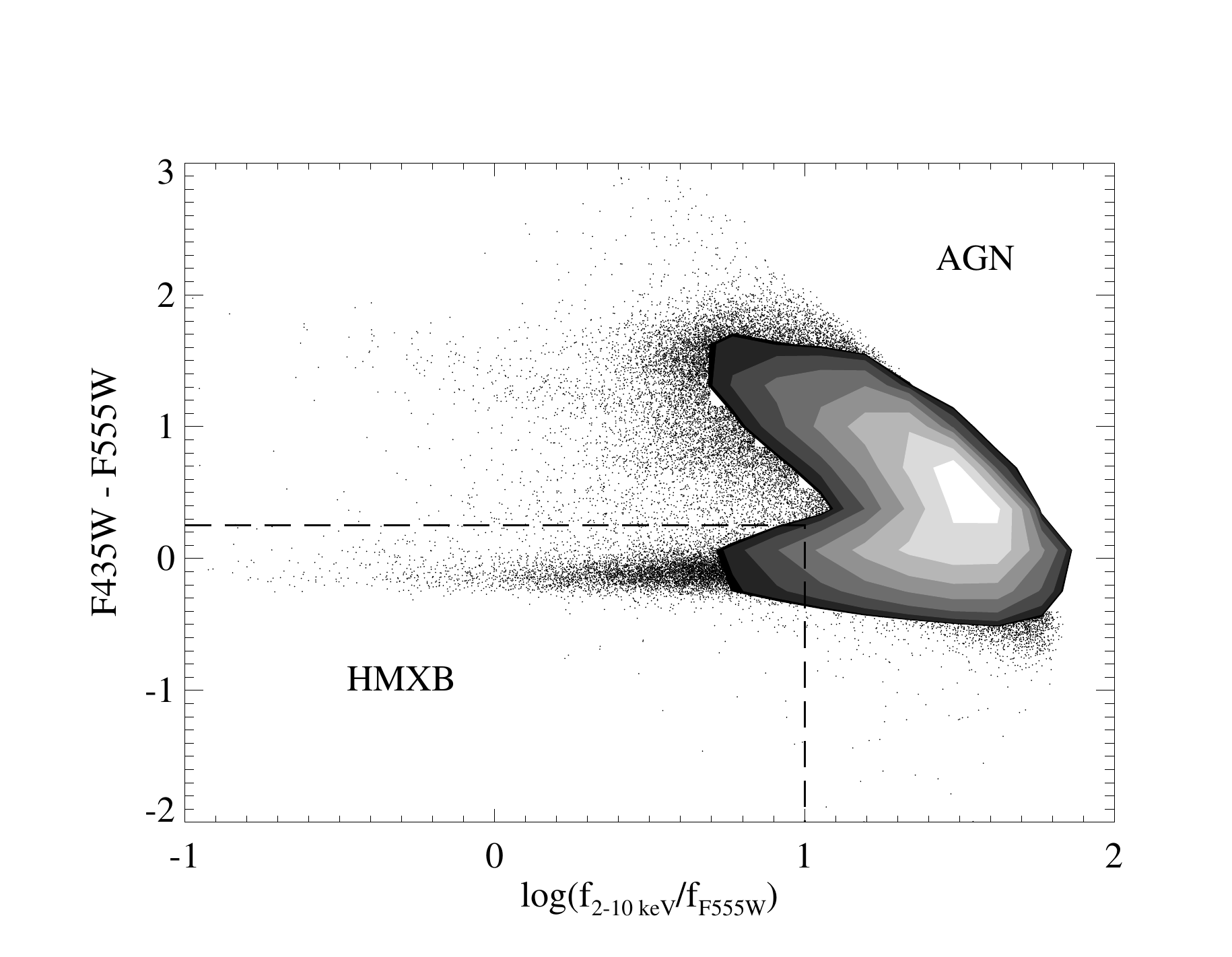} \\
\includegraphics[width=0.5\linewidth,clip=true,trim=2cm 12.5cm 1cm 1cm]{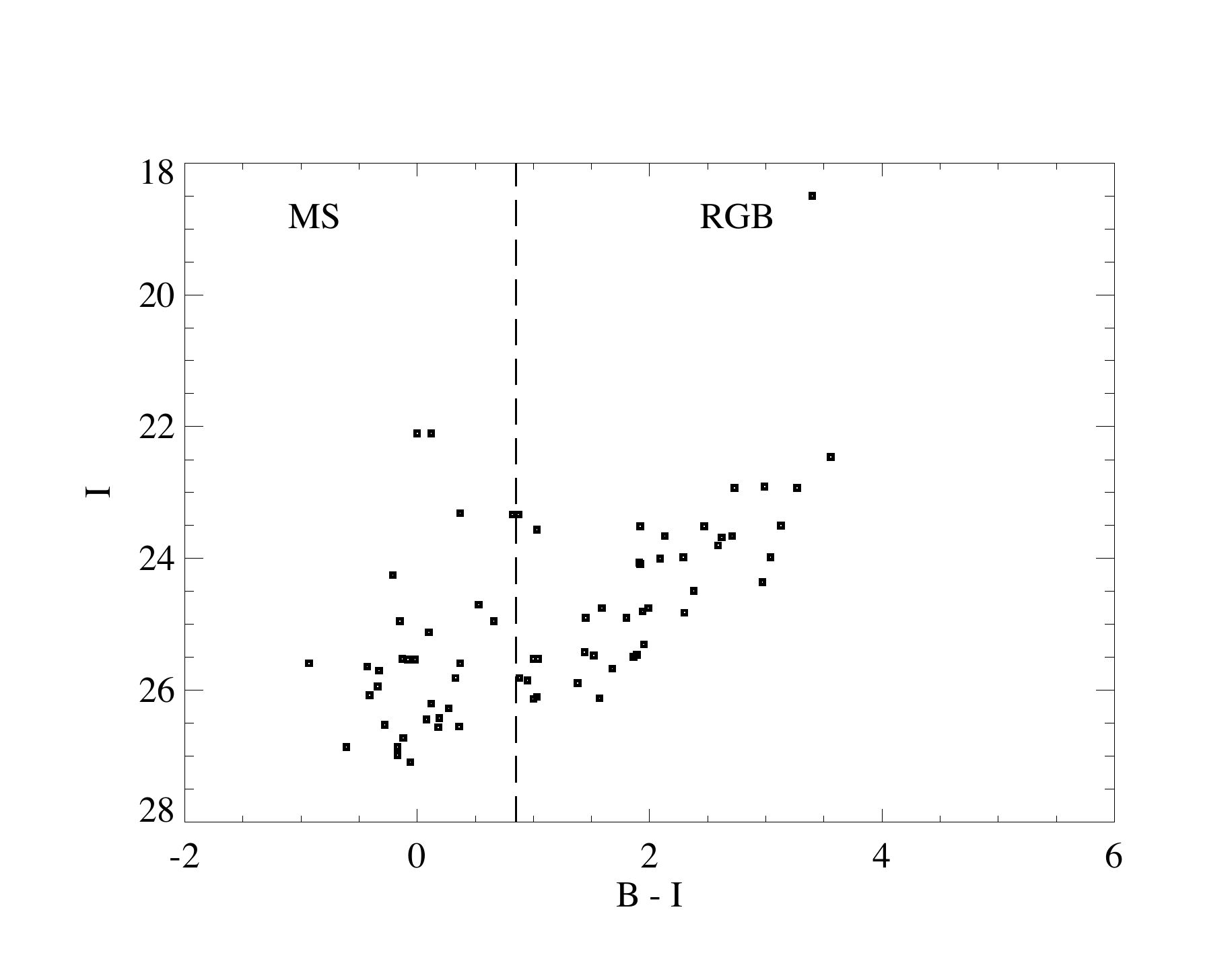} &
\includegraphics[width=0.5\linewidth,clip=true,trim=2cm 12.5cm 1cm 1cm]{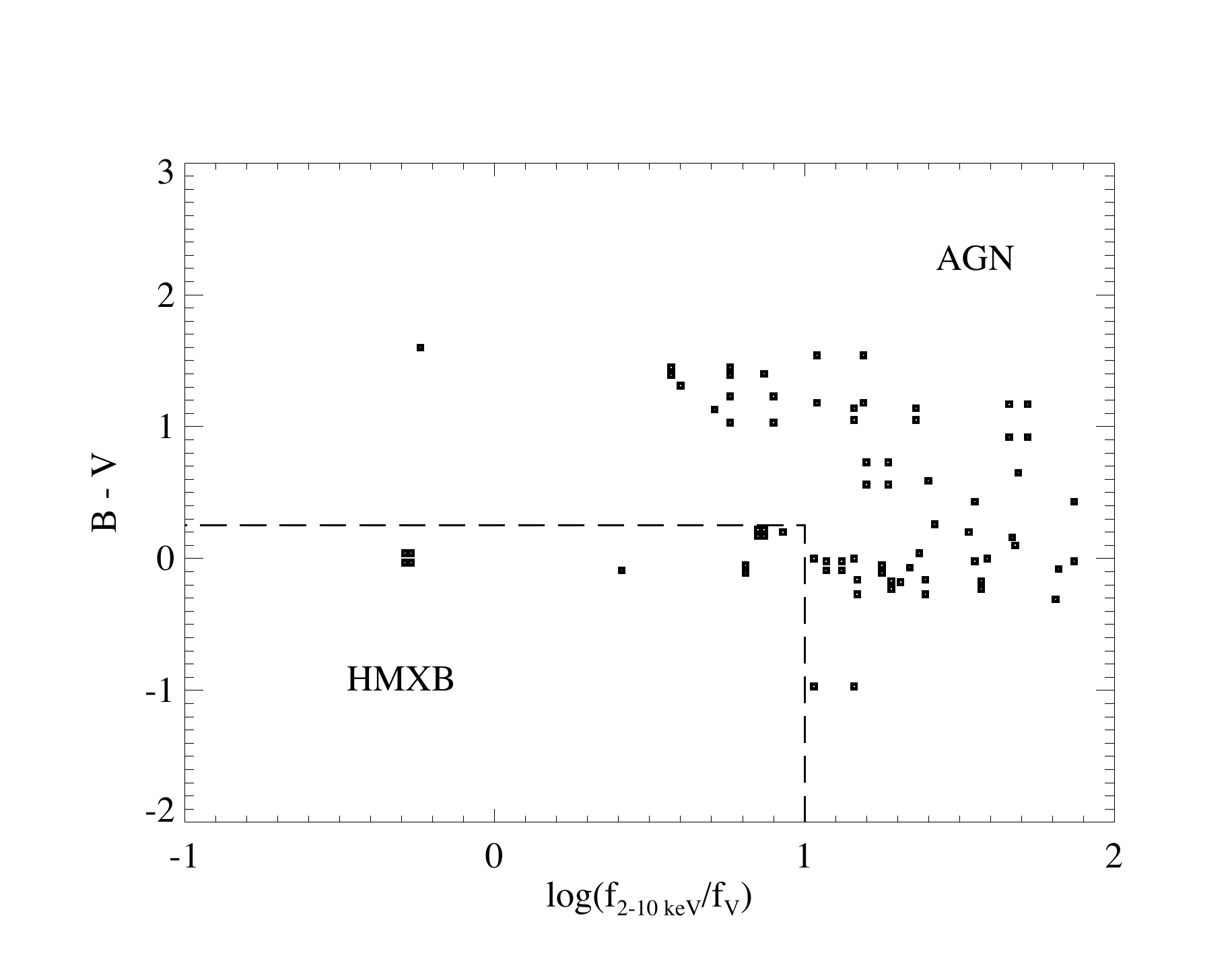} \\
\end{tabular}
\caption{Top row: a sample color-magnitude diagram (CMD; left) and X-ray color-magnitude diagram (XCMD; right). The MS and RGB regions are marked on the CMD, and the HMXB and AGN regions are marked on the XCMD. Bottom row: the locations of our candidate optical counterparts as either MS or RGB stars on a CMD (left), and an XCMD (right) showing the actual values of the X-ray/optical flux ratio for our candidate optical counterparts to each X-ray source.}
\label{CMDs}
\end{figure*}

Optical counterpart candidates were then compared to the XCMD generated for each \HST field. Sources with blue optical colors ($F435W-F555W$ or $F475W-F606W\lesssim0.2$) and $log(f_X/f_V)<1$ are likely to match a young massive main sequence star and are therefore designated as HMXB candidates. Sources falling in the other region of the XCMD are associated with either red giant branch (RGB) stars or background galaxies, and are therefore classified as either AGNs or LMXBs. For illustration, Figure~\ref{CMDs} presents a sample CMD and XCMD for the field `NGC300-5' (top panel) as well as a CMD and XCMD of our X-ray/optical source candidates. The sample CMD was generated using the \HST $F435W$ and $F814W$ catalog magnitudes for all optical sources in the `NGC300-5' field. For illustrative purposes, we have created a sample XCMD using the mean X-ray flux of sources in the NGC~300 catalog ($\sim5\times10^{-15}$ \flux, corresponding to $\sim2\times10^{36}$ \lum at the distance of NGC~300). Using this X-ray flux, we calculate a $log(f_X/f_V)$ for every optical source in the \HST `NGC300-5' field. The bottom panels show the magnitudes, colors, and X-ray/optical flux ratios for the candidate optical counterparts only. Regions corresponding to likely main sequence (MS) and RGB stars are labeled on the CMD, and the regions defining likely HMXBs and AGNs are labeled on the XCMD.

In order to combine all the relevant X-ray and optical information available from our \Chandra and \HST observations, we constructed an IDL routine called \texttt{xclass.pro} to automatically process each X-ray source. The routine reads in an input source file containing all X-ray and optical information (if available), and returns a preliminary source classification. The routine additionally tracks which classification channels were utilized by the input sources. The \texttt{xclass.pro} routine is publicly available\footnote{See \url{http://www.astro.washington.edu/users/bbinder/xclass/}}.

The source classification reported by the \texttt{xclass} routine was then compared to the classification determined by our hardness ratio analysis. In some cases, the multiwavelength and hardness ratio classifications disagreed. This was most commonly seen with sources classified as XRBs by our hardness ratio analysis but with multiwavelength properties consistent with a background AGN. In these cases, we favor the AGN classification. Sources that are initially classified as AGN but which exhibit a soft X-ray spectrum (i.e., classified as `SOFT' in our HR analysis) may be potential LMXBs. Only one of our HR-classified SNRs shows X-ray and optical characteristics inconsistent with a SNR origin, and may be an unusually soft background AGN or a foreground star. Our final source classification yields 46 background AGN candidates, 26 HMXB candidates, 11 LMXB candidates, 10 SNRs, and one foreground star. A summary of all X-ray and optical data and our final source classification is present in Table~\ref{optical_id}. 

\begin{table*}[ht]
\centering
\caption{X-ray Point Source Classifications}
\begin{tabular}{ccccccccccc}
\hline \hline
Source	& $r^a$	& Location$^b$	& Variability$^c$ & In \HST	& \# Optical	& Location& Location	& Prelim.	& ID. from	& Final	\\
No.		& (kpc)	& in NGC~300	&			& footprint?	& Counterparts	& on CMD	& on XCMD	& ID		& HRs	& ID		\\
(1)		& (2)		& (3)			& (4)			& (5)			& (6)			& (7)			& (8)		& (9)		& (10) 	& (11)	\\
\hline
1		& 5.2		& bkg?	& none		& no			& 			&		&			& AGN?		& ABS 	&	AGN \\
2		& 6.7		& bkg?	& long		& no			&			&		&			& AGN		& XRB	&	AGN	 \\
3		& 5.5		& bkg?	& none		& yes		& 2 			& RGB 	& AGN		& AGN 		& SOFT	&	LMXB? \\
4		& 6.7		& bkg?	& long		& yes		& 1 			&  		&  			& Galaxy		& ABS	&	Galaxy/FS? \\
5		& 4.6		& bkg	& none		& yes		& 10		 	& MS, RGB 	& HMXB, AGN	& HMXB? & XRB	&	HMXB \\
\hline \hline
\label{optical_id}
\end{tabular}
\tablecomments{
$^a$Inclination-corrected galactocentric radius; the X-ray point source population at $r>4$ kpc is expected to be dominated by background AGN.\newline
$^b$Location in NGC~300 was determined by looking at ground-based $R$-band and H$\alpha$ imaging. Locations are described as spiral arm, inter-arm regions, or background. X-ray sources that fall beyond the extent of the optical images are labeled as `background?'\newline
$^c$Variability is classified as `long'-term, `rapid,' or `none.'\newline
$^d$SN 2010da, see \cite{Binder+11a}.\newline
$^e$NGC~300 X-1, see \cite{Binder+11b}.\newline
Only the first five entries are shown.}
\end{table*}

We use the \cite{Cappelluti+09} AGN \lognlogs distribution to estimate the expected number of AGN in our X-ray source catalog to be $\sim$50. The similarities in both the X-ray and optical properties of AGN and LMXBs (especially at the distance of NGC~300) is likely to cause confusion in separating these two populations. With 47 AGN and 11 LMXB candidates obtained using our source classification scheme, we estimate our X-ray/optical classified \lognlogs distribution to differ from the statistical by less than 10\%. Our source classification scheme correctly classifies both NGC~300 X-1 (source 80) and SN 2010da (source 73) as HMXBs, which were previously identified as HMXBs in earlier works \citep[see][]{Binder+11a,Binder+11b}. We are therefore confident that our source classification scheme provides the most reliable indicator of the physical nature of the NGC~300 X-ray sources possible with the available data.

\subsubsection{Source 4: Foreground or Background?}\label{src4}
As can be seen in Figure~\ref{opticalIDs}, X-ray source 4 is coincident with the disk of a background galaxy. This source was not detected by \XMM, suggesting a large brightening by a factor of $\gtrsim22$. The source has hardness ratios consistent with those of other XRBs or background AGN. We consider three possible scenarios for the physical nature of source 4: (1) a background, high-redshift quasar, (2) an ultra-luminous X-ray (ULX) source located within the disk of the background galaxy, and (3) a flaring foreground M star.

We first consider the high-redshift quasar interpretation. Source 4 would have a typical quasar X-ray luminosity of $\sim10^{45}$ \lum at a luminosity distance of $\sim26$ Gpc, roughly $z\approx3$. However, high-redshift quasars typically exhibit variations in X-ray luminosity on the order of $\sim15$\%, and extreme changes (such as seen in source 4) are rare \citep{Gibson+12}.

To evaluate the likelihood of this source being a ULX, we assume the background spiral galaxy has a physical diameter of $\sim$22 kpc. We measure the angular diameter of the galaxy to be $\sim$6\as5 across, implying the distance to the host galaxy is $\sim$700 Mpc. At this distance, source 4 would have a 0.35-8 keV luminosity of $L_X\sim7\times10^{41}$ \lum. For a source radiating at the Eddington luminosity, this implies a mass of $\sim$5000 \Msun. The factor of $\sim22$ brightening observed in this source is not typical of ULXs, with the notable exception of HLX-1 in the galaxy ESO 243-49 which has exhibited variations in X-ray luminosity of a factor of $\sim$20-50 \citep{Lasota+11}. The high luminosity of HLX-1 is interpreted as mass accretion via Roche-lobe overflow onto a $\sim10^4$ \Msun BH. We therefore find the ULX interpretation plausible.

Young, active M dwarfs in the Milky Way undergo X-ray flares, and may provide a natural explanation for the factor of $\sim22$ increase in X-ray luminosity exhibited by source 4. Assuming a typical M-dwarf $M_V\sim10$ and our estimated $m_V\sim23.8$ for source 4 implies a distance of $\sim60$ pc, yielding an X-ray luminosity of $\sim5\times10^{27}$ \lum. The implied X-ray luminosity and color ($F435W-F555W = 0.9$) is typical of a flaring late-type star \citep[typically $10^{27-29}$ \lum with $B-V\sim0.5-1$,][]{Singh+99}. Using XSPEC, we simulate the spectrum of a stellar corona using the \texttt{MEKAL} model with similar signal-to-noise as source 4 and estimate the hardness ratios produced. We find the hardness ratios of source 4 to be consistent with the predicted hardness ratios, although at low signal-to-noise the spectrum cannot be distinguished from a power law. While we cannot rule out the possibility of an AGN or ULX origin, our observations are most consistent with being a flaring, late-type star within the Milky Way.

\subsection{Comparison to Other X-ray Point Source Populations}
It is interesting to compare the properties of the NGC~300 point source population to other galaxies with a range of SFHs, morphologies, and stellar masses ($M_*$). Since HMXBs serve as a tracer of young stellar populations, their numbers should correlate with the recent SFH of their host galaxy \citep{Grimm+03}, while the number of longer-lived LMXBs should correspondingly correlate with the total stellar mass of the host galaxy \citep{Gilfanov04}. Typically, the ratio $L_{\rm 2-10 keV}$/SFR is taken as a measure of the HMXBs in a galaxy, while LMXBs are measured through $L_{\rm 2-10 keV}/M_*$. Both these ratios are found to correlate with the morphology of the host galaxy \citep{Grimm+02,Gilfanov04}.

We estimate the 2-10 keV luminosity for our X-ray/optically classified HMXB candidates to be $\sim7\times10^{37}$ \lum and, using equation (22) of \cite{Grimm+03}, we derive a predicted SFR for NGC~300 of 0.12 \Msun yr$^{-1}$. This predicted SFR is in excellent agreement with the average recent SFR of 0.12$\pm$0.05 \Msun yr$^{-1}$ measured by \cite{Gogarten+10}. Normalized to the SFR, the HMXBs in NGC~300 emit $\sim6\times10^{38}$ \lum/(\Msun yr$^{-1}$) in the 2-10 keV band, consistent with the value measured for the Milky Way by \cite{Grimm+02}. 

We find luminosity of our LMXB candidates normalized to the stellar mass of NGC~300 to be $\sim9\times10^{27}$ \lum \Msun$^{-1}$, a factor of $\sim$5 lower than is observed in the Milky Way \citep{Grimm+02}. This value of LMXB luminosity normalized by stellar mass is a lower limit, as our source classification scheme is unable to discriminate very well between LMXBs and background AGN. We can provide an upper limit to LMXB luminosity per unit stellar mass by assuming all AGN candidates are actually LMXBs -- in this case, the luminosity normalized by the stellar mass of NGC~300 is $\sim6.5\times10^{28}$ \lum \Msun$^{-1}$. NGC~300 lacks the prominent nuclear bulge present in the Milky Way (where older LMXBs dominate the X-ray source population), and the gas in NGC~300 (from which new HMXBs could be formed) accounts for $\sim$26\% of the baryonic mass of NGC~300 \citep{Westmeier+11}. Only $\sim15$\% of the baryonic mass of the Milky Way \citep{Flynn+06} consists of gas that could potentially form stars. Given the differences in stellar mass, gas content, morphology, and star formation history of NGC~300 and the Milky Way, this observed differences in LMXB populations is not surprising.

When we consider the total XRB population, we find the number of X-ray sources above 10$^{37}$ \lum ($N_X$) and the total X-ray luminosity ($L_X$, in units of 10$^{38}$ \lum) normalized by the stellar mass (in units of $10^{11}$ \Msun) of NGC~300 to be $N_X/M_*=93\pm10$ and $L_X/M_*=0.47\pm0.05$, respectively. \cite{Gilfanov04} reported similar ratios of $L_X/M_*$ for other late-type galaxies, although the average value of $N_X/M_*$ for the same galaxies was found to be $\sim160\pm50$. It should be noted that the comparison between the \cite{Gilfanov04} sample and NGC~300 comes with two significant caveats: first, the limiting X-ray luminosities of the galaxies examined by \cite{Gilfanov04} were $\gtrsim10^{37}$ \lum, and the total number of X-ray sources above $10^{37}$ \lum often had to be estimated; second, the stellar masses of the galaxies sampled were significantly larger than NGC~300 (often by more than an order of magnitude). Therefore, the ratios derived by \cite{Gilfanov04} may require modifications for deeper X-ray data (below $\sim10^{37}$ \lum), and for galaxies with stellar masses comparable to NGC~300.

\section{The \lognlogs Relation and the X-ray Luminosity Function}\label{XLF}
Numerous studies have shown a correlation between the shape of the XLF of a point source population and the SFH of the host galaxy \citep{Grimm+03,Eracleous+06}. Those galaxies with a high degree of recent SF have their X-ray emission dominated by young, luminous HMXBs, while older stellar populations show LMXBs as the dominant source of X-ray emission, although this relationship may not be monotonic over a few hundred Myr \citep{Belczynski+04}. Using the \Chandra X-ray point source catalog and overlapping \HST exposures, we aim to directly test the XLF slope-age correlation in NGC~300.

The cumulative number of sources above a given flux ($N(>S)$, where $S$ is in units of \flux) found in a survey covering a total geometric area $A$ can be evaluated as:

\begin{equation}
N(>S) = \sum\limits_{i,S_i>S}\frac{1}{A(S_i)}.
\end{equation}

\noindent $N(>S)$ has units of sources deg$^{-2}$ and is weighted by the survey area over which each source could have been detected. The $A(S)$ factor would account for survey completeness if the sources were distributed uniformly, although we do not expect this to be true for sources associated with NGC~300. We terminate the low flux end of the distribution at the 90\% limiting luminosity for each energy band considered. The sensitivity maps calculated in Section~\ref{sensmapsection} allow the area function $A(S)$ (the area over which a source with flux $S$ would be detectable) to be directly evaluated for each source.

We use two approaches to construct the XLF for the NGC~300 point source population. First, we calculate the cumulative \lognlogs distributions for the soft (0.5-2 keV) and hard (2-8 keV) bands using all 95 X-ray sources detected in our \Chandra observation. Using the background AGN distribution of \cite{Cappelluti+09}, we then statistically correct for the anticipated background contamination -- we hereafter refer to these XLFs as our ``statistical XLFs.'' 

\begin{figure*}
\centering
\begin{tabular}{cc}
\includegraphics[width=0.48\linewidth,clip=true,trim=2.5cm 12.5cm 2cm 2cm]{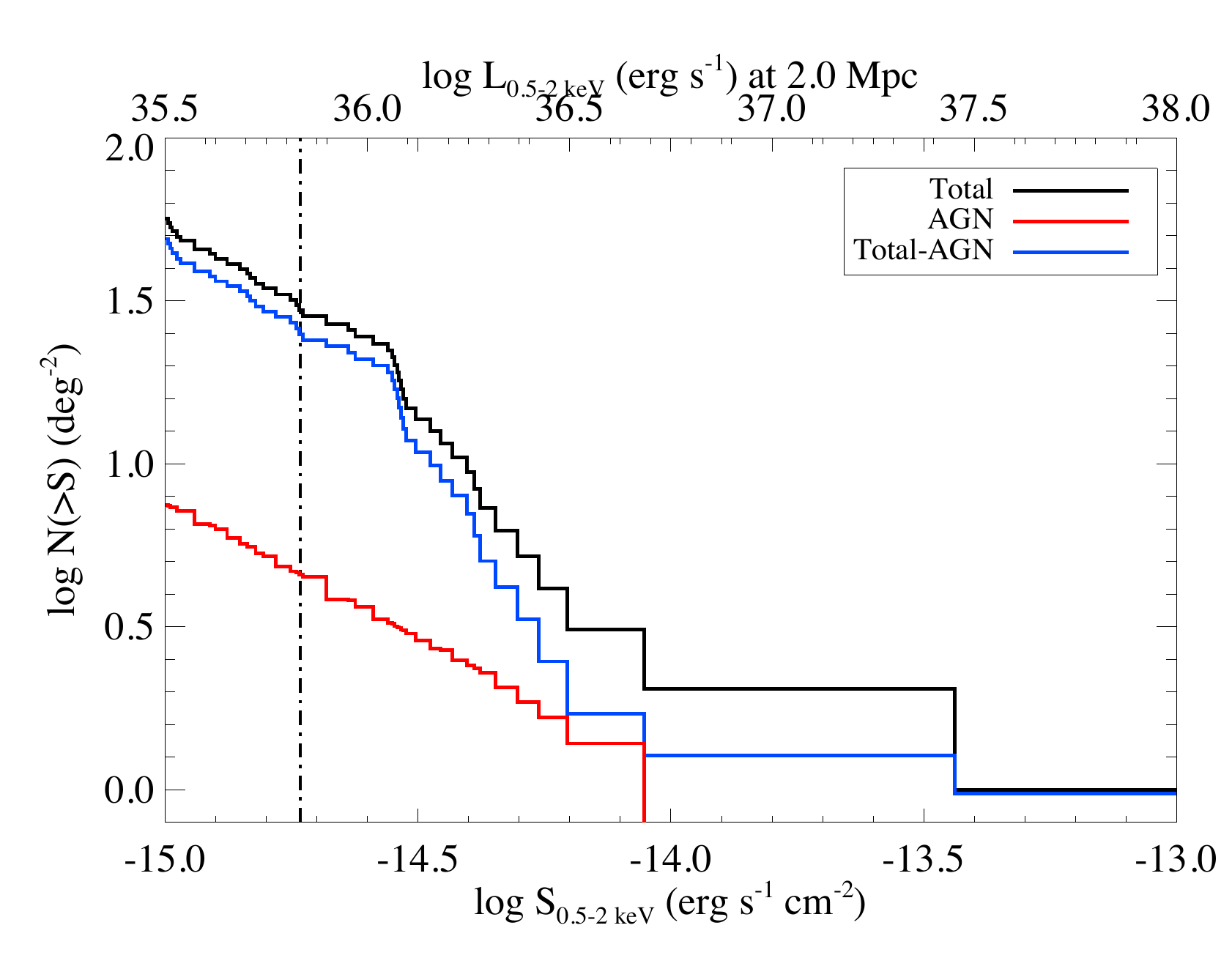} &
\includegraphics[width=0.48\linewidth,clip=true,trim=2.5cm 12.5cm 2cm 2cm]{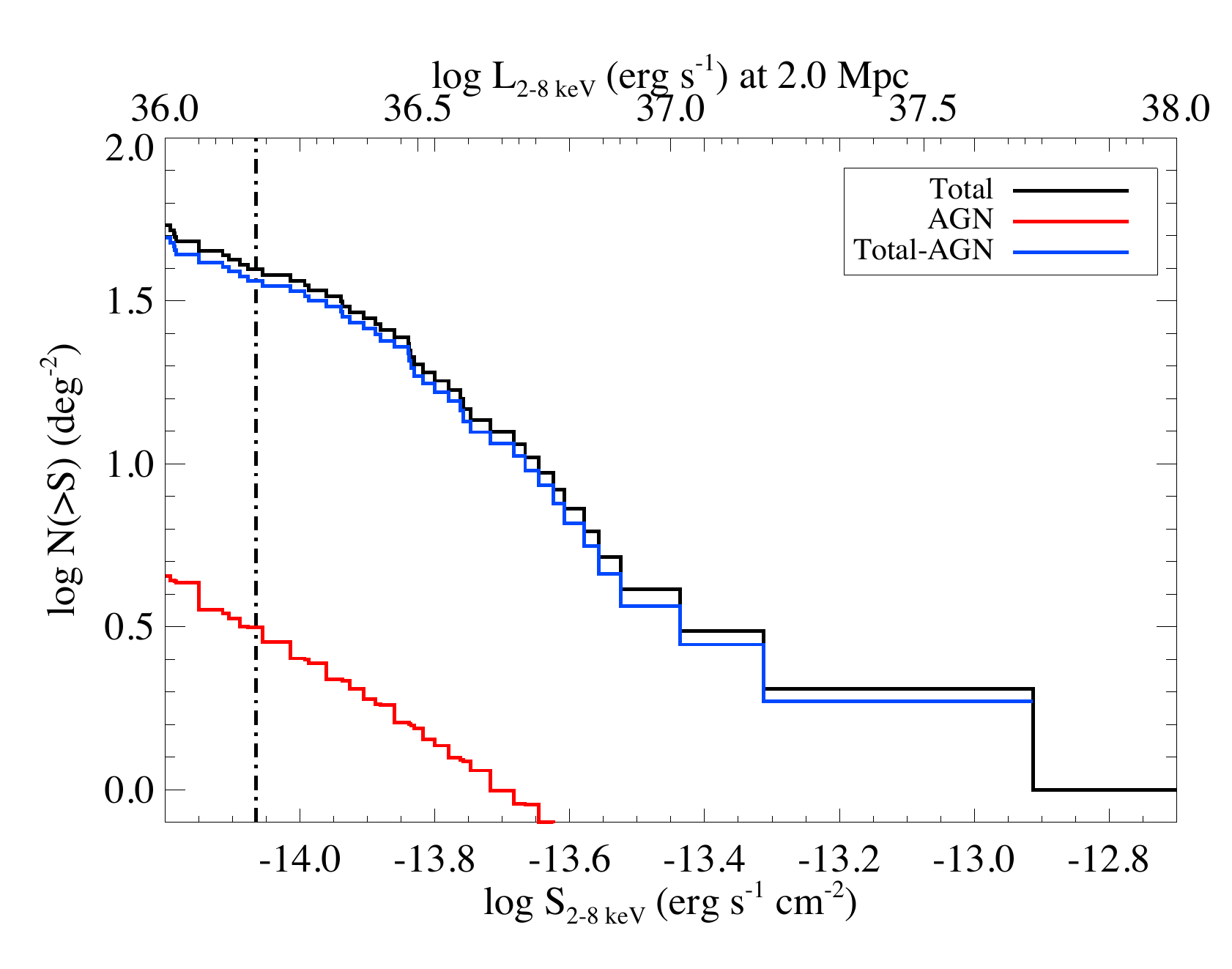} \\
\end{tabular}
\caption{The cumulative 0.5-2 keV (left) and 2-8 keV (right) X-ray \lognlogs distributions for the NGC~300 X-ray sources. The dot-dashed lines indicate the 90\% limiting flux for our survey in each band. The AGN \lognlogs model from \cite{Cappelluti+09} was assumed (red). The AGN-subtracted \lognlogs distribution is shown in blue.} 
\label{XLF_stat}
\end{figure*}

The statistical XLFs are shown in Figure~\ref{XLF_stat}. For each energy band, we model the \lognlogs distribution as a power law. For the 0.5-2 keV distribution, we find a best-fit power law index of $\gamma=2.03\pm0.10$; there is a small bend in the XLF at $\sim1.6\times10^{36}$ \lum, although it is not statistically significant. Our 2-8 keV \lognlogs distribution power law fit yields an index of $\gamma=1.91\pm0.05$.

\begin{figure*}
\centering
\begin{tabular}{cc}
\includegraphics[width=0.48\linewidth,clip=true,trim=2.5cm 12.5cm 2cm 2cm]{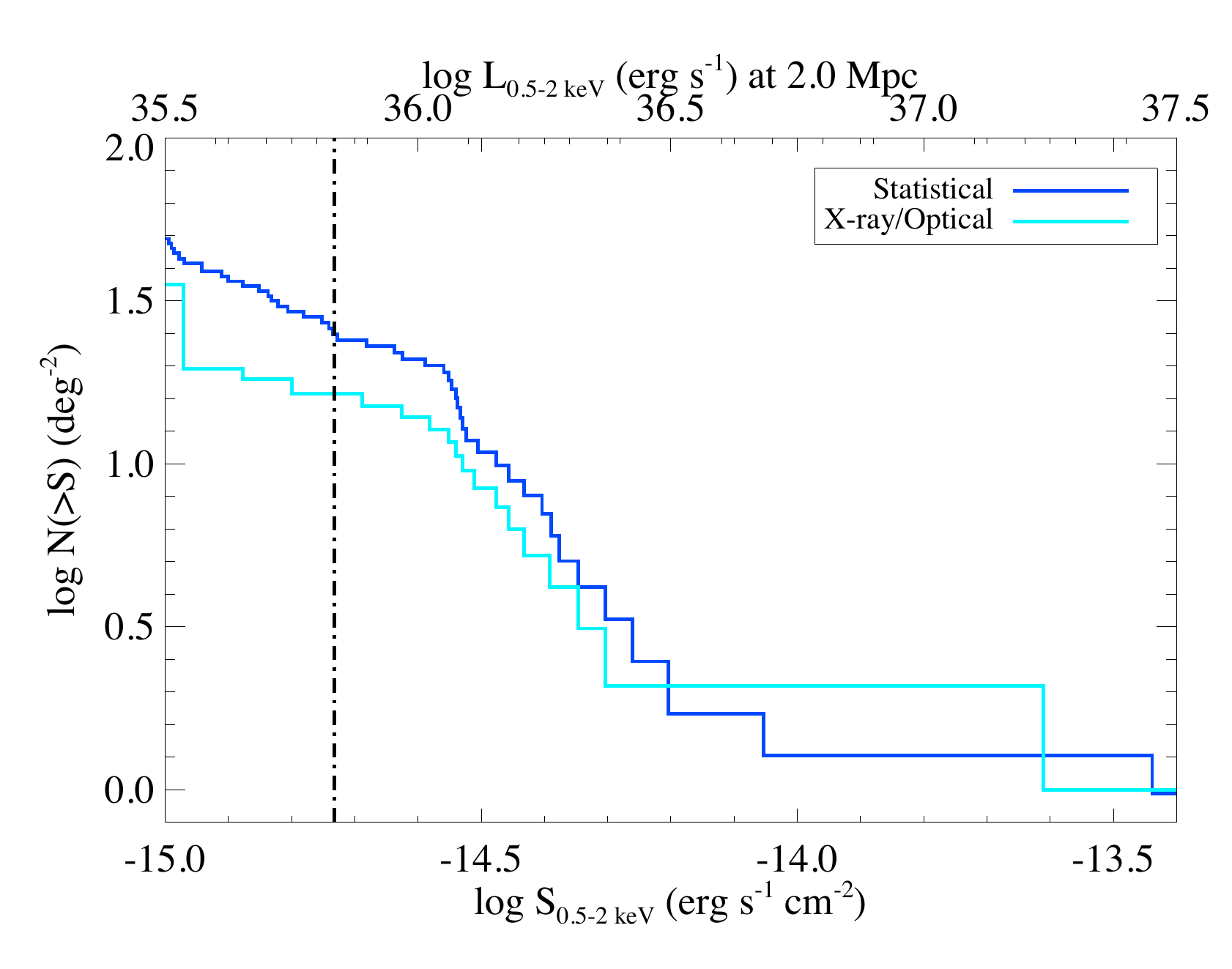} &
\includegraphics[width=0.48\linewidth,clip=true,trim=2.5cm 12.5cm 2cm 2cm]{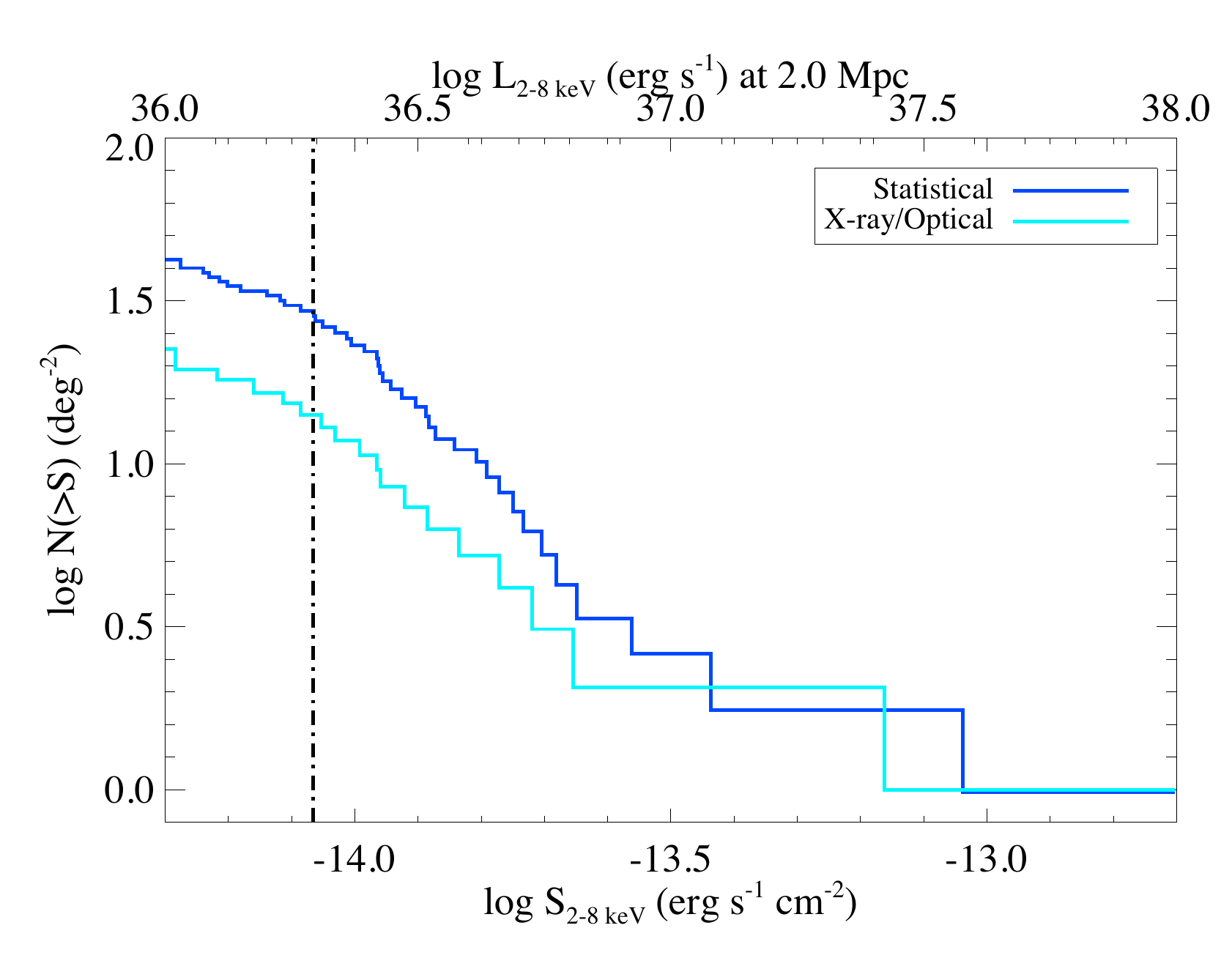} \\
\end{tabular}
\caption{The 0.5-2 keV (left) and 2-8 keV (right) \lognlogs distributions for our background-subtracted total \lognlogs distributions (blue) and our X-ray/optically classified XRB candidates (cyan).}
\label{XLF_ID}
\end{figure*}

We next compute the cumulative \lognlogs distribution for objects classified as XRBs from our X-ray and optical classification scheme, hereafter referred to as our ``X-ray/optically classified XLFs.'' We find the \lognlogs distributions of our X-ray/optically classified XLFs agree with the statistical XLFs, especially in the 2-8 keV energy range (due to less contamination by soft X-ray emitters such as SNRs). Figure~\ref{XLF_ID} shows the X-ray/optically classified XLFs (red) and our statistical XLFs (black) in both soft and hard bands. Due to their similarity in X-ray spectral shape and optical colors, the mis-classifications are likely to be LMXBs that are incorrectly identified as background AGN (or vice versa), while our HMXB sample is likely to have significantly fewer mis-classifications. The XLFs of X-ray/optically classified AGN are also consistent with the \cite{Cappelluti+09} AGN distribution.

We fit the X-ray/optically classified XLF in both the soft and hard bands with power laws. The 0.5-2 keV XLF has a slope $\gamma=2.42\pm0.12$, while the 2-8 keV XLF has $\gamma=1.80\pm0.07$. A low-significance kink is observed in the soft XLF, similar to the feature observed in the soft statistical XLF. Our fits are consistent with those derived from the \XMM point source catalog of NGC~300 above $\sim1.5\times10^{36}$ \lum \citep[see][their Figure~5]{Carpano+05}. All the XLF slopes derived here are typical of spiral galaxies with recent star formation activity \citep{Fabbiano06,Colbert+04,Grimm+03}.

\begin{figure*}
\centering
\includegraphics[width=0.6\linewidth,clip=true,trim=2.5cm 12.5cm 2cm 2cm]{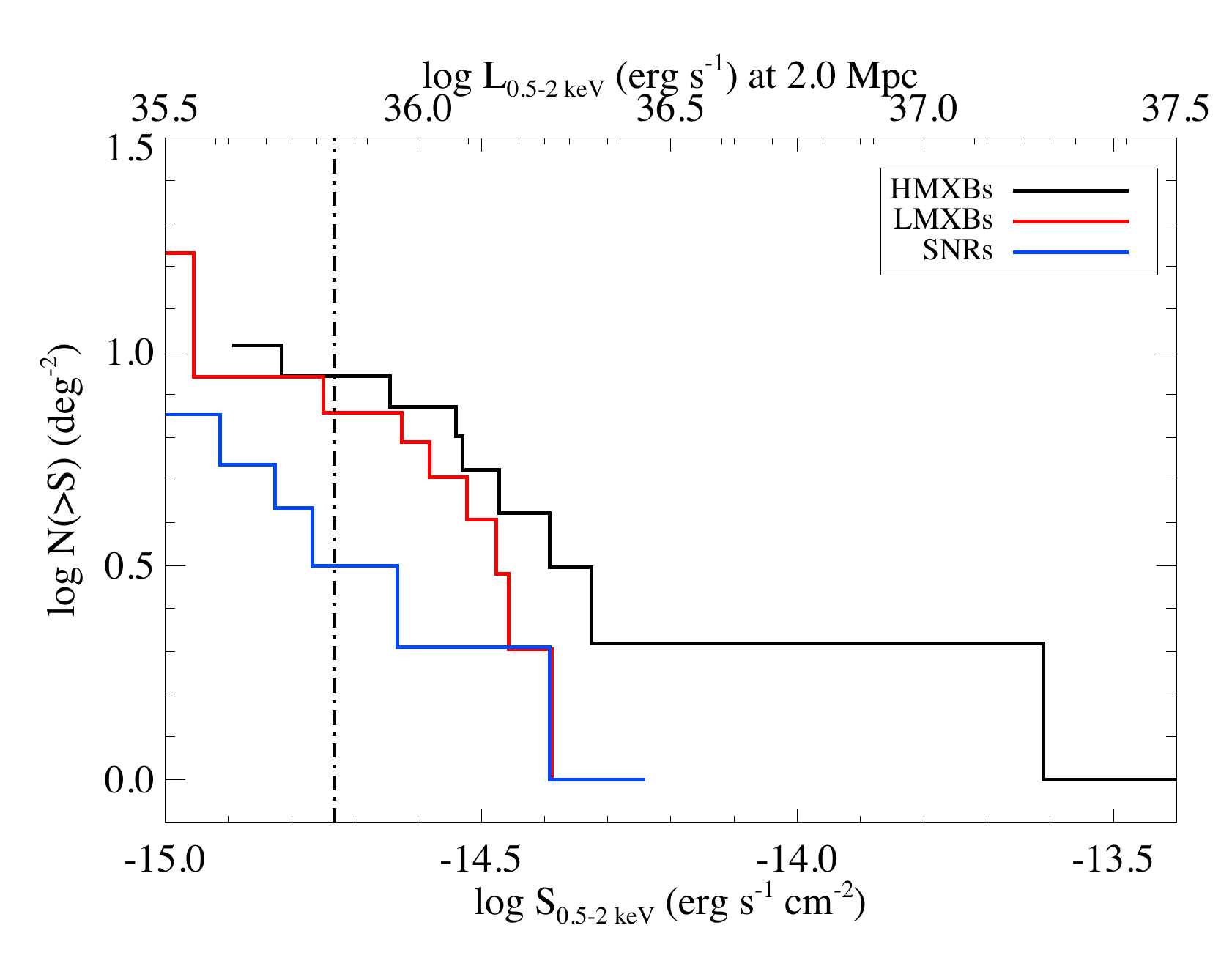}
\caption{The 0.5-2 keV \lognlogs distribution for HMXBs (solid black), LMXBs (red), and SNRs (blue) as classified based on our X-ray and optical observations. The dashed line shows the statistical XRB \lognlogs using the \cite{Cappelluti+09} background AGN distribution. We find HMXBs are the only sub-group without a significant number of sources detected at fainter luminosities, potentially producing the break near $3\times10^{-15}$ \flux.}
\label{XLF_class}
\end{figure*}

To examine whether the bend in the soft XLF near 3$\times10^{-15}$ \flux (equivalent to $\sim1.6\times10^{36}$ \lum at the distance of NGC~300) may be due to a specific X-ray population we constructed \lognlogs distributions for HMXBs, LMXBs, and SNRs separately. Figure~\ref{XLF_class} shows the soft \lognlogs distributions for each of these three groups, as well as the statistical XLF. We find that the HMXBs (black line) are the only population without a significant number of sources at fainter luminosities. It is possible that there exists a substantial population of lower luminosity, wind-fed HMXBs that would require deeper observations to detect at high significance. Alternatively, the change in the XLF slope may be due to XRTs that are observable only during their more luminous outbursting phase (and fall below our detection threshold during quiescence). Table~\ref{XLF_model_fits} summarizes our best-fit power law slopes for each of our XLFs.

\begin{table*}[ht]
\centering
\caption{Power Law Fits to the NGC~300 \lognlogs Distributions}
\begin{tabular}{ccccccc}
\hline \hline
\lognlogs Distribution		& 0.5-2 keV		& 2-8 keV \\
\hline
Statistically XLF		& 2.03$\pm$0.10	& 1.91$\pm$0.05 \\
X-ray/optically Classified XLF	& 2.42$\pm$0.12	& 1.80$\pm$0.07 \\
X-ray/optical XLF, $r<2.7$ kpc		& 1.68$\pm$0.31	& 1.63$\pm$0.18 \\
X-ray/optical XLF, $2.7<r<5.4$ kpc	& 2.65$\pm$0.55	& 1.23$\pm$0.07 \\
HMXBs					& 1.86$\pm$0.19	& 1.56$\pm$0.09 \\
LMXBs					& 2.73$\pm$0.30	& 1.69$\pm$0.29 \\
SNRs					& 1.04$\pm$0.01	& N/A \\
\hline \hline
\end{tabular}
\label{XLF_model_fits}
\end{table*}

\subsection{Radial \lognlogs Distributions}
\begin{figure*}
\centering
\begin{tabular}{cc}
\includegraphics[width=0.48\linewidth,clip=true,trim=2.5cm 12.5cm 1cm 2cm]{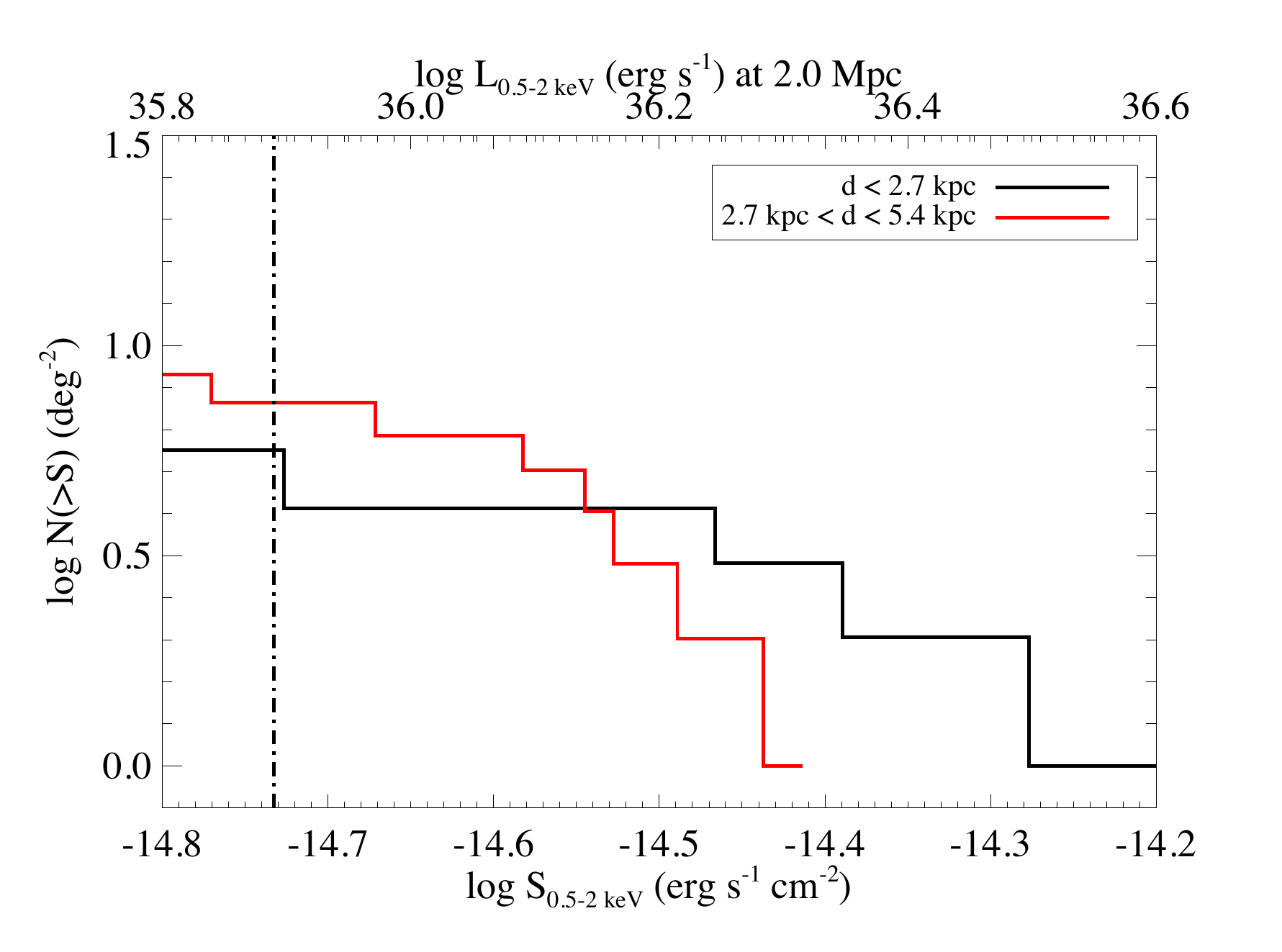} &
\includegraphics[width=0.48\linewidth,clip=true,trim=2.5cm 12.5cm 1cm 2cm]{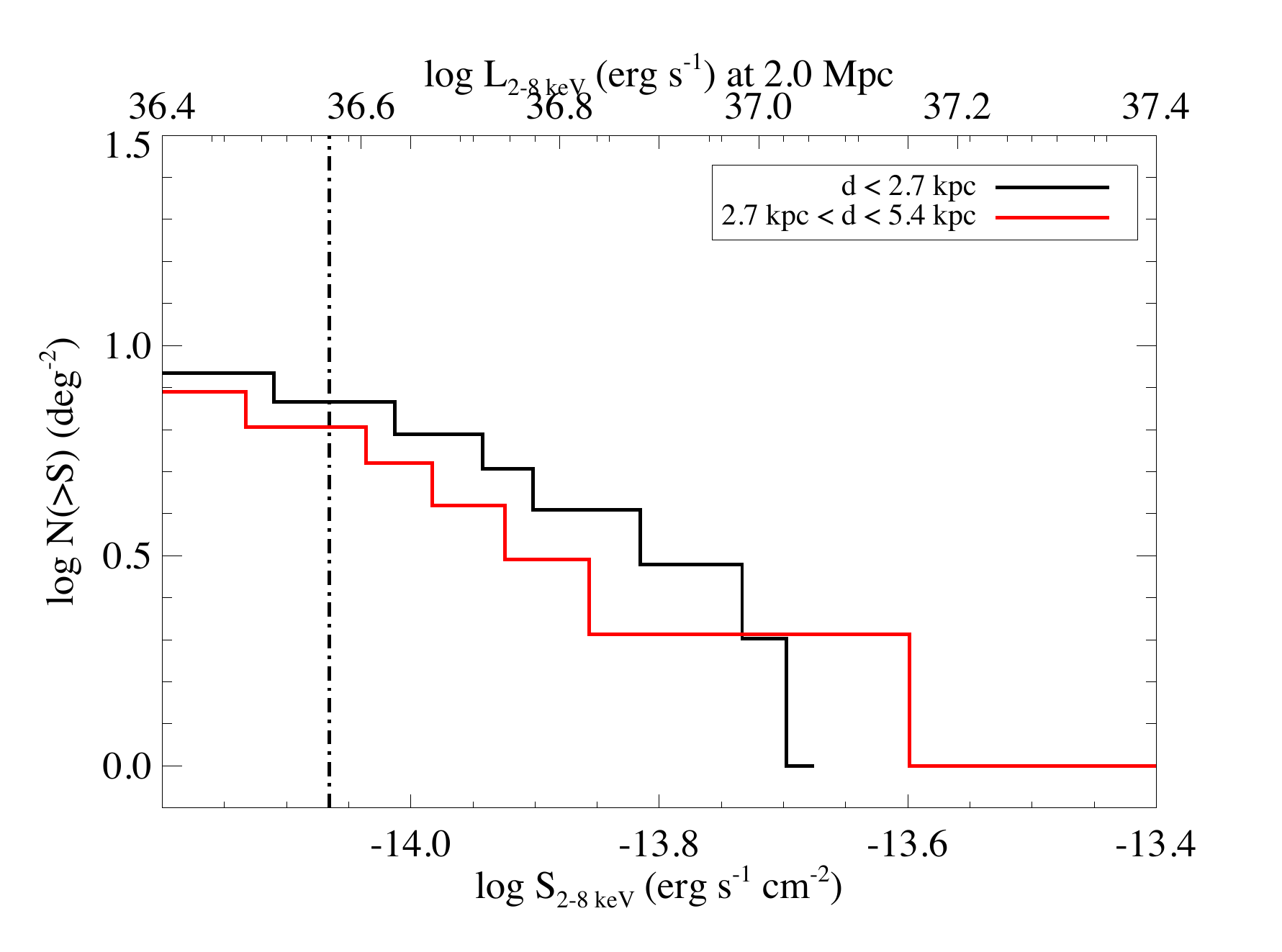} \\
\end{tabular}
\caption{The 0.5-2 keV (left) and 2-8 keV (right) \lognlogs distributions for X-ray/optically identified XRBs detected within the inner 2.7 kpc (black) and outer 2.7-5.4 kpc (red) regions.} 
\label{XLF_radial}
\end{figure*}

We next searched for radial trends in our NGC~300 X-ray source population. We divided our X-ray/optically classified XRB candidates into two radial bins: an ``inner'' region ($d<2.7$ kpc) and ``outer'' region (2.7-5.4 kpc) so that we could directly compare our \lognlogs distributions to the radial SFHs measured by \cite{Gogarten+10}. Using the SFHs at different epochs \citep{Gogarten+10}, we assign an approximate, weighted average age to each radial bin. The radial \lognlogs distributions in both the soft and hard bands are shown in Figure~\ref{XLF_radial}.

\begin{figure*}
\centering
\includegraphics[width=0.7\linewidth,clip=true,trim=2.5cm 12.5cm 2cm 3cm]{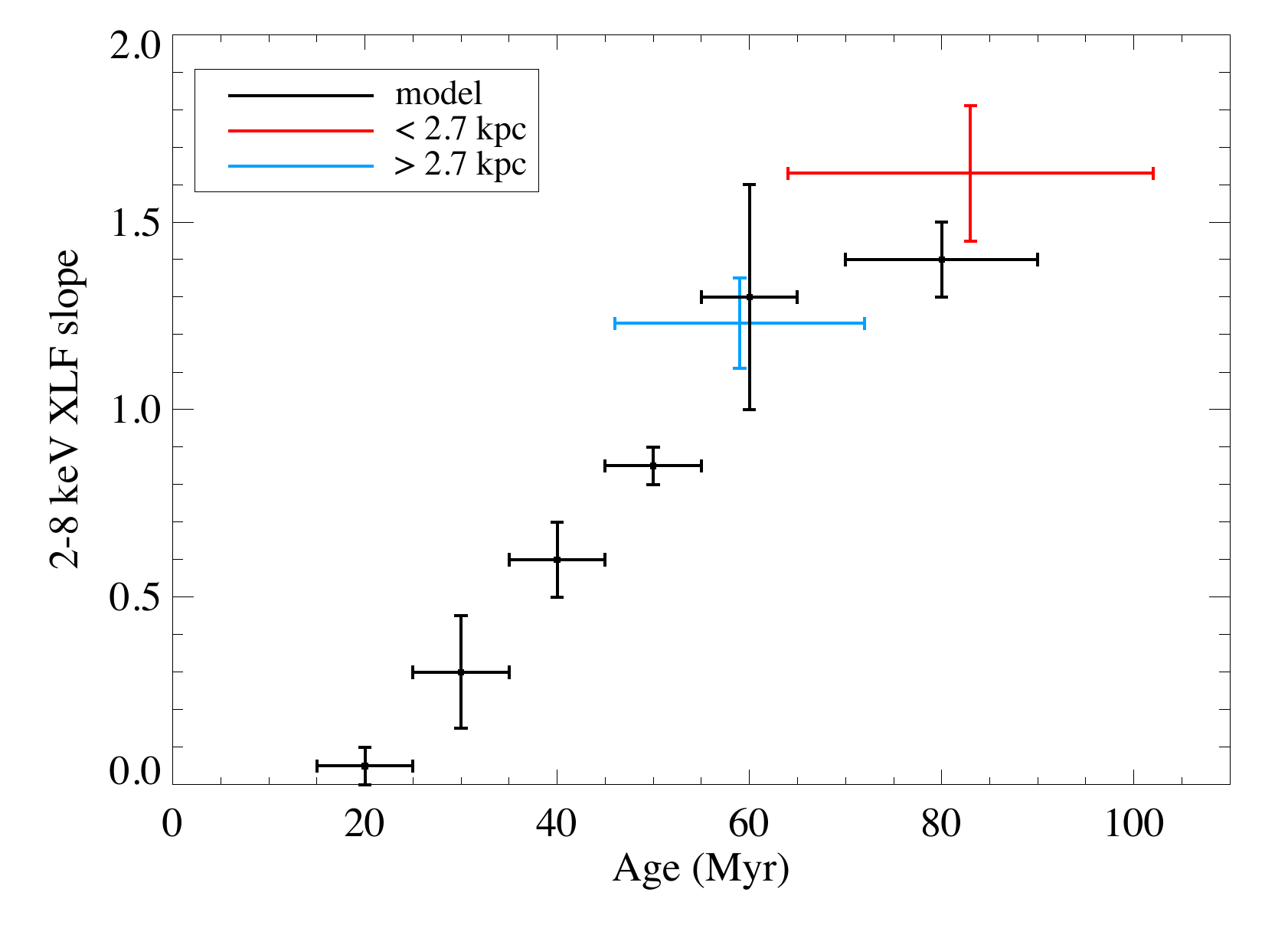}
\caption{The slope of the cumulative 2-8 keV XLF as a function of stellar age. Black points indicate model predictions. The red cross indicates our data for radial bins less than 2.7 kpc, while the blue cross shows our data for X-ray sources from 2.7 - 5.4 kpc. Our combined \Chandra/\HST data of NGC~300 agree very well with predictions, and further analysis of the remaining CLV galaxies will provide further tests of these model predictions.}
\label{age_slope_radial}
\end{figure*}

We observe a preliminary correlation towards flatter 2-8 keV XLF slope with increasing galactocentric radius and younger average age of the underlying stellar population. This trend is not observed in the soft band, which may suffer more severely from contamination from SNRs or foreground stars or from obscuration. We therefore consider the 2-8 keV XLF a more reliable diagnostic of the XRB population. In Figure~\ref{age_slope_radial}, we compare our radial XLF/age data to model predictions \citep{Eracleous+06}. X-ray sources detected at smaller galactocentric radii ($<$2.7 kpc) show an average older age, $\sim$83 Myr, and a steeper XLF slope ($\gamma\sim1.6$) than those sources found in the outer regions of the NGC~300 disk (59 Myr in age with $\gamma\sim1.2$). 

Our X-ray/optical data of NGC~300 thus constitutes one of the first tests of such model predictions relating the nature of the X-ray sources to the underlying stellar population of a galaxy; additional XLF-SFH measurements will become available when the remaining four CLVS galaxies (NGC~404, NGC~55, NGC~2403, and NGC~4214) are analyzed in upcoming works.

\section{Summary and Conclusions}\label{end}
We have presented the X-ray point source catalog for a new, deep \Chandra observation of NGC~300 as part of the \Chandra Local Volume Survey. A total of 95 sources were detected in NGC~300 down to a 90\% limiting 0.35-8 keV luminosity of $\sim1.5\times10^{36}$ \lum. By comparing our X-ray source catalog to earlier observations of NGC~300 performed by \XMM, we were able to place long-term variability constraints on each source and identify 25 XRT candidates. We fit the radial X-ray source distribution to an exponential profile and were able to derive a scale length of 1.7$\pm$0.2 kpc, in good agreement with measurements taken at other wavelengths. We evaluated hardness ratios for each of our X-ray sources, and performed spectral fitting for those sources with more than 50 counts.

We utilized nine deep {\it Hubble Space Telescope} fields to search for optical counterparts to each of our X-ray sources; 32 of our X-ray sources were observed in at least one of the overlapping \HST fields. We developed a new source classification method which combined both X-ray and optical data to classify each source as a HMXB, LMXB, or background AGN, and have made our IDL routine which automates our source classification technique publicly available. We classified 47 sources as background AGN, 26 as HMXBs, 11 as LMXBs, 10 as SNRs, and one as a foreground star.

The cumulative \lognlogs distributions were presented in both 0.5-2 and 2-8 keV bands and found to be consistent with those of other spiral galaxies with a dominant HMXB population. Additionally, we divided our X-ray sources into two radial bins and modeled the radially-resolved \lognlogs distributions with power laws. Using the SFHs of \cite{Gogarten+10}, we were able to assign an average to the underlying stellar populations, and we find the measured XLF slopes and stellar population ages to be in good agreement with model predictions. The remaining four CLVS galaxies (NGC~55, NGC~404, NGC~2403, and NGC~4214) will provide additional constraints on XLF-age models with upcoming works.

\acknowledgements
Support for this work was provided by the National Aeronautics and Space Administration through Chandra Award Number G01-12118X issued by the Chandra X-ray Observatory Center, which is operated by the Smithsonian Astrophysical Observatory for and on behalf of the National Aeronautics Space Administration under contract NAS8-03060. T.J.G and P.P.P acknowledge support under NASA contract NAS8-03060. This research has made use of the NASA/IPAC Extragalactic Database (NED) which is operated by the Jet Propulsion Laboratory, California Institute of Technology, under contract with the National Aeronautics and Space Administration. 

\bibliography{apjmnemonic,ms}
\bibliographystyle{apj}
                                                                                             
\end{document}